\documentclass[a4paper,11pt]{article}
\pdfoutput=1
\usepackage{jcappub}
\usepackage[latin1]{inputenc}
\usepackage{amsmath}
\usepackage{amsfonts}
\usepackage{hyperref}
\hypersetup{colorlinks=True,citecolor=blue}
\usepackage{amssymb}
\usepackage{graphicx}
\usepackage{subfigure}

\usepackage[usenames,dvipsnames,svgnames,table]{xcolor}

\newcommand{\beq}{\begin{equation}}
\newcommand{\eeq}{\end{equation}}
\newcommand{\beqry}{\begin{eqnarray}}
\newcommand{\eeqry}{\end{eqnarray}}
\newcommand{\hn}{\hat{n}}

\begin{document}

\title{Isotropy-Violation Diagnostics for $B$-mode Polarization Foregrounds to the Cosmic Microwave Background}

\author[]{Aditya Rotti and}
\author[]{Kevin Huffenberger}
\affiliation[]{Florida State University, Tallahassee, FL 32306, USA}

\emailAdd{adityarotti@gmail.com}
\emailAdd{khuffenberger@fsu.edu}

\abstract{
  \noindent Isotropy-violation statistics can highlight polarized galactic foregrounds that contaminate primordial $B$-modes in the Cosmic Microwave Background (CMB).  We propose a particular isotropy-violation test and apply it to polarized Planck 353 GHz data, constructing a map that indicates $B$-mode foreground dust power over the sky.   We build our main isotropy test in harmonic space via the bipolar spherical harmonic basis, and our method helps us to identify the least-contaminated directions. By this measure, there are regions of low foreground in and around the BICEP field, near the South Galactic Pole, and in the Northern Galactic Hemisphere.  There is also a possible foreground feature in the BICEP field.  {We compare our results to those based on the local power spectrum, which is computed on discs using a version of the method of Planck Int.~XXX (2016).}  The discs method is closely related to our isotropy-violation diagnostic.  We pay special care to the treatment of noise, including chance correlations with the foregrounds.
Currently we use our isotropy tool to assess the cleanest portions of the sky, but in the future such methods will allow isotropy-based null tests for foreground contamination in maps purported to measure primordial $B$-modes, particularly in cases of limited frequency coverage. }
\maketitle

\section{Introduction}
Measurements of the Cosmic Microwave Background (CMB) that aim to detect signatures of primordial $B$-mode polarization require very tight control over foreground contamination, much better than in the case of temperature anisotropies.  While the temperature anisotropies exceed the galactic foreground over much of the sky at 30--200 GHz \citep{wmap_frg,planckdiffusefrgs2015a}, the polarization $B$-mode signal may be smaller than the galactic foreground at all frequencies everywhere on the sky \citep{BICEP2014,bicepplanck, planck-intermediate-xxx}.  We cannot know for sure the size of the $B$-mode signal until we find it; its amplitude is proportional to the ratio of tensor-to-scalar perturbations, and therefore to the unknown energy scale of inflation.

Frequency information helps to distinguish the CMB from foregrounds, but with multiple mechanisms producing galactic emission, it is not easy to separate the components with spectral information from a small number of frequency channels.  For this reason several next generation CMB experiments will employ broad spectral coverage, either singly or in concert with other experiments:
Advanced Atacama Cosmology Telescope, 
SPT-3G, 
POLARBEAR/Simons Array, 
Keck/BICEP3,
CLASS,
PIXIE,
COrE+, 
LiteBIRD, 
CMB-S4, and others{\interfootnotelinepenalty=10000\footnote{
\url{http://www.princeton.edu/act/};
\url{http://pole.uchicago.edu/}; 
\url{http://cosmology.ucsd.edu/simonsarray.html}; 
\url{https://www.cfa.harvard.edu/CMB/bicep3/}; 
CLASS, \cite{CLASS_2014};
PIXIE, \cite{PIXIE_2011};
\url{http://www.core-mission.org/}; 
\url{http://litebird.jp/eng/};
CMB-S4 in DOE P5 report  (\url{http://science.energy.gov/hep/hepap/reports}).
}}.
For every experiment, there is a trade-off between depth and frequency coverage, so our first indications of a primordial $B$-mode signal may come from the deepest single frequency from such experiments.  Hence it is important to develop additional techniques to distinguish the primordial CMB from foregrounds in the case of limited spectral coverage, and to enhance the power of spectral information when it is present.

Many near-term $B$-mode searches seek to identify the portions of the sky with the least contamination from Galactic foregrounds, using analysis of multi-frequency Planck and WMAP maps.  These cleanest portions will then be targeted by ground-based telescopes to hunt for primordial $B$-modes, subject to the constraints of geography and sky accessibility.



The CMB distinguishes itself by its blackbody spectrum, statistical isotropy, and near-Gaussianity.  Since the primordial $B$-mode signatures may be very small, it will always be important to use these to test for galactic contamination.  In this paper we exploit the CMB's statistical isotropy. The foregrounds, as noted by Kamionkowski and Kovetz (2014) \cite{kamionkowski_kovetz_2014}, are by contrast statistically anisotropic, since they are determined by the physics of the galactic disc.  Motivated by this fact, we propose a blind isotropy-violation test as a diagnostic of CMB foreground contamination.

There is no single way to be statistically anisotropic, so to construct blind tests---using no physical model of foreground---we focus on several techniques that study the square of the $B$-mode foreground, a notion we will make more precise. In the ensemble average, information in the statistically isotropic part of the map is confined to the monopole of the squared map, while all the higher multipoles capture the information on the statistically anisotropic components.  We show that analysis of the local power spectrum on discs \citep{planck-intermediate-xxx,Krachmalnicoff2015} can also be thought of as an isotropy test of this type.

We build our most sophisticated estimators using the Bipolar Spherical Harmonic (BipoSH) basis, a convenient means to express deviations from statistical isotropy for two-point statistics on the sphere.  This basis has been extensively used to test for violations of isotropy in CMB maps \citep{wmap7_anomalies, planck2013_I&S, Kumar2015}. They have also been used to model the small but important deviations from isotropy due to instrument beam ellipticities \citep{febecop, Joshi2013, Das2014}.  The BipoSH basis applies to CMB polarization when expressed in terms of the scalar $E$-modes and pseudo-scalar $B$-modes \citep{Basak2006}.

The technique we propose can be used to analyze any foreground, but in this work we focus on galactic dust emission, and employ Planck polarization data at 353 GHz.
Emission from galactic dust is one of the major contaminants to CMB maps, and exceeds the CMB intensity measurement above $\sim 200 $ GHz.  Its polarized component is poorly understood. Analysis of Planck data shows that the power spectrum of the polarized dust emission follows a power law ($C_\ell \propto \ell^\alpha$) with a spectral index $\alpha \approx -2.5$, while the amplitude varies in different portions of the sky \cite{planck-intermediate-xxx}.

In Sec.~\ref{data} we describe the data used in the analysis and the simulations used to estimate the biases, uncertainties, and statistics of various estimators.  In Sec.~\ref{analysis} we discuss the details of the analysis, briefly describing disc-based methods, the bipolar spherical harmonic basis, and the construction of our estimators. In Sec.~\ref{thebestplaces} we discuss what regions of the sky may be the cleanest.  In Sec.~\ref{sec:pam_power_spectrum} we examine our diagnostic's power spectrum, a four-point statistic.  We discuss the results and draw conclusions in Sec.~\ref{conclusions}.


\section{Data and Simulations}\label{data}

We use the Planck polarization maps at 353 GHz from the 2015 Planck data release.\footnote{\url{http://irsa.ipac.caltech.edu/data/Planck/release_2/all-sky-maps/}}  In all analysis the portions of the sky with the strongest galactic emission are masked either using the GAL40 mask provided by the Planck team (leaving 40\% of the sky available), the GAL60 Planck mask, or a $|b|<35^\circ$ band mask.  For Planck 353 GHz polarization data, the CMB is negligible.  Galactic foreground dominates at large scales, while noise dominates at small scales: autocorrelation power in foregrounds and noise is equal for $\ell \approx 150$ for the $f_{\rm sky}=0.6$ mask. We use apodized masks in all our analysis to avoid complications in harmonic space operations arising from sharp boundaries. 


%
%
%
We have developed a Stokes $Q/U$ to $E/B$ pipeline based on the methods of \cite{Ferte2013, Kim2010a}.  We have thoroughly tested our pipeline and find that the residual power from masking-induced $E$ to $B$ leakage of CMB power is at a level comparable to $r=10^{-7}$, well below $r \gtrsim 0.001$ targeted by various current and planned experiments.  We also test that we recover the power law power spectra quantifying the dust polarization emission in $E$-mode and $B$-mode maps and that it is characterized by the parameters specified in Table~1 of \cite{planck-intermediate-xxx} for $f_{\rm sky}=0.4$ and $0.6$. As a further test, we below repeat the exercise of finding power in discs and find results generally consistent with those presented in \cite{planck-intermediate-xxx}.


We require a set of simulations to characterize noise and biases in our foreground estimates. We simulate the CMB maps using the best fit Planck spectra \citep{planck2015spectrum} with tensor to scalar power spectra ratio $r=0$, smoothing with the 353 GHz beams.  (We include lensing $B$-modes.) For noise,  we take into account the $TQU$ covariances for Planck 353 GHz data, yielding spatially varying but uncorrelated noise.  On large angular scales we partially account for correlated noise by rescaling noise maps with an effective window function that is designed to mimic the large scale noise power spectrum in  difference maps between year 1 and year 2 Planck data.  We do not consider this to be an important correction since in all our analysis we restrict to multipoles $\ell > 40$.  For other tests we also generate isotropic Gaussian simulations of polarized foregrounds assuming a power law model for the dust polarization power spectrum: $[ \ell(\ell+1)C_{\ell}/2\pi = D_{\ell} = A   \ell(\ell+1) \ell^\alpha/2\pi$, $A^{BB}/A^{EE}= \gamma ]$, where we use $[A^{EE}_{(\ell=80)} = 51.0 \mu K^2, \alpha = -2.38 , \gamma = 0.48 ]$ while working with the GAL40 mask and $[A^{EE}_{(\ell=80)} = 124.2 \mu K^2 , \alpha = -2.40 , \gamma = 0.54]$ while working with GAL60 mask, parameters given in \cite{planck-intermediate-xxx}.

\section{Methods} \label{analysis}

We model the observed CMB map as a linear combination of CMB, foreground, and noise for each pixel $i$,
\begin{equation}\label{data_model}
  X_i = X_i^{\rm CMB} + F^X_i + n^X_i,
\end{equation}
where $X = T,E,B$.  Here we focus on the $B$-mode component.  We treat the data such that the CMB is statistically isotropic, while the foregrounds are at least partially anisotropic.

We use quadratic statistics, and our basic idea relies on the fact that a map of the variance---or equivalently the ensemble mean squared signal---is a constant for a statistically isotropic signal. Thus the monopole of the mean-squared map has a mixture of information from the statistically isotropic CMB and the isotropic component of the foreground and noise.  As in the power spectrum, it is difficult to disentangle these without knowledge of the amplitude of the primordial $B$-mode spectrum or a thorough understanding of $B$-mode foregrounds (perhaps from multi-frequency information).  However, the higher multipoles of the mean-squared map we may expect to contain information about the anisotropic, spatially varying component of the map, namely the foregrounds.  If they are detected, these modes can then be used to probe foregrounds.  If they are not detected, they can be used to set limits on foreground contamination.

In the following sections we examine the Planck 353 GHz $B$-mode maps with a series of statistics based on the  variance of the map.  
Note that since we always work with mean-subtracted CMB maps, the variance of the map is equivalent to just the square of the map and hence we use these terms interchangeably.
We discuss a set of different estimators and explore their sensitivities and the connection between them.

\subsection{Mean field bias, errors, and effective $r$}

Some foreground estimators implemented in this work are akin to CMB lensing estimators \citep{Okamoto2003, planck2013_lensing} and our analysis strategy is similar. {(We discuss details of this analogy in Appendix~\ref{lensing_analogy}.)} For that reason, our quadratic isotropy estimators---written generically like $\hat E_X$ and computed from a pair of maps--- will have a mean field bias.  The bias is imposed by the mask and anisotropic noise.  These also break the isotropy of the sky, but in a way that we can simulate. We use a set of 1000 simulations to measure these mean biases and also quantify errors in the isotropy estimates.  In all our analysis we work with three combination of simulations: CMB plus noise ($CN$), CMB plus noise plus (isotropic, Gaussian) foregrounds ($CNF$) and foregrounds-only ($F$). 

To evaluate the mean bias, we evaluate $\langle \hat{E}_{CN} \rangle$ from the simulation.  We then duly subtract the bias from all the maps resulting from the respective estimators. {Since here we only cross-correlate year-1 and year-2 Planck maps, the anisotropic noise bias is expected to be absent and the only mean field bias is due to the CMB $B$-mode (consisting only of lensing $B$-modes in our simulations).}

Although we subtract the bias, it is tiny in 353 GHz data since the noise and foregrounds are significantly larger than the CMB $B$-mode.  However this bias will be important for future analysis on maps in which the CMB, noise, and foregrounds are comparable. In that case the proposed estimators can only reconstruct reliably the spatially varying components of the foregrounds.   The monopole will have contribution from both CMB and foregrounds and disentangling them will require more information.

Simulated maps also help us assess the statistics of the bias-corrected estimators, and establish our errors. Specifically we use the expression ${\rm Var}(E_{CN}) = \langle (\hat E_{CN}-\langle \hat{E}_{CN} \rangle)^2 \rangle$ to estimate the variance per pixel.
 We treat the foregrounds as a \textit{fixed} field and do not include the cosmic variance for the foreground in our accounting of the error. Foregrounds do enhance errors through their chance cross-correlations with CMB and noise.  We estimate this with the expression $\langle (\hat E_{CNF}- \hat E_{F} - \langle \hat{E}_{CN} \rangle)^2 \rangle$ to evaluate the variance. Note that by evaluating $\hat E_{CNF}- \hat E_{F}$ before the averaging operation, we take care to remove the statistical auto-correlation for foregrounds in each random simulation. Subtracting foreground auto-power is analogous to the method adopted in \cite{planck-intermediate-xxx} to estimate errors.  Alternatively we can get the same result by cross-correlating ($CNF$) simulations with ($CN$) simulations, picking up chance correlations without the foreground auto-power.

 Estimating noise this way assumes a foreground model that is Gaussian and isotropic.  We can make a more realistic (non-Gaussian and non-isotropic) estimate of the chance correlations between foregrounds and noise by computing our quadratic estimator using the real data map and randomized ($CN$) simulations.  We denote this cross-correlation as Data\,$\otimes\, CN$.
This cross-correlation accounts for two sets of chance correlations: (1) between foreground and noise, and (2) between year-1 noise and year 2 noise.
We expect the correction induced due to missing CMB auto-correlations---the CMB in data and the simulated realization are different---to be negligibly small. Hence this procedure robustly estimates noise in the estimated foreground tracer maps, accounting for the uncertainty contributions of the actual foregrounds without having to model them.  We use these more realistic error estimates to carry out several of the statistical  studies on the estimated maps.

All estimated foreground maps are put in terms of an effective tensor-to-scalar ratio.  They are multiplied by a factor ($f_{\rm eff}$) such that the map returns (on average) the value of the ratio $r$ of power in the tensor to scalar perturbations, for similar analysis carried out on a full sky map composed only of primordial $B$-modes. {This factor is generally evaluated as $f_{\rm eff} = \langle \hat{E}_{C}[r=1]\rangle^{-1}$, calculated on the full sky. The specifics of evaluating this for each estimator vary, and are given in the respective sections below.  In addition to the above rescaling, following \cite{planck-intermediate-xxx}, we extrapolate the results to 150 GHz with the dust spectral energy distribution, rescaling our quadratic statistics by a factor of $0.0395^2$.
The rescaled, estimated foreground maps are hence presented in terms of the effective tensor-to-scalar ratio, $r_{\rm eff}$, at 150 GHz.


\subsection{Pixel space variance statistics}

\label{sec:pixelspace}

All our estimators examine the local variance as a function of position on the sky.  For a concrete analogy, we begin with the simplest way to estimate local variance in real space.  We can make the following cross-correlation (evaluated on patches),
\begin{equation}
\hat \sigma^2_B(\hat n) = \frac{1}{N_{\rm pix}} \sum_{i} (B_{1,i} - \bar B_{1,i})(B_{2,i} - \bar B_{2,i}) \,,
\end{equation}
%
where the sum $i$ is over the $N_{\rm pix}$ pixels in the patch centered on direction $\hat n$, using two different years of data.   For a pure CMB signal, the variance map should be a constant (monopole) map giving $\langle B^2 \rangle$ within the errors.  In real space, some care is required with the subtraction of the mean from the map.  If the monopole of the map is reliable, the foreground mean value contains useful information, and should not be subtracted.  However, data acquisition and map-making may leave the data without a meaningful monopole, making it sensible to subtract the data monopole.   Alternatively, subtracting the mean value of each patch acts like a high-pass filter based on the patch size.

The pixel space operation has the advantage that it is simple and easy to work with disjoint subsets of pixels, and is immune to ringing due to harmonic space operations.  
The disadvantage to this approach is the inability to control the angular scales that contribute (by comparison, \cite{planck-intermediate-xxx} used $\ell = 40$--$370$ only) or filter. Examining the local variance in this way is really a form of isotropy-violation test, as will become clearer in the following sections.   

%



\label{harmonic_space_variance_statistics}

\subsection{Local power spectrum analysis}
\label{sec:planck-disc-analysis}

We can also evaluate local variance from the power spectrum, in analogy to the foreground analysis of \cite{planck-intermediate-xxx}.  The variance in the $B$-mode map in a particular multipole range is given by the following expression,
\begin{equation}
 \hat \sigma_B^2  =  \frac{1}{4\pi} \sum_{\ell=\ell_{\rm min}}^{\ell_{\rm max}} (2 \ell +1 ) \hat{C}^{BB}_\ell. \label{pow_monopole}
\end{equation}
This also corresponds to the value of the monopole, the only non-zero multipole, of the appropriately filtered $\langle B^2 \rangle$ map.  If the variance changes as a function of position, the local variance can be evaluated in terms of a local estimate for the power spectrum.

In \cite{planck-intermediate-xxx} they follow this approach, computing the local $B$-mode power spectrum in disc-shaped patches of 400 deg$^2$ at galactic latitude $|b| > 35^\circ$, fitting the amplitude to a power law to $l(l+1)C_l/2\pi$ for each patch over a restricted range $ 40 < \ell < 370$.  To express in terms of a tensor-to-scalar ratio, they compare the amplitude of the power-law fit to the amplitude of primordial $B$-mode power spectrum (with $r=1$) at $\ell=80$.

This method is really an isotropy test, because they used the direction-dependent power in $B$-mode maps as an indicator of foregrounds.

\begin{figure}
\centering
\subfigure[Local power disc analysis]{\label{pda_data}\includegraphics[width=0.49\columnwidth]{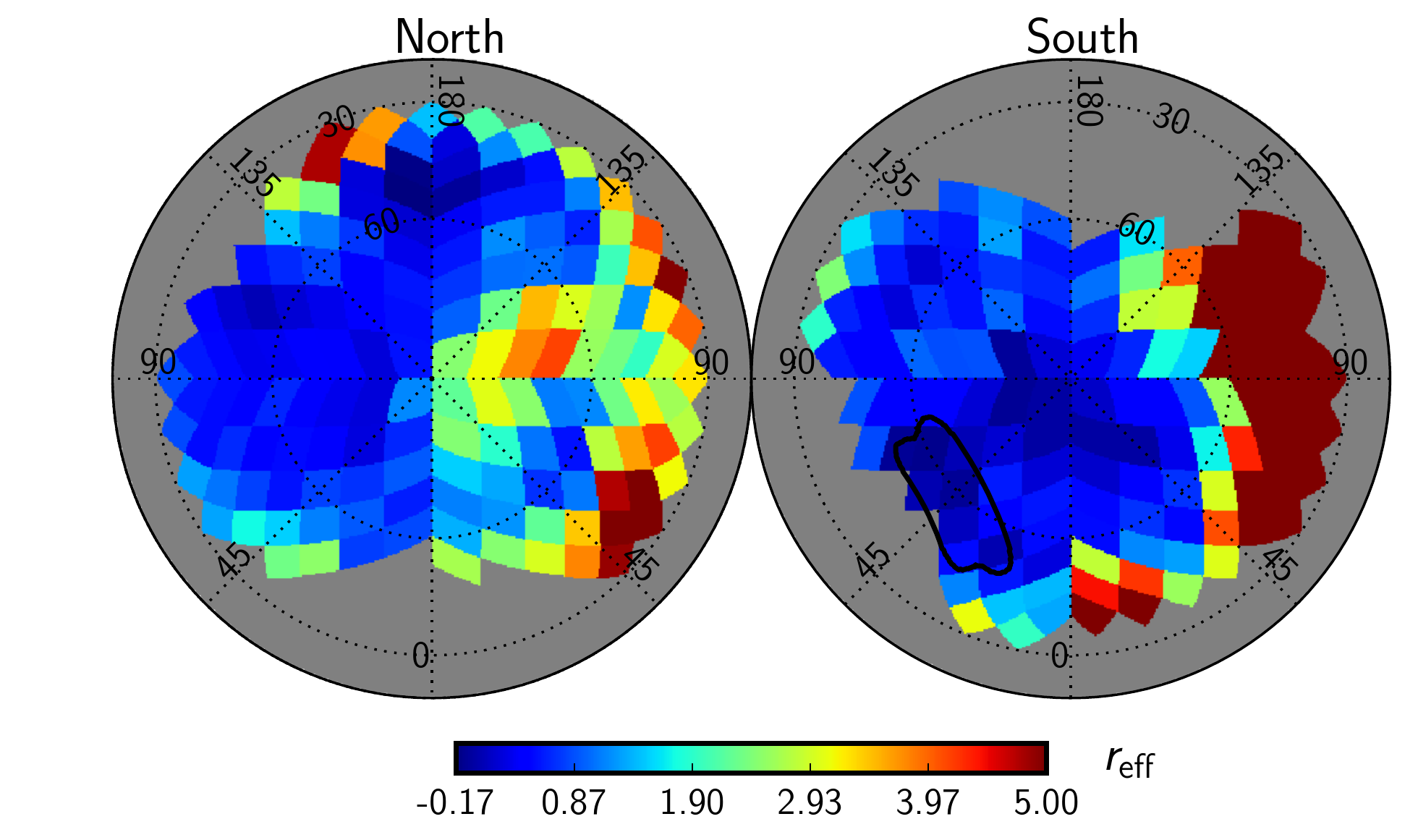}}
\subfigure[Error map (Data $\otimes$ $CN$)]{\label{pda_ocn_err}\includegraphics[width=0.49\columnwidth]{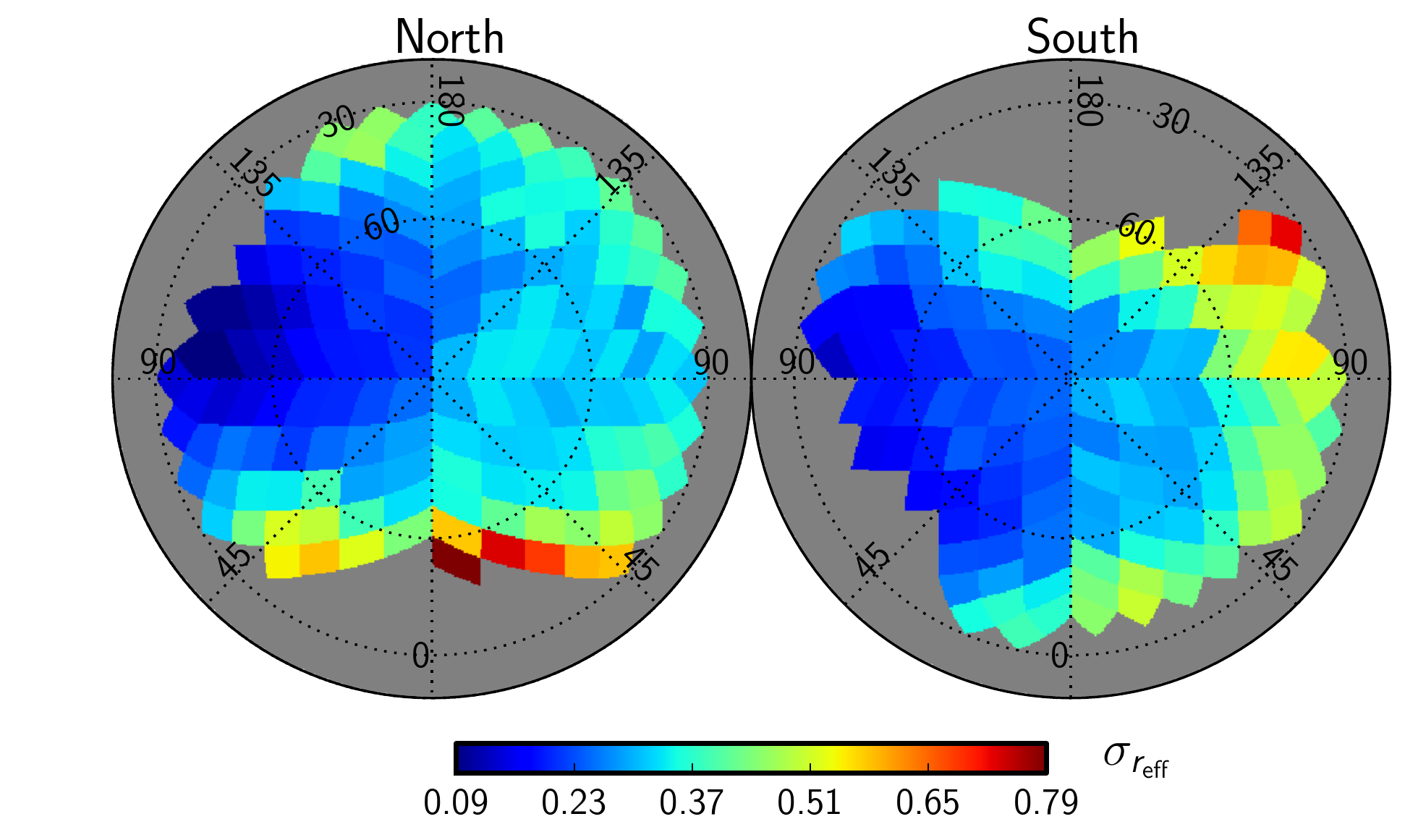}}
\caption{ Local $B$-mode power in $11.3^\circ$ discs (plus $2^\circ$ additional apodization) estimated from 353 GHz Planck maps, extrapolated to 150 GHz, and scaled to an effective tensor-to-scalar ratio. Discs are centered on pixels of an $N_{\rm side}=8$ map.  The North and South Galactic hemispheres are displayed separately.  \textit{Left:} Power in the $B$-mode map as a function of direction. This map closely resembles the Planck team's analysis \cite[][Fig.~8]{planck-intermediate-xxx}. {The color scale is saturated at $\rm r_{\rm eff }=5$ to highlight the key features in the maps. At native scale the maximum in the map is at $\rm r_{\rm eff} =15.28 $.} \textit{Right:} Error per pixel, estimated by cross-correlating data with randomized simulations.}
\label{fig:planck-disc-analysis}
\end{figure}
To test our Q/U to E/B pipeline, here we make a similar analysis, but with slight changes to make it more similar to our later estimators.  Full details of the procedure are in \cite{planck-intermediate-xxx} and here we highlight only the differences.  While they cross-correlated Planck maps based on distinct sets of detectors,  here we perform our analysis by cross-correlating year 1 and year 2 maps to reduce  detector cross-talk. They performed a $Q/U$ to $E/B$ translation on each sky patch defined by the local disc  mask.  Here we  use the GAL60 mask while translating from Q/U to E/B maps, which needs to be done only once.   Following this step, we apply the disc masks to these $B$-mode maps to evaluate the local power spectrum. While  \cite{planck-intermediate-xxx} uses the \textit{MASTER} algorithm \citep{Hivon2002} to correct for power  loss due to masking, here we simply correct for the power by rescaling the spectra by the inverse sky fraction  $f_{sky}^{-1}$. In addition, here we do not perform any power law fits to the local power spectrum, instead we evaluate the total power using Eq.~\ref{pow_monopole} in the multipole range $\ell \in [40,370]$.

To compare with primordial CMB $B$-modes, \cite{planck-intermediate-xxx} report the value of the fitted power law at $\ell = 80$, divided by a $r=1$ model for inflationary $B$-modes.  This is slightly different than our method of rescaling.  Here we sum up the power over $\ell \in [40,370]$ and normalize by the factor 
\begin{equation}f_{\rm eff} = \left[ \sum_{\ell_{min}}^{\ell_{max}} \frac{2\ell+1}{4\pi} C^{BB}_{\ell}(r=1) \right]^{-1}\,.
\label{rescale_disc_analysis}
\end{equation}
Here the factor is {$f_{\rm eff} = 13.073$}.  The map recovered from our analysis, seen in Fig.~\ref{pda_data}, closely resembles the map seen in Fig. 8 of \cite{planck-intermediate-xxx}, though there are subtle differences which may be attributed to this modified analysis procedure and using other data map sets. 


{ As we noted,  foregrounds enhance errors through their chance cross-correlations with CMB and noise. In our robust methodology to account for these chance correlations, we estimate errors by cross correlating the data maps with the (CN) simulations.
  The resultant error map for the disc analysis, depicted in Fig.~\ref{pda_ocn_err}, shows that the spatial variation of errors resembles with inverse detector hit-count map and is further modulated by regions of high foregrounds, as expected. In \cite{planck-intermediate-xxx}, they estimate errors by evaluating the variance of the cross data power spectrum ($\sigma^2_{C_{\ell}}=2 C_{\ell}^2/(2 \ell+1)$), and subtract the error contributed by foreground auto correlation, estimated from the fitted empirical power law foreground model.  Subtracting the auto correlation is important to ensure that the foreground auto-correlations are not accounted for in the error budget. 
  Since the pixels are not all independent one should take caution while interpreting these error maps.}

This method implicitly assumes that the foreground is statistically isotropic within each disc.  The MASTER technique \citep{Hivon2002} robustly corrects the power spectrum measured on masked sky, but assumes that the sky is isotropic.  Hence measuring the power spectrum measurement from discs may not completely characterizes the region under study. Only in the limit of very small discs may it be safe to assume that the foreground field is statistically isotropic. Finally, the disc-based methods are limited in resolution by the size of the disc mask.  The optimal disc size is not clear \textit{a priori}  and may even vary as a function of sky position.

 In the following section we discuss harmonic space based estimators which overcome these limitations, and also allow for optimal filtering.

\subsection{Bipolar spherical harmonic analysis}
\label{biposh-analysis}
The position-dependent variance can also be written in harmonic space, using the  Bipolar Spherical Harmonic (BipoSH) basis \cite{hajian2003,hajian2005}.  The coefficients of this basis are analogous to the angular power spectrum $C_{\ell}$, but are generalized to non-isotropic skies where the harmonic space covariance is non-diagonal. The most general two-point correlation function can be expressed as,
\beq
C(\hn_{1},\hn_{2}) = \sum_{LM \ell_1 \ell_2} A^{LM}_{\ell_1 \ell_2} \lbrace Y_{\ell_1}(\hn_{1}) \otimes Y_{\ell_2}(\hn_{2}) \rbrace_{LM} \,,
\eeq
\noindent where the new harmonic spectra $A^{LM}_{\ell_1 \ell_2}$, like $C_\ell$, are related to the ensemble average of a pair of spherical harmonic coefficients, and in fact 
$A^{00}_{\ell \ell} = (-1)^{\ell} \sqrt{2 \ell +1} C_{\ell}$ 
is a special case. The basis functions $\{ Y_{\ell_1} \otimes Y_{\ell_2}  \}_{LM}$ are again related to a pair of spherical harmonic functions \citep{varshalovich}.

This BipoSH basis is a convenient language to describe deviations from statistical isotropy on the sky.  For the statistically isotropic signal, only the $A^{00}_{\ell \ell}$ are non-vanishing, and the rest are zero.  Hence searching for non-zero power in the $L > 0$ BipoSH spectra forms the basic criteria to test for deviations from isotropy; they should be consistent with zero within errors for the statistically isotropic case.  


Although they are spectra, the $A^{LM}_{\ell_1 \ell_2}$ coefficients share some mathematical properties with spherical harmonic coefficients: for a fixed $\ell_1, \ell_2$, the $A^{LM}_{\dots}$ can be thought of as the spherical harmonic coefficient of some map, the \textit{BipoSH map}. Therefore these spectra carry spectro-spatial information of the sky. The monopole of such a map, $A^{00}_{\dots} \propto C_{\ell},$ encodes information about the isotropic component of the map while the higher multipoles $A^{LM}_{\dots}$ encode information on the anisotropic component of the map. Hence searches for statistically significant power in higher multipoles of these BipoSH maps can indicate a deviation from statistical isotropy, and in our case, evidence of foreground contamination.



The BipoSH spectra from a CMB map $X$, can be evaluated using the following estimator,
\begin{equation} \label{biposh_est}
\hat{A}^{LM}_{\ell_1 \ell_2} = \sum_{m_1 m_2} a_{\ell_1 m_1} a_{\ell_2 m_2} \mathcal{C}^{LM}_{\ell_1 m_1 \ell_2 m_2} \,,
\end{equation}
where $\mathcal{C}^{LM}_{\ell_1 m_1 \ell_2 m_2}$ denote the Clebsch-Gordon coefficients and $a_{\ell m}$  are the spherical harmonic coefficients of the maps, given by the following expression,
\begin{equation}
a_{\ell m} =  \int X(\hat{n}) Y_{\ell m}(\hat{n}) ~d \hat{n} \,.
\end{equation}
For brevity, here onwards we follow the notation that variables with $\tilde{Z}$ denote estimates from masked skies, while variables without tildes denote estimates made from full skies.

The estimator for the BipoSH spectra given in Eq.~\ref{biposh_est} is unbiased while working with full sky maps.  However in the presence of a mask, these spectra are biased. To correct for this, it is necessary to estimate the bias $\mathcal{B}$ by evaluating the estimator on accurately-simulated and masked CMB skies (including noise),
\begin{equation} \label{biposh_mask_bias}
\tilde {\mathcal{B}}^{LM}_{\ell_1 \ell_2} = \sum_{m_1 m_2} \langle \tilde a^{}_{\ell_1 m_1} \tilde a^{}_{\ell_2 m_2} \rangle_{\rm sims} \mathcal{C}^{LM}_{\ell_1 m_1 \ell_2 m_2} \,,
\end{equation}
where the $\langle \cdots \rangle_{\rm sims}$ denotes an average over the ensemble of simulations.

Finally we work with the bias-corrected BipoSH spectra given by the following expression,
\begin{equation} \label{bias_corrected_biposh}
\tilde A^{LM}_{\ell_1 \ell_2} = \tilde D^{LM}_{\ell_1 \ell_2}  - \tilde{\mathcal{B}}^{LM}_{\ell_1 \ell_2} \,,
\end{equation}
where $\tilde D$ denotes the BipoSH spectra measured from masked data maps.  {Because of the relatively localized type of statistic we build below, we do not need a mode-coupling correction or even a correction for sky area.  (We demonstrate in Appendix \ref{sec:mask_dependence} that the results do not depend on the mask in this way.)}

%


\subsubsection{Distinguishing CMB from foregrounds}
Given the data model in Eq.~\ref{data_model}, the BipoSH spectra derived from data maps contain the following components,
\begin{eqnarray} \label{data_biposh_components}
D^{LM}_{\ell_1 \ell_2} &=& X^{LM}_{\ell_1 \ell_2} \delta_{L0} \delta_{M0} + F^{LM}_{\ell_1 \ell_2} + N^{LM}_{\ell_1 \ell_2}  \nonumber \\ &+& [XN]^{LM}_{\ell_1 \ell_2} + [XF]^{LM}_{\ell_1 \ell_2} + [NF]^{LM}_{\ell_1 \ell_2}  \,,
\end{eqnarray} 
where  $X$, $F$ and $N$ denote the BipoSH spectra evaluated from the auto-correlation of the components CMB, foregrounds and noise respectively, while the terms in $[\cdots]$ denote the BipoSH spectra resulting from the cross-correlation of these different components and are expected to vanish on average. Note that though the cross-correlation terms do not affect the ensemble mean, they do contribute to the variance on the measured BipoSH spectra.

The bias $\tilde{\mathcal{B}}$, which is subtracted from the data BipoSH spectra, generally accounts for the anisotropic noise model and the known CMB which in the present study only includes the weak lensing induced $B$-modes,
\begin{equation} \label{bias_terms}
\tilde{\mathcal{B}}^{LM}_{\ell_1 \ell_2} = \langle \tilde{X}^{LM}_{\ell_1 \ell_2} \rangle + \langle \tilde{N}^{LM}_{\ell_1 \ell_2} \rangle \,.
\end{equation}
On substituting the masked form of Eq.~\ref{data_biposh_components} and Eq.~\ref{bias_terms} in Eq.~\ref{bias_corrected_biposh} and averaging over CMB and noise we are left with the following terms for the BipoSH spectra measured from data,
%
\begin{eqnarray} \label{biposh_f_r}
\tilde A^{LM}_{\ell_1 \ell_2} &=& \tilde D^{LM}_{\ell_1 \ell_2} - \tilde{\mathcal{B}}^{LM}_{\ell_1 \ell_2} \,, \\
&=& \tilde F^{LM}_{\ell_1 \ell_2}  +  \tilde{ \mathcal{R}}^{LM}_{\ell_1 \ell_2} + \tilde{\mathcal{N}}^{LM}_{\ell_1 \ell_2}  \,,  \nonumber\\
\langle \tilde A^{LM}_{\ell_1 \ell_2} \rangle &=& \tilde F^{LM}_{\ell_1 \ell_2}  +  \langle \tilde{\mathcal{R}}^{LM}_{\ell_1 \ell_2} \rangle \,,\label{biposh_f_r2}
\end{eqnarray}
%

where $\tilde{\mathcal{N}} = \tilde{N} - \langle \tilde N \rangle$ denotes the noise fluctuations minus the mean field bias.  Noise fluctuations result from the CMB, instrument, and all the chance cross-correlations between various components of the map.  There is a small residual bias,  $\tilde{\mathcal{R}}$, arising from the fact that we have not accounted for the primordial $B$-mode spectrum in our simulations, as it is unknown. The noise term vanishes in the ensemble average.  We reiterate that in all our analysis we cross correlate year-1 and year-2 data maps, which have independent noise realizations.  Hence the noise bias vanishes on average: $\langle \tilde{N}^{LM}_{\ell_1 \ell_2} \rangle=0$. 

 Written this way, Eq.~\ref{biposh_f_r2} makes clear that the BipoSH spectra measurements provide a direct probe of foregrounds.  Since the foregrounds are a fixed field, their BipoSH spectra do not appear with ensemble averages $\langle \cdots \rangle$.


\subsubsection{Non-optimal estimator of foregrounds squared: $\hat F^2$} 
\label{f2_estimator}
\begin{figure}
\centering
\subfigure[Local power disc analysis]{\label{pda_data2}\includegraphics[width=0.49\columnwidth]{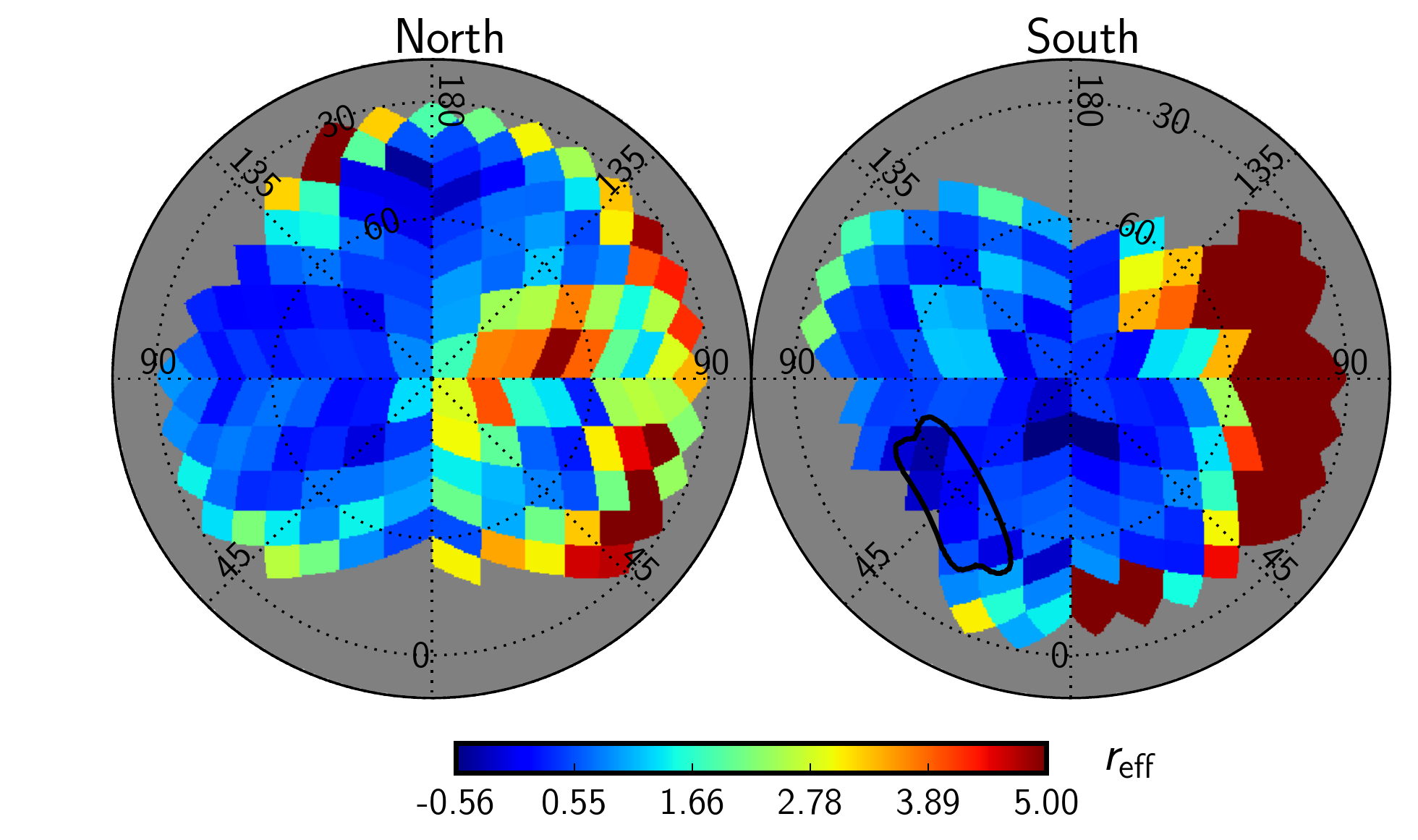}}
\subfigure[$\hat F^2$ smoothed with $9.3^\circ$ disc and downgraded]{\label{pam2pda_nop_data}\includegraphics[width=0.49\columnwidth]{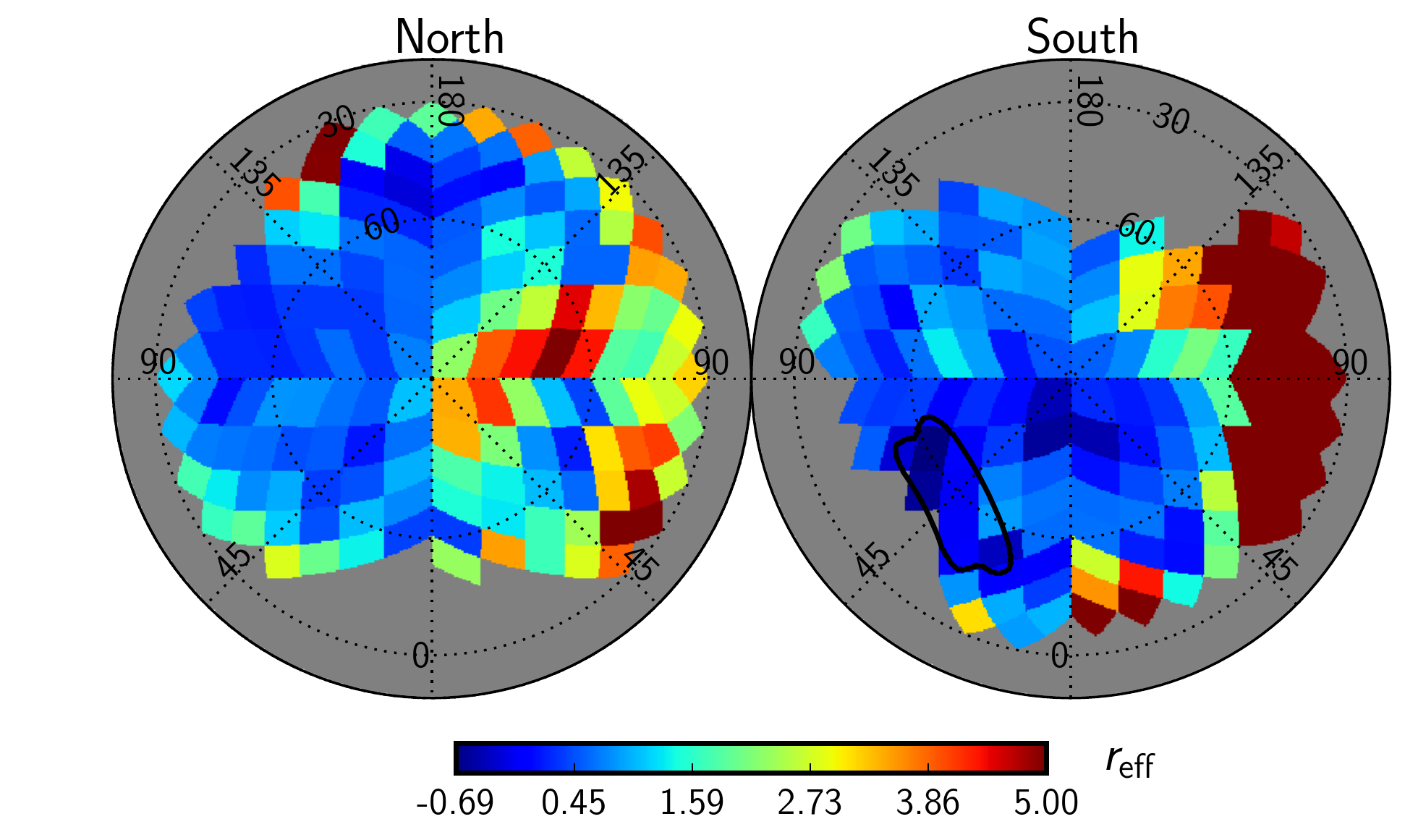}}

\caption{Local $B$-mode power as a function of direction in {$9.3^\circ$} discs ($2^\circ$ additional apodization), estimated from 353 GHz Planck maps, extrapolated to 150 GHz, and scaled to an effective tensor-to-scalar ratio. {The color scale is saturated at $\rm r_{\rm eff }=5$ to highlight the key features in the maps. At native scale the maximum in both the maps is at $\rm r_{\rm eff} \simeq19$.} \textit{Left:} Power estimated from the local disc power spectrum analysis, centered around pixels of an $N_{\rm side}=8$ map.   \textit{Right:} Power estimated from the non-optimal Bipolar Spherical Harmonic estimator $\hat F^2$, smoothed and downgraded in resolution with the same discs. }
\label{fig:planck-disc-analysis}
\end{figure}

We construct linear combinations of the bias-corrected BipoSH spectra to make a map of the foreground squared in $B$-modes, evaluated at a single position on the map,
\beqry \label{nonoptimal_estimator}
\hat F^2(\hat n) &=&  \sum_{LM} P_{LM} Y_{LM}(\hn)  \,, \\
& = & \sum_{LM}  \left[\sum_{\ell_1 \ell_2} \tilde A^{LM}_{\ell_1 \ell_2}\frac{\Pi_{\ell_1}\Pi_{\ell_2}}{\Pi_{L} \sqrt{4\pi}} \mathcal{C}^{L0}_{\ell_1 0 \ell_2 0} \right] Y_{LM}(\hn) \nonumber \,.
\eeqry
where $ \Pi_{\ell} =  \sqrt{2 \ell +1}$. In the ensemble mean, this weighting gives equivalently the local variance or local total power at point $\hat n$.

Assuming that the residual biases $\mathcal{R}$ are small, this estimator provides an unbiased measurement of foregrounds squared: on average the isotropic, primordial $B$-mode contributes only to the monopole.  The bias subtraction step removes this, and so for simulations without foregrounds our procedure returns a null map on average. Therefore any unexpected and statistically significant power measured in this map indicates foreground contamination.

While this estimator offers the advantage of being unbiased, it is non-optimal in the sense that we treat all harmonic modes the same, without regard to their noise level.
Note that it is easy to control the multipoles $\ell_1,\ell_2$ that contribute to the map by restricting the sum in Eq.~(\ref{nonoptimal_estimator}). This is particularly relevant for primordial $B$-mode searches since we most care about the $B$-modes in particular $\ell$-ranges. This also allows us to probe levels of foreground contamination as a function of multipole bin. 

%

 
We evaluate this estimator on B-mode maps derived from Planck 353 GHz data, specifically cross correlating year-1 and year-2 maps. In this process we evaluate Eq.~\ref{nonoptimal_estimator}, where the summations are performed over the following ranges: $L \in [0, 64]$, $M \in [0,L]$ and $ \ell_1,\ell_2 \in [40, 370]$. While filtering in $\ell$-space we use an apodized filter function to lessen ringing.  The filter varies smoothly from 0 to 1 with a cosine-squared profile over the interval $\ell = [40,90]$, remains set to 1 for the interval $[90,320]$ and finally varies smoothly from 1 to 0 in the interval $[320,370]$, again with a cosine-squared profile.

Although we use the same multipole range as the disc analysis, the harmonic coefficients here are significantly different.  Here they are extracted from maps masked with the large GAL40 or GAL60 foreground mask, where as for the disc analysis the harmonic coefficients are repeatedly estimated from the regions left un-masked by each small, local disc mask. 

To convert this map in terms of an effective value for tensor to scalar ratio, $r_{\rm eff}$, we multiply by the factor $f_{\rm eff}$, like the one  given in Eq.~\ref{rescale_disc_analysis} for the local power disc analysis, but we account for the harmonic space filter described above. At full resolution, the non-optimal estimator closely resembles the related, optimal estimator shown later in Fig.~\ref{fig:op_data}, though the non-optimal map is roughly twice as noisy.

This estimator is akin to the local power disc analysis, but at higher resolution. To demonstrate this we downgrade the resolution of the reconstructed $\hat F^2$ map in real space, using 9.3 degree discs (with 2 degrees additional apodization) centered on pixels of $N_{\rm side}=8$ map. The resultant low-resolution, smoothed map is shown in Fig.~\ref{pam2pda_nop_data} alongside the map estimated from the disc analysis. The maps closely resemble each other, though derived from very different analysis procedures.


%
\begin{figure}
\centering
\subfigure[Non-optimal $\hat F^2 $ estimator]{\label{fig:noppam-pixel-histogram}\includegraphics[width=0.49\columnwidth]{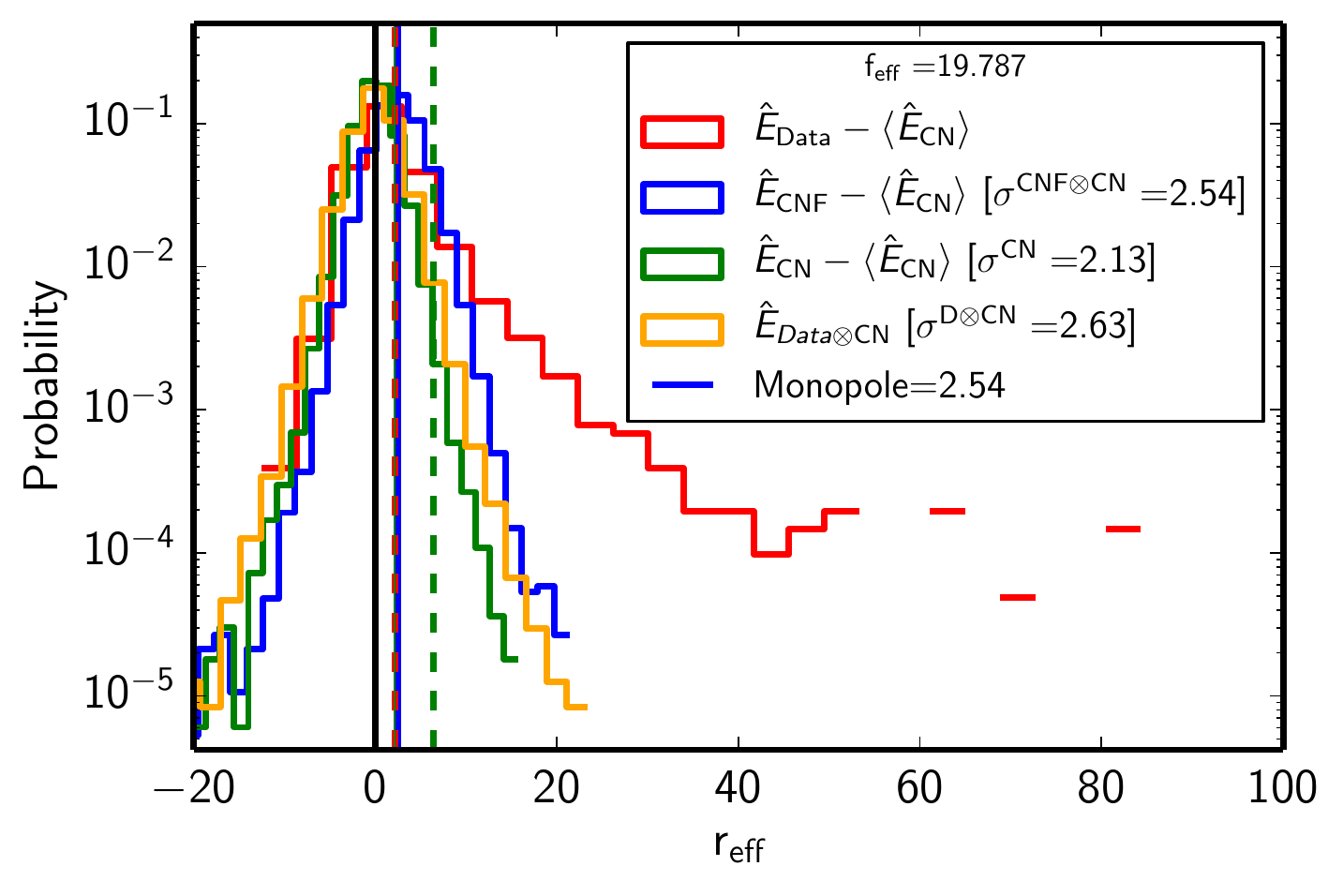}}
\subfigure[Optimal $\hat P $ estimator]{\label{fig:oppam-pixel-histogram}\includegraphics[width=0.49\columnwidth]{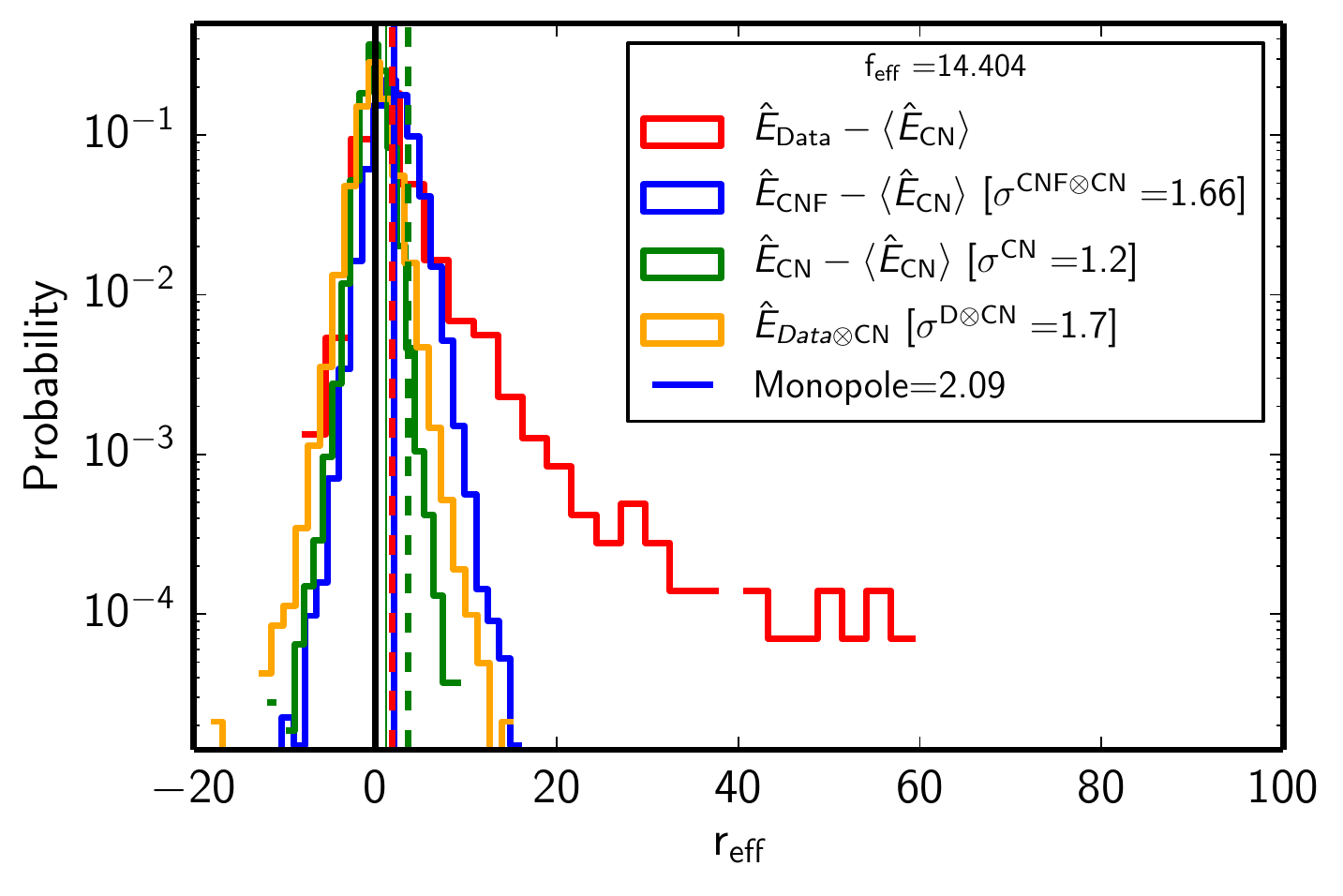}}
\caption{Histograms of pixel values for the non-optimal foreground estimator $\hat E = \hat F^2$ and the corresponding optimal estimator $\hat E = \hat P$. We compare results from data (red) to simulations. The blue histogram shows the distribution of simulations that include isotropically-modeled foregrounds ($CNF$). While the green shows the errors from CMB plus noise ($CN$) simulations, orange shows the errors given from cross-correlation between data and simulations ($CN \otimes$~Data).   The mean estimated monopole derived from data is dashed red; the same for $CNF$ simulations is solid blue.  The standard deviation of the ($CN$) distribution is solid green; three time the standard deviation is dashed green. The optimal estimator detects the monopole at higher significance. Direct comparisons between the right and left sets of histograms should be avoided since each estimator requires a different rescaling factor $f_{\rm eff}$.}
\label{fig:pixel-histogram}
\end{figure}
We also study the histogram of pixel values for this map, in Fig~\ref{fig:noppam-pixel-histogram}, and compare them against histograms of pixel values from simulations.  The $\hat F^2$ maps derived from data and the ($CNF$) simulation both have histograms shifted to the right of zero by the same amount.  We expect the monopoles to match in this way, since both the data and these simulations have similar power spectra by construction. We see that the histogram for data has a tail extending to higher values as compared to the ($CNF$) simulations. The high excursion pixels values correspond to high SNR peaks 
and are driven by foregrounds.  
These traits are not captured by the Gaussian, isotropic foreground simulations produced by the empirical power law power spectrum model. 

It is important to note that the histograms only reflect the mean statistical properties of the map, since the information of varying noise as a function of direction is lost in this representation of the data. The standard deviation per pixel on the power anisotropy maps, estimated from ($CN$) simulations, varies in the range [0.33,6.57] for the $\hat F^2 $ estimator, indicating that it is possible to get better statistics in certain regions of the sky than indicated by the mean statistics seen in the histograms.



\subsubsection{Optimal estimator of foregrounds squared: $\hat P$}
\label{power_anisotropy_estimator}

We may substantially improve our estimate by downweighting noisy modes, at the cost of downweighting the signal in the same modes, and so introducing a bias.
The BipoSH spectra measured from the data are composed of two terms,
\begin{equation}
\tilde A^{LM}_{\ell_1 \ell_2} = \tilde F^{LM}_{\ell_1 \ell_2} + \tilde{\mathcal{N}}^{LM}_{\ell_1 \ell_2} \,,
\end{equation}
where $F$ denotes the BipoSH spectra for the foregrounds, $\mathcal{N}$ denotes the zero mean noise fluctuations and we have ignored the residual spectra $\mathcal{R}$.
We now consider a generalization of  Eq.~\ref{nonoptimal_estimator} in order to construct a minimum variance estimator. This minimum variance estimator has a general form given by the following expression,
\beqry \label{optimal_estimator}
\hat P(\hn ) &=&  \sum_{LM}  \hat P_{LM} Y_{LM}(\hn)  \,, \\
& = & \sum_{LM}  \left[ \sum_{\ell_1 \ell_2} w^L_{\ell_1 \ell_2} \tilde A^{LM}_{\ell_1 \ell_2}\frac{\Pi_{\ell_1}\Pi_{\ell_2}}{\Pi_{L} \sqrt{4\pi}} \mathcal{C}^{L0}_{\ell_1 0 \ell_2 0} \right]  Y_{LM}(\hn) \nonumber
\eeqry
where the weights $w^L_{\ell_1 \ell_2}$ are to be chosen such that they minimize the variance from the scaled noisy modes $ \mathcal{N}^{LM}_{\ell_1 \ell_2} \frac{\Pi_{\ell_1}\Pi_{\ell_2}}{\Pi_{L} \sqrt{4\pi}} \mathcal{C}^{L0}_{\ell_1 0 \ell_2 0} $.
The solution to the weights that meet this criterion can be arrived at by solving the constrained minimization problem, under the constraint $\sum_{\ell_1 \ell_2} w^L_{\ell_1 \ell_2} = n_L$, {where $n_L$ is the number of BipoSH $\ell_1,\ell_2$ modes available for a given BipoSH multipole $L$.} The weights that minimize the variance of the estimator for $\hat P_{LM}$ given in Eq.~\ref{optimal_estimator} in the absence of foregrounds are given by the following expression,
\begin{equation}
 w^L_{\ell_1 \ell_2} = \frac{n_L(G^L_{\ell_1 \ell_2})^2 }{ C_{\ell_1} C_{\ell_2}} \left[ \sum_{\ell_1 \ell_2}\frac{(G^L_{\ell_1 \ell_2})^2 }{ C_{\ell_1} C_{\ell_2}} \right]^{-1}\,,
\end{equation}
where $C_{\ell}$ is the mean power spectrum characterizing the CMB and detector noise, estimated from simulations, and an independent geometric factor given by the following expression,
\begin{equation}
  G^L_{\ell_1 \ell_2} = \left[\frac{1}{\sqrt{4\pi}}\frac{\Pi_{\ell_1}\Pi_{\ell_2}}{\Pi_{L}} \mathcal{C}^{L0}_{\ell_1 0 \ell_2 0}\right]^{-1}. \label{geom_factor}
\end{equation} 
We explore these weights further in Appendix \ref{weights}.  Note that the weights use no prior knowledge of the foregrounds; they rely solely on isotropy violation compared to the statistically isotropic CMB and yield a map that traces where the foregrounds are most prominent.

Since the weights also operate on the foreground BipoSH spectra, the resultant map is a biased estimate of $F^2$.
{However, the weights do not depend on the azimuthal index $M$, which encodes the phase of the mode at angular scale L.  Hence the bias of the reconstructed $F^2$ only affects the relative amplitude of foregrounds on different angular scales.  While this may present a biased view as compared to the estimates from $\hat F^2$ estimator, the inferences we make about the statistical significance of foregrounds are not biased, since the noise maps are evaluated using the same weights.}
To distinguish this from the $\hat F^2$ map  discussed in the previous section, we call this the power anisotropy map and denote it by $\hat P$, as it still is a measure of direction-dependent power in the measured sky. 

We evaluate this optimal estimator on $B$-mode maps derived from Planck 353 GHz data, again cross-correlating year-1 and year-2 maps. In evaluating the power anisotropy estimator given in  Eq.~\ref{optimal_estimator}, the summations are performed over identical multipole ranges as the non-optimal estimator, $\ell_1,\ell_2 \in [40,370]$. 
{The results from an identical analysis on simulated maps, where foregrounds are sub-dominant to CMB and noise, are presented in Appendix \ref{forecast}}.

Again we recast the maps in terms of an effective $r$, this time multiplying by a factor that accounts for the weights and also incorporates the harmonic space apodization described in the previous section,
\begin{equation}f_{\mathrm{\rm eff}} = \left [\sum_{\ell_{min}}^{\ell_{max}} w^{0}_{\ell \ell} \frac{2 \ell +1}{4 \pi}C^{BB}_{\ell}[r=1] \right]^{-1}.
\end{equation}
Thus $f_{\rm eff}^{-1}$ corresponds to the \emph{now weighted} ensemble mean power in the primordial $B$-mode sky corresponding to $r=1$ from the same multipole range.

\begin{figure}
\centering
\subfigure[]
          {\label{fig:op_data}\includegraphics[width=0.49\columnwidth]{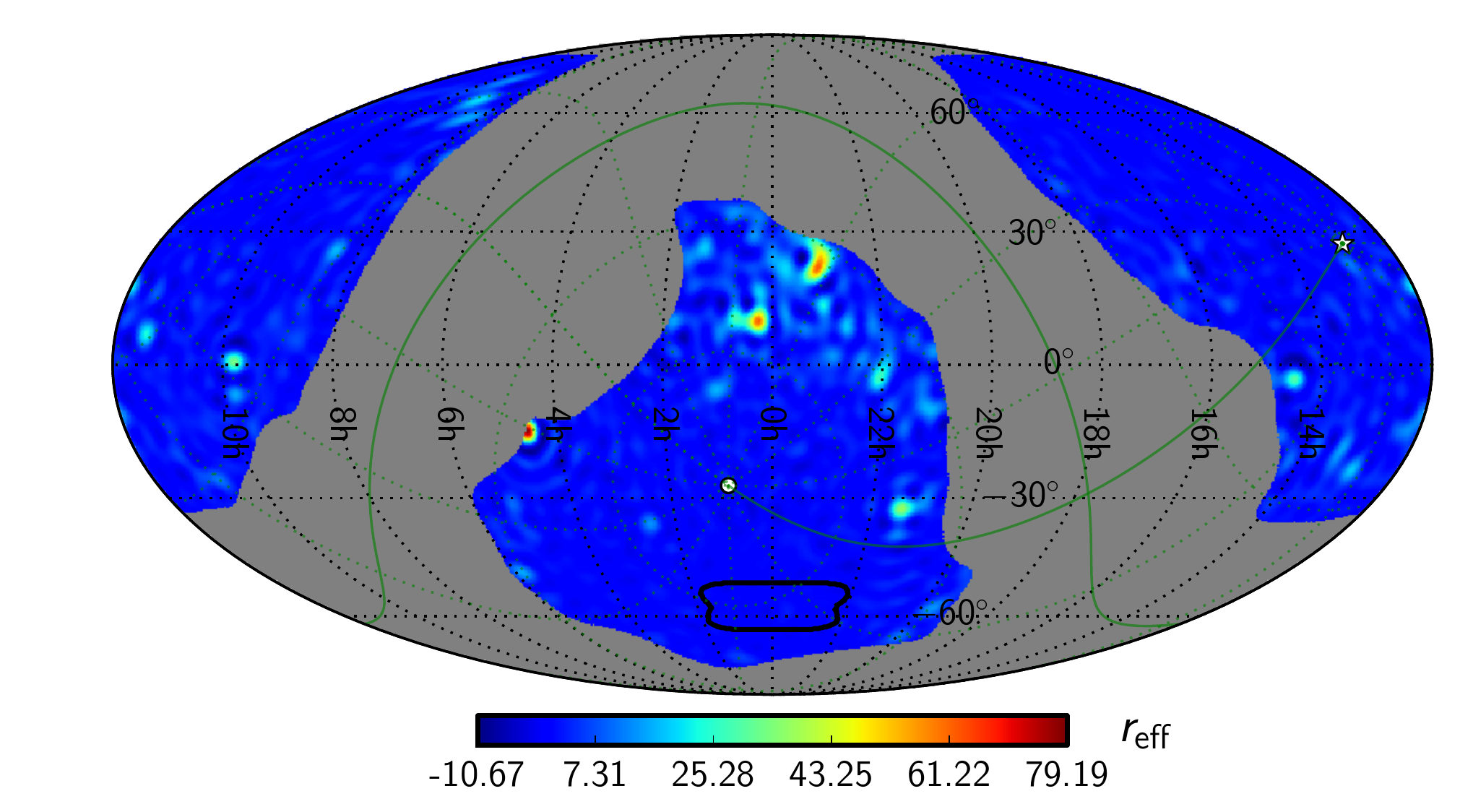}}
\subfigure[]
          {\label{fig:op_cn_err_snr}\includegraphics[width=0.49\columnwidth]{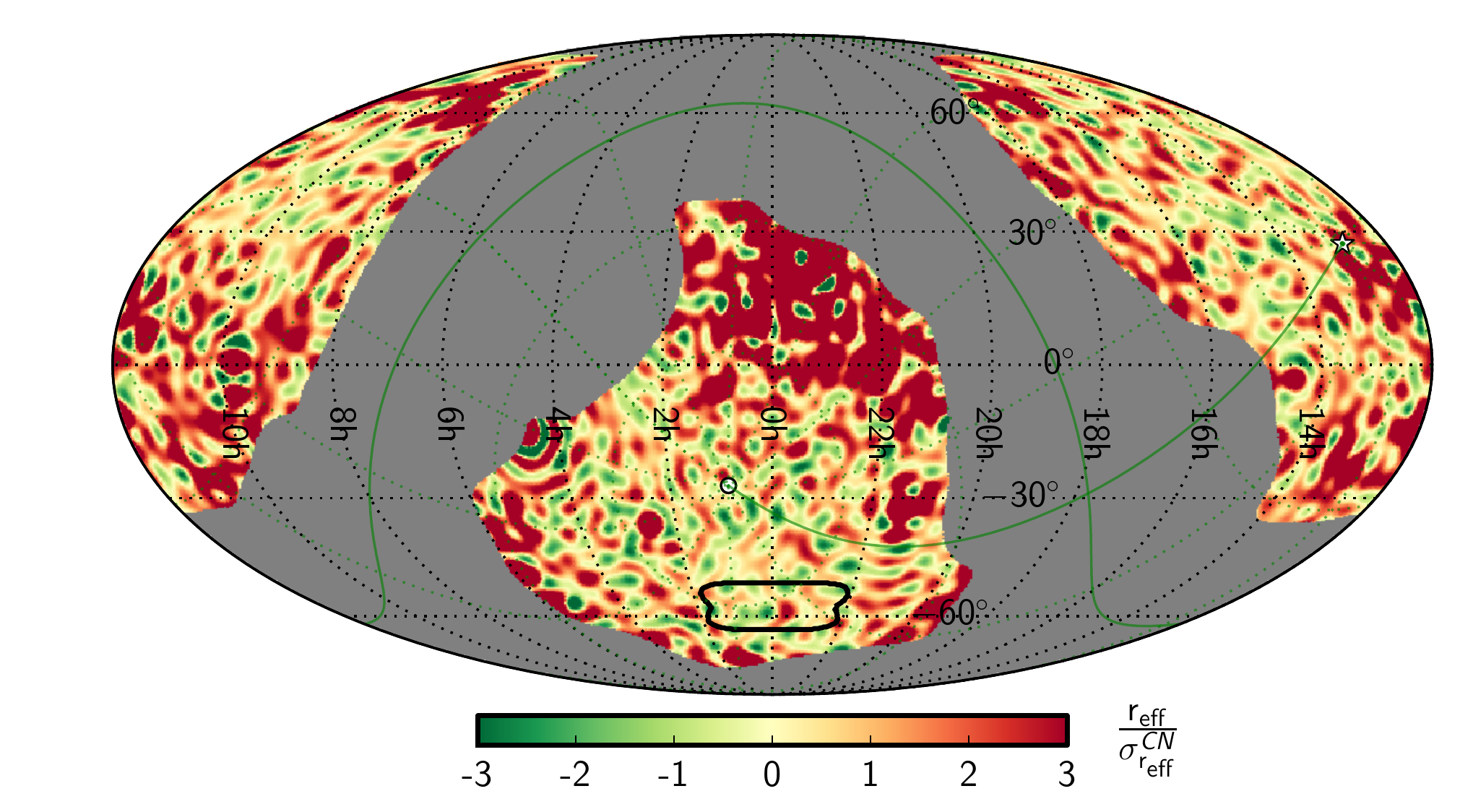}}
\subfigure[]
          {\label{fig:op_ocn_err}\includegraphics[width=0.49\columnwidth]{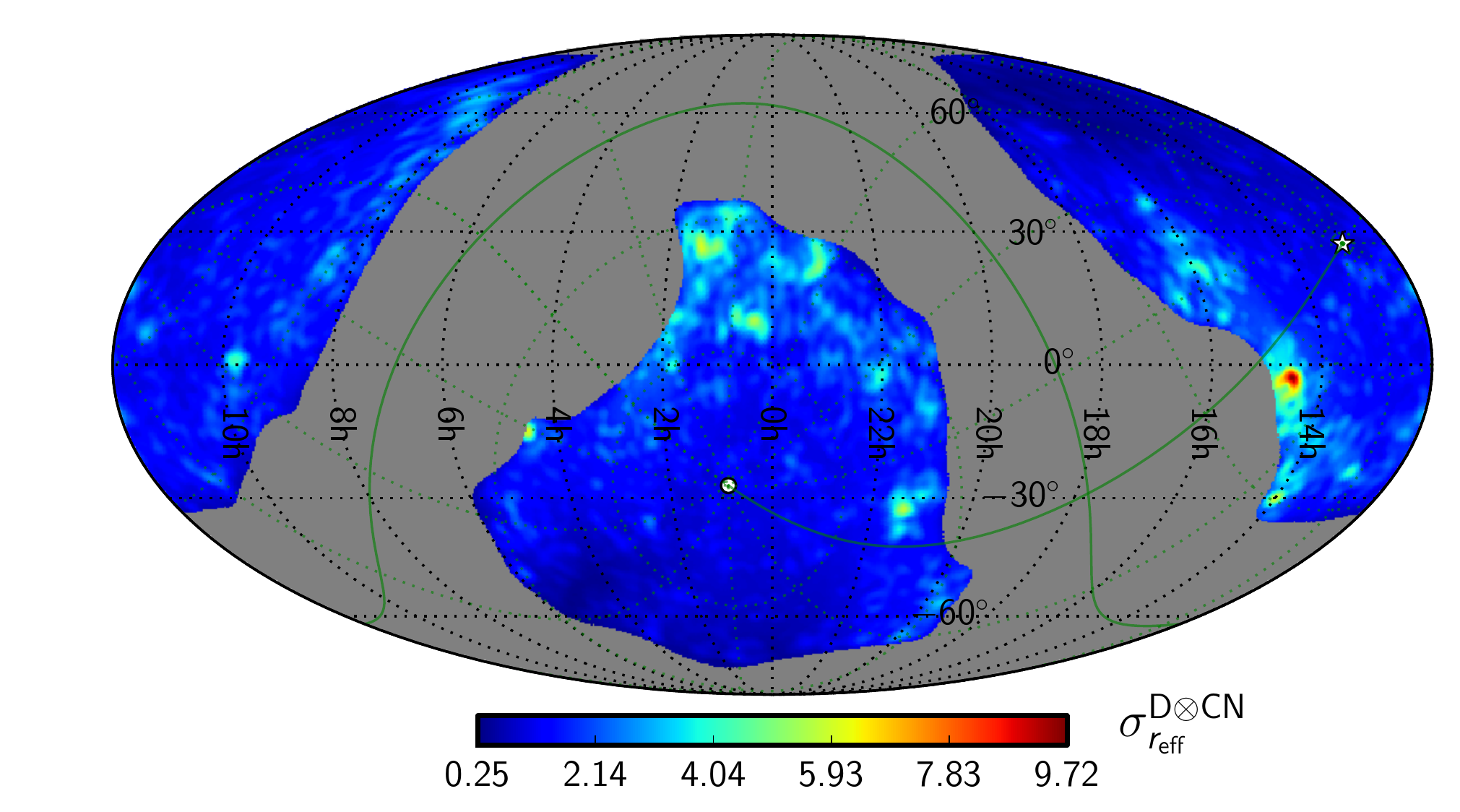}}
\subfigure[]
          {\label{fig:op_cn_err}\includegraphics[width=0.49\columnwidth]{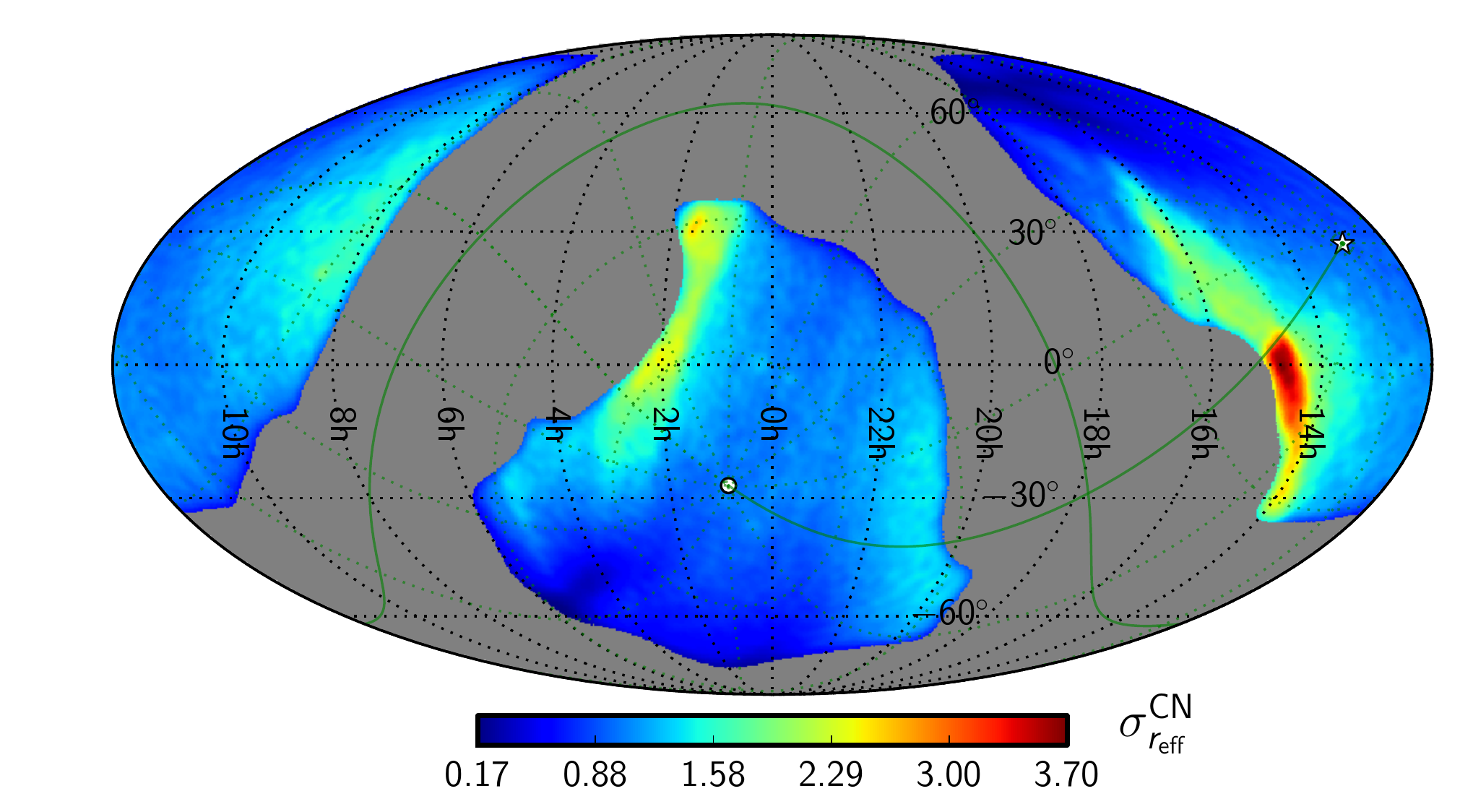}}
\caption{Results derived from the optimal Bipolar Spherical Harmonic estimator $\hat P$ from 353 GHz Planck maps, extrapolated to 150 GHz, and rescaled to an effective tensor-to-scalar ratio. \textit{Top left:} Estimate of foreground-squared from Year 1 $\times$ Year 2 combination of Planck 353 GHz data.
  {\textit{Top right:} Signal-to-noise ratio per pixel for the estimated $F^2$ field, evaluated using errors estimated from ($CN$) simulations. The significant negative excursions seen here are due to two reason, (1) systematic artifacts due to ringing around sharp foreground features, and (2) chance correlations of noise with foregrounds. \textit{Bottom left:} Error map estimated from cross correlating data with simulations (Data~$\otimes CN$). The errors expectedly trace the foreground, as they are modulated by the foreground due to chance correlations. \textit{Bottom right:} Error map estimated from cross correlating year-1 and year-2 CMB plus noise simulations ($CN$).}
  All results use the GAL60 mask and a band limit $L_{\rm max} = 64$, and are shown in equatorial coordinates at a Healpix resolution of $N_{\rm side} = 256$. The green contours at top left mark the galactic coordinates in 30$^\circ$ spacing and the solid green curves mark the zeros of the coordinate system.
}
 \end{figure}

We show the power anisotropy map $\hat P$ in Fig~\ref{fig:op_data}. The foreground features are clearly discernible in this map, where the red regions are those with strong foreground emission and the blue regions indicate relatively low foreground emissions.

We run an identical analysis on simulations, and use the resultant ensemble of $\hat P$ maps to derive the error per pixel. We evaluate the error maps using three different procedures.  First, we make an error estimate ($\sigma^{CN}_{r _{\rm eff}}$) by cross correlating year-1 year-2 CMB plus noise ($CN$) simulations. In this case the error map only traces the observation depth of Planck as seen in Fig.~\ref{fig:op_cn_err}. {These errors are used to assess the presence of foreground features at any given pixel on the map,  over and above the noise fluctuations.} Second, we make an error estimate ($\sigma^{\rm D\otimes CN}_{r _{\rm eff}}$) by cross correlating the data maps with $CN$ simulations, which accounts for chance correlations with the actual foregrounds. Here again the error maps trace the observation depth for Planck, but in addition are also modulated by the actual foregrounds as seen in Fig.~\ref{fig:op_ocn_err}. Inclusion of these chance correlations increases the mean error by a factor of 1.42 for the GAL60 mask and 1.23 for GAL40, but the increases are localized to the foreground features. {These errors are used to assess the uncertainty on the inferred $r_{\rm eff}$, most relavant when there is a definite foreground feature detected. Note that $\sigma^{\rm D\otimes CN}_{r _{\rm eff}} = \sigma^{CN}_{r _{\rm eff}}$ in regions where there are no foregrounds.} { Finally we also estimate an error map by cross correlating ($CNF$) simulations with ($CN$) simulations, using the isotropic Gaussian empirical power law model for foregrounds. In this case the mean error is raised due to chance correlations by a similar factor. However the isotropic simulations do not carry information on the true morphology of foregrounds, and hence the pattern in the error map only traces the observation depth of Planck and is similar to Fig.~\ref{fig:op_cn_err}, though with a higher mean error.}

To detect excess power due to foregrounds we use $\sigma^{CN}_{r _{\rm eff}}$ errors to compute the signal to noise ratio shown in Fig.~\ref{fig:op_cn_err_snr}. We use errors that include chance correlation ($\sigma^{\rm D\otimes CN}_{r _{\rm eff}}$) to assess the uncertainty on the measured value of $r_{\rm eff}$. Since we are reconstructing the foreground-squared field, we do not normally expect statistically significant negative excursions.  The largest negative excursions we do see (Fig.~\ref{fig:op_cn_err_snr} at $4$h,$-15^\circ$) are due to ringing around sharp, highly significant foreground features.  These arise from the band limit imposed in the analysis.
We make a more thorough statistical assessment of these maps, to uncover the least-foreground-contaminated regions, in Section~\ref{thebestplaces}. Also note that significant negative excursion can occur due to chance correlations with foregrounds, since $\sigma^{CN}_{r _{\rm eff}}$ do not account for these chance correlations.

In Fig.~\ref{fig:pixel-histogram}, we also study the histogram of pixel values for the optimal estimator.  Though the histogram for $r_{\rm eff}$ noise corresponding to the $\hat P^2 $ estimator is narrower than for the $\hat F^2 $ estimator, part of that comes from the normalization with $f _{\rm eff}$, so we cannot directly compare them.  Accounting for the normalization, we find that $\sigma^{CN}_{r _{\rm eff}}$ are reduced by 23\% while  $\sigma^{D\otimes CN}_{r _{\rm eff}}$ are reduced only by 11\%, on an average: the weights are only designed to downweight the noise portion, not chance correlations.  Though the optimal estimator is biased, the bias is not large: the monopole amplitude $\hat P$ is larger by 13\% than for $\hat F^2$, accounting for $f_{\rm eff}$. More important the optimal estimator is more efficient at detecting foregrounds: the SNR on the foreground  monopole is 1.46 times better than the non-optimal estimator.

  The error per pixel $\sigma_{r_{\rm eff}}$ varies over a wide range of values as seen in Fig.~\ref{fig:op_ocn_err} and Fig.~\ref{fig:op_cn_err} indicating that it is possible to get significantly better statistics in certain regions of the sky than indicated by the mean statistical properties seen in the histogram.

\section{Where are the best sky regions for primordial $B$-mode searches? } \label{thebestplaces}
Based on our analysis, where should $B$-mode observatories search? Any figure-of-merit should favor low measured values of the effective tensor-to-scalar ratio, but must also account for our confidence in that measurement, as determined by the noise level. In regions where noise dominates the signal, we must decide whether one area can be sensibly favored over another.  These judgments are difficult to make simply by inspecting the signal and noise maps.

To characterize our results, and to set limits on  potentially clean regions of the sky, we construct a posterior distribution for $r_{\rm eff}$.  We use $r_{\rm eff}$ measurements from the more conservative GAL40 mask to reduce ringing from bright regions near the galactic plane (Appendix \ref{sec:mask_dependence}).  From our simulations, we derive a map of the error, and build a Gaussian likelihood.  Then using Bayes' theorem we define the posterior probability function for $r_{\rm eff}$ given the measured data as follows,
\begin{equation}
\mathcal{P}(r_{\rm eff} | r_{\rm eff}^{\rm meas.}, \sigma_{r_{\rm eff}}) \propto \exp \left[ -\frac{(r_{\rm eff}^{\rm meas.} - r_{\rm eff})^2}{2 \sigma^2_{r_{\rm eff}}}\right] \mathcal{P}(r _{\rm eff})\,,
\end{equation}
where $\sigma_{r_{\rm eff}}$ is the error estimated per pixel (incorporating the chance correlation with foregrounds) and $\mathcal{P}(r _{\rm eff})$ is the prior on $r_{\rm eff}$.
We assume a simple prior that restricts the physical tensor-to-scalar ratio to be positive: it is zero for $r_{\rm eff} < 0$ and unity for $r_{\rm eff} \geq 0$.  The formulation is strictly valid only when pixels are statistically uncorrelated.  We examined the correlation function and find the $1/e$ correlation length is about $2^\circ$, about the size of an $N_{\rm side}=32$ pixel.

Normalizing the posterior probability distribution, we can evaluate 95 percent upper limits on $r_{\rm eff}$.  We write the limit $r^{95}_{\rm eff}$ as an integral over the posterior such that 
\begin{equation}
  \int_0^{r^{95}_{\rm eff}} \mathcal{P}(r_{\rm eff} | r_{\rm eff}^{\rm meas.}, \sigma_{r_{\rm eff}}) \, d r_{\rm eff} = 0.95.
\end{equation}
A high measurement with a large error will make the $r_{\rm eff}^{95}$ limit high; a low measurement with a small error will make it low.  Since we are trying to make a statement about the value of $r_{\rm eff}$, this error estimate $\sigma_{r_{\rm eff}}$ should include the chance correlation with foregrounds, and so the inferred $r^{95}_{\rm eff}$ upper limit is especially enhanced in places with high foregrounds.
Hence regions where $r^{95}_{\rm eff}$ is the lowest probably have the least foreground contamination.

\begin{figure}[!t]
\centering
\subfigure[]{\label{fig:r95_smth0}\includegraphics[width=0.85\columnwidth]{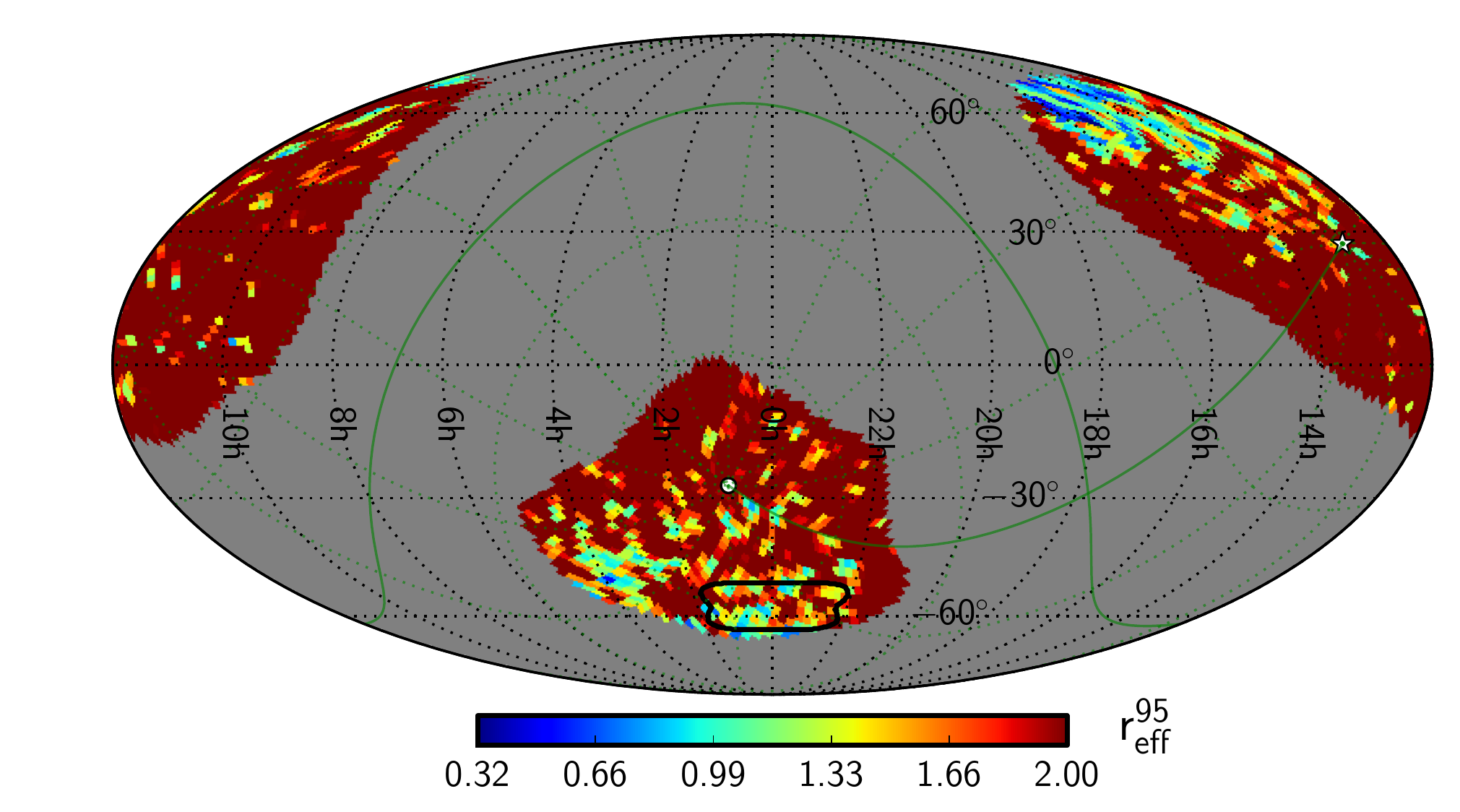}}
\subfigure[]{\label{fig:snr_smth0}\includegraphics[width=0.85\columnwidth]{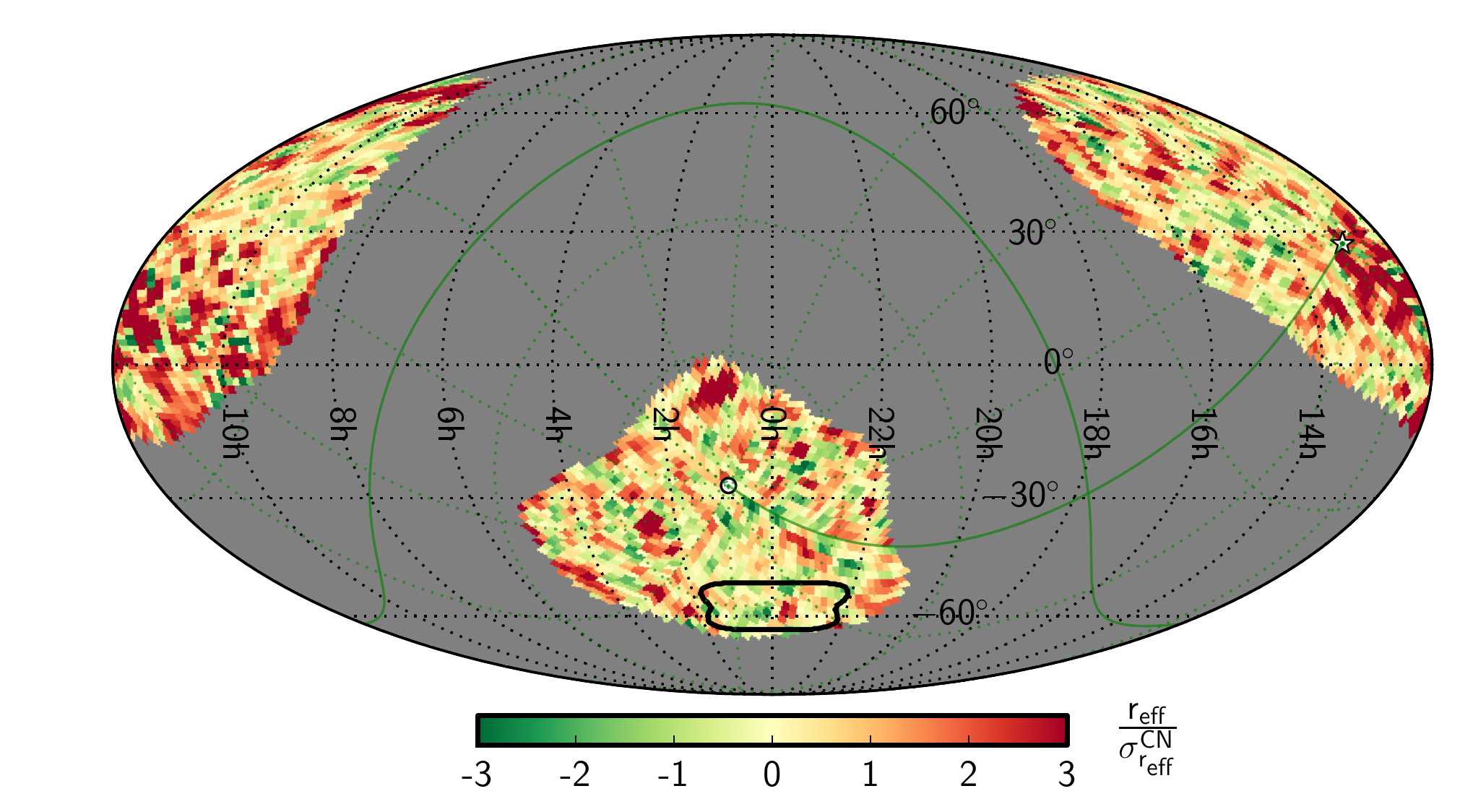}}
  \caption{Statistics on $r_{\rm eff}$ maps evaluated using the optimal estimator $\hat P$, using the GAL40 mask. Maps shown at a Healpix resolution of $N_{\rm side}=32$. These maps do not have any additional smoothing, other than $L_{\rm max}=64$ cutoff used while evaluating the estimator. \textit{Top:} Depicts the $r_{\rm eff}^{95}$, estimated using $r_{\rm eff}$ measured from data and $\sigma^{\rm D \otimes CN}_{\rm r_{\rm eff}}$ estimated from cross correlating data with (CN) simulations. Since we are interested in least contaminated regions, this map is saturated at $r^{95}_{\rm eff}=2$.  \textit{Bottom:} Significance of the excess power due to the presence of foreground contamination. Note the significant foreground detection in the BICEP region.}
 \label{fig:fom_smth0}
 \end{figure}
The $r^{95}_{\rm eff}$ map in Fig.~\ref{fig:r95_smth0} depicts 95 percent confidence limits on the foreground contamination after accounting for the measured foregrounds and noise fluctuations. 
The region with the lowest $r^{95}_{\rm eff}$ limit is in the Northern Hemisphere, at the northern ecliptic pole where the Planck noise is low.  In the Southern Hemisphere, there are also regions with low  $r^{95}_{\rm eff}$ in and near the BICEP patch, but near the center of the BICEP patch we see signs of contamination.  Below we take a closer look.

Regions with a low upper limit can nonetheless have a significant foreground detection, if the errors there are small.  We assess the significance of pixels in the  signal-to-noise map of Fig~\ref{fig:snr_smth0}.  In this case it is important that the errors used to estimate this SNR \textit{not} include chance correlations with foregrounds, because we are trying to compare against the null hypothesis that foregrounds are absent.  Thus here we only incorporate the fluctuations in power estimated from ($CN$) simulations.

In the signal-to-noise map, the Northern Ecliptic pole looks less remarkable.  This means that although the foreground there is low (from $r^{95}_{\rm eff}$), the errors are also small and so we have some significant detections of that foreground.  In the South, we observe that the BICEP patch and its surroundings do have low signal-to-noise pixels, but again there is a notable feature in the middle of the BICEP patch.

\begin{figure}[!t]
\centering
\subfigure{\label{fig:r95_north_smth2}\includegraphics[width=0.49\columnwidth]{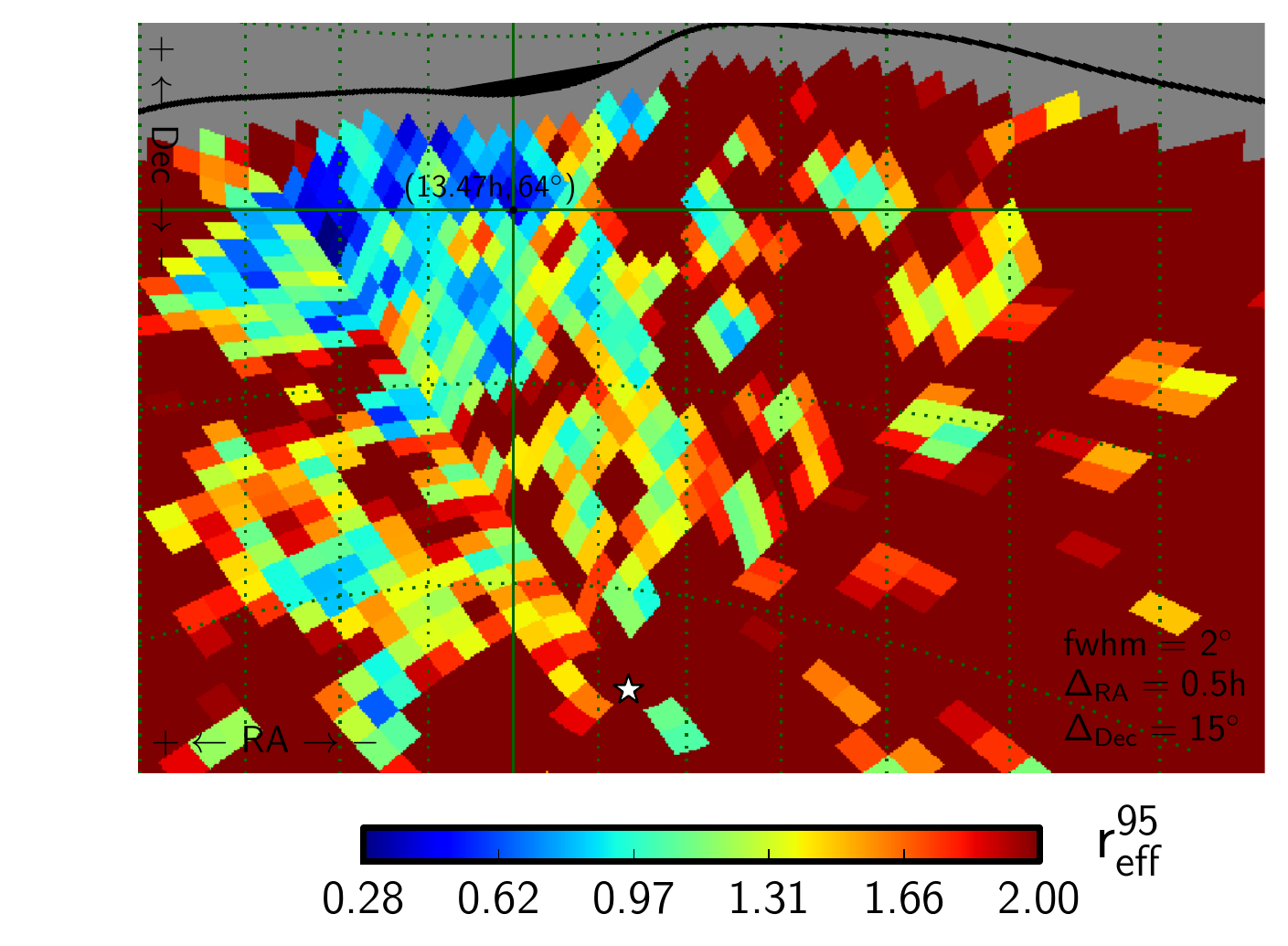}}
\subfigure{\label{fig:snr_north_smth2}\includegraphics[width=0.49\columnwidth]{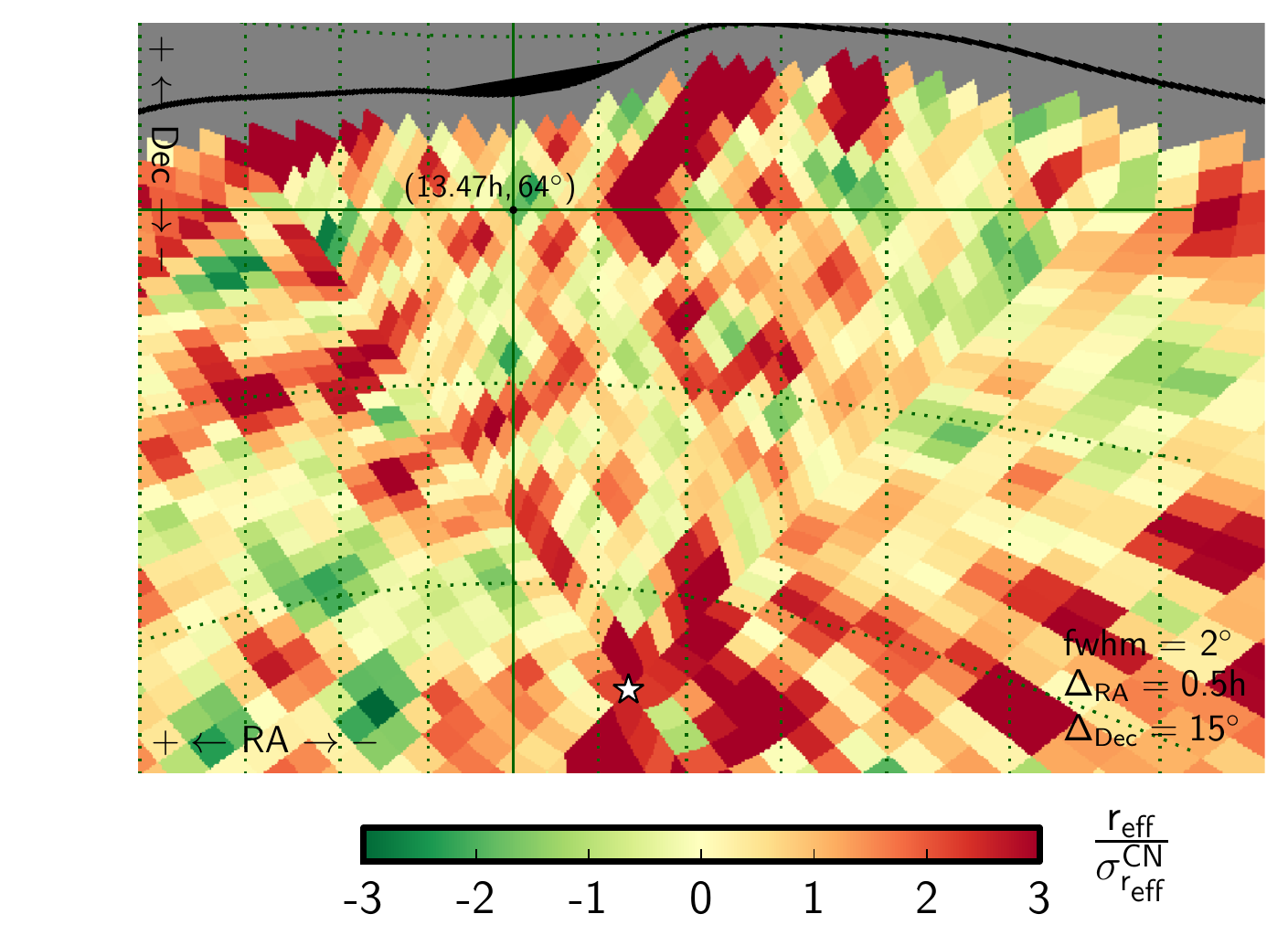}}
\subfigure{\label{fig:r95_north_smth4}\includegraphics[width=0.49\columnwidth]{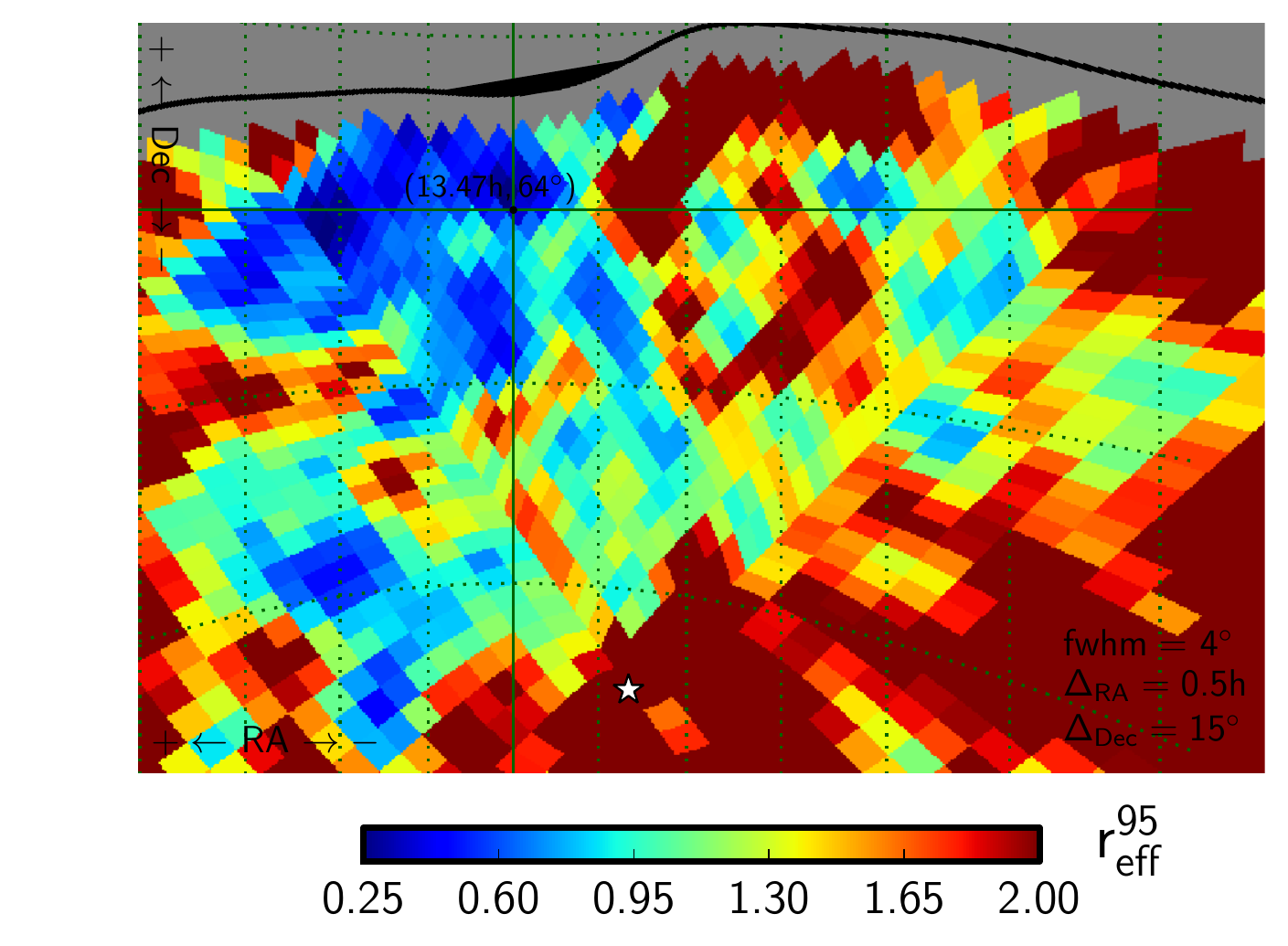}}
\subfigure{\label{fig:snr_north_smth4}\includegraphics[width=0.49\columnwidth]{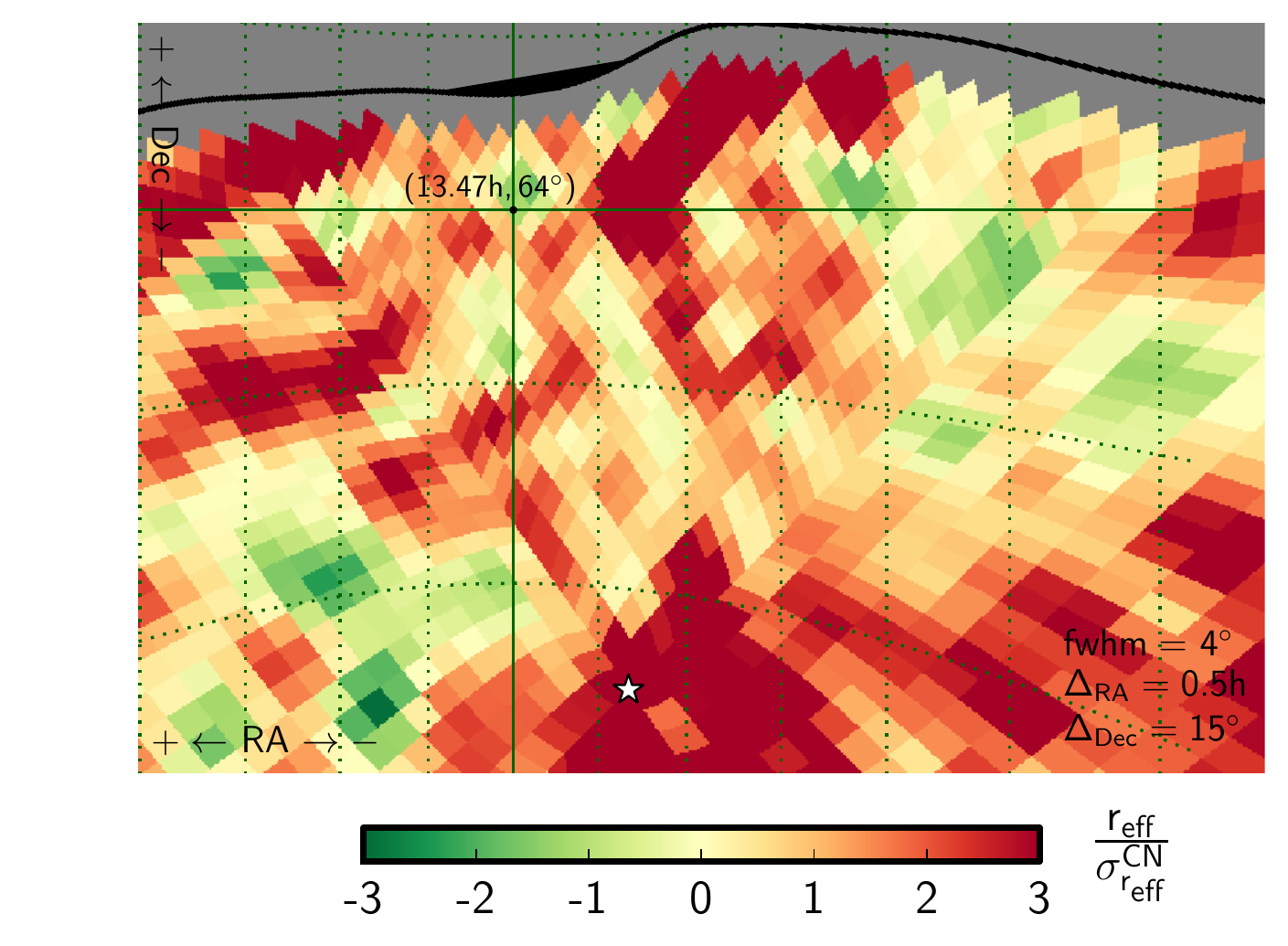}}
\subfigure{\label{fig:r95_north_smth6}\includegraphics[width=0.49\columnwidth]{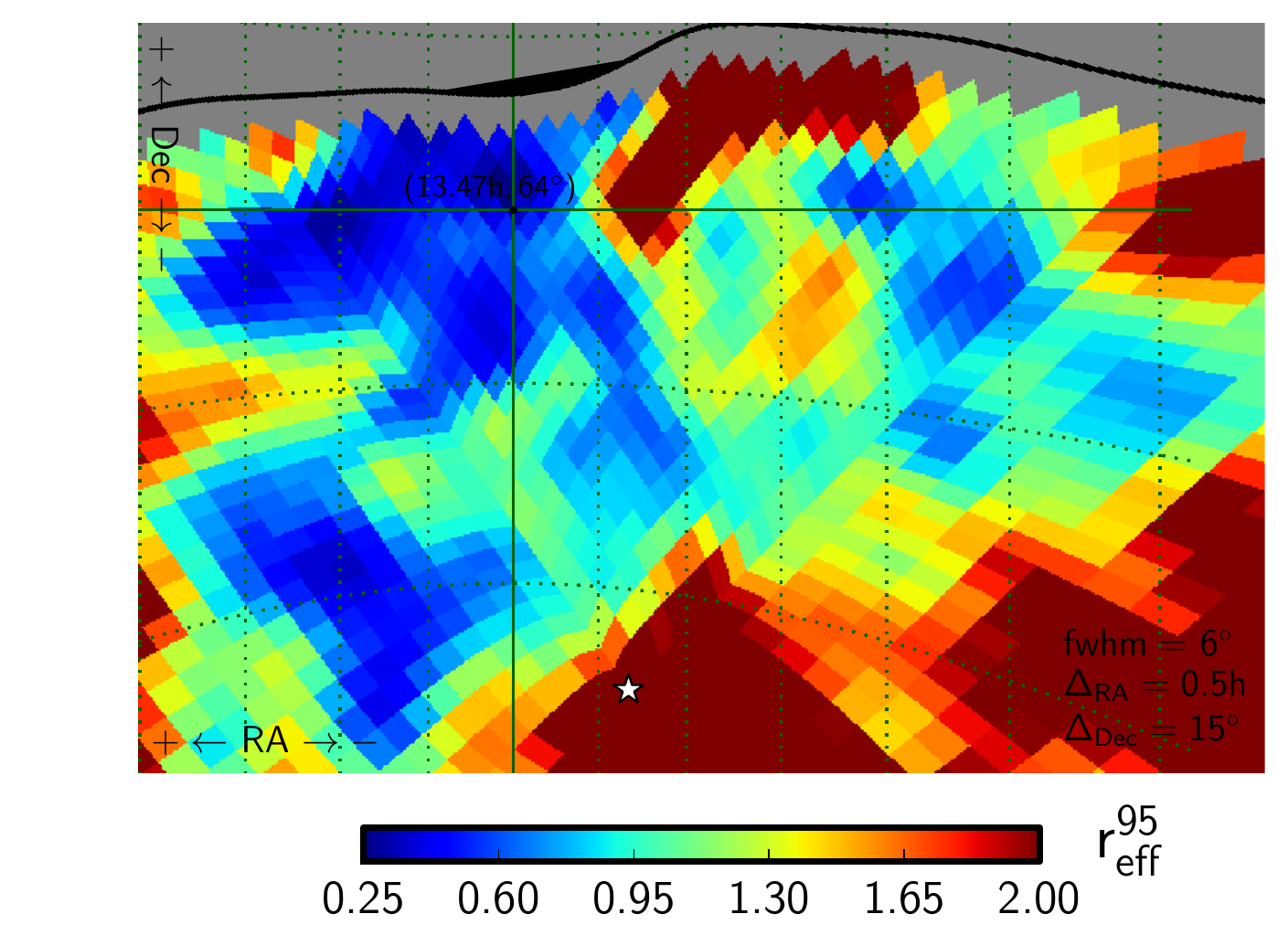}}
\subfigure{\label{fig:snr_north_smth6}\includegraphics[width=0.49\columnwidth]{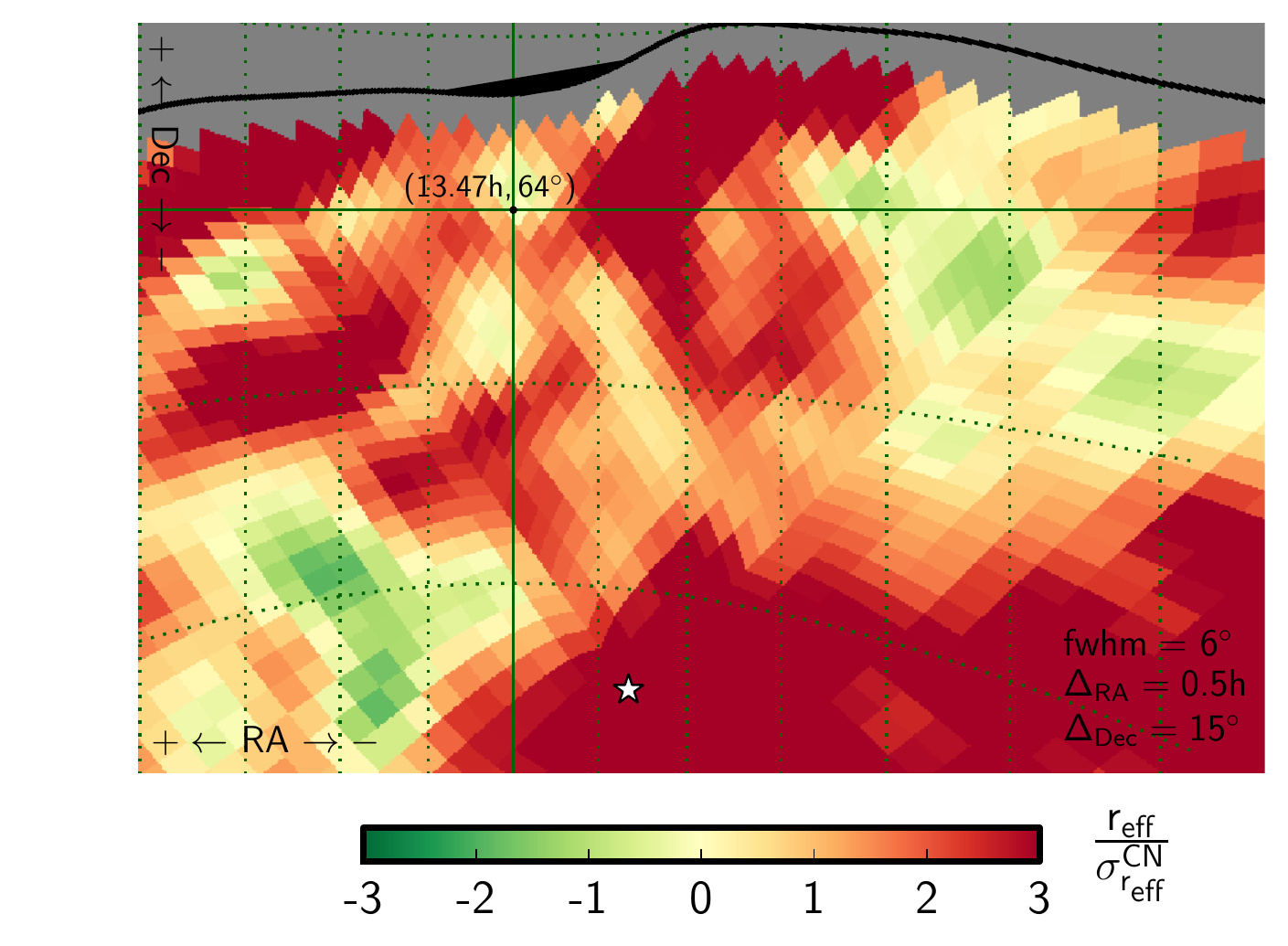}}
\caption{The potentially least contaminated regions near the north Galactic pole---marked by the white star. The top, middle and bottom rows show statistics derived from maps smoothed with a Gaussian of width fwhm$=2^\circ$, $4^\circ$, and $6^\circ$ respectively. The black contours mark the edges of the mask. The figures in the left column show $r^{95}_{\rm eff}$  estimates and those on the right show the SNR of excess power due to foregrounds. Note that $r^{95}_{\rm eff}$ does not evolve on increasing the smoothing scale, while the SNR steadily increases.}\label{fig:fom_smth_north}
 \end{figure}
\begin{figure}[!t]
\centering
\subfigure{\label{fig:r95_south_smth2}\includegraphics[width=0.49\columnwidth]{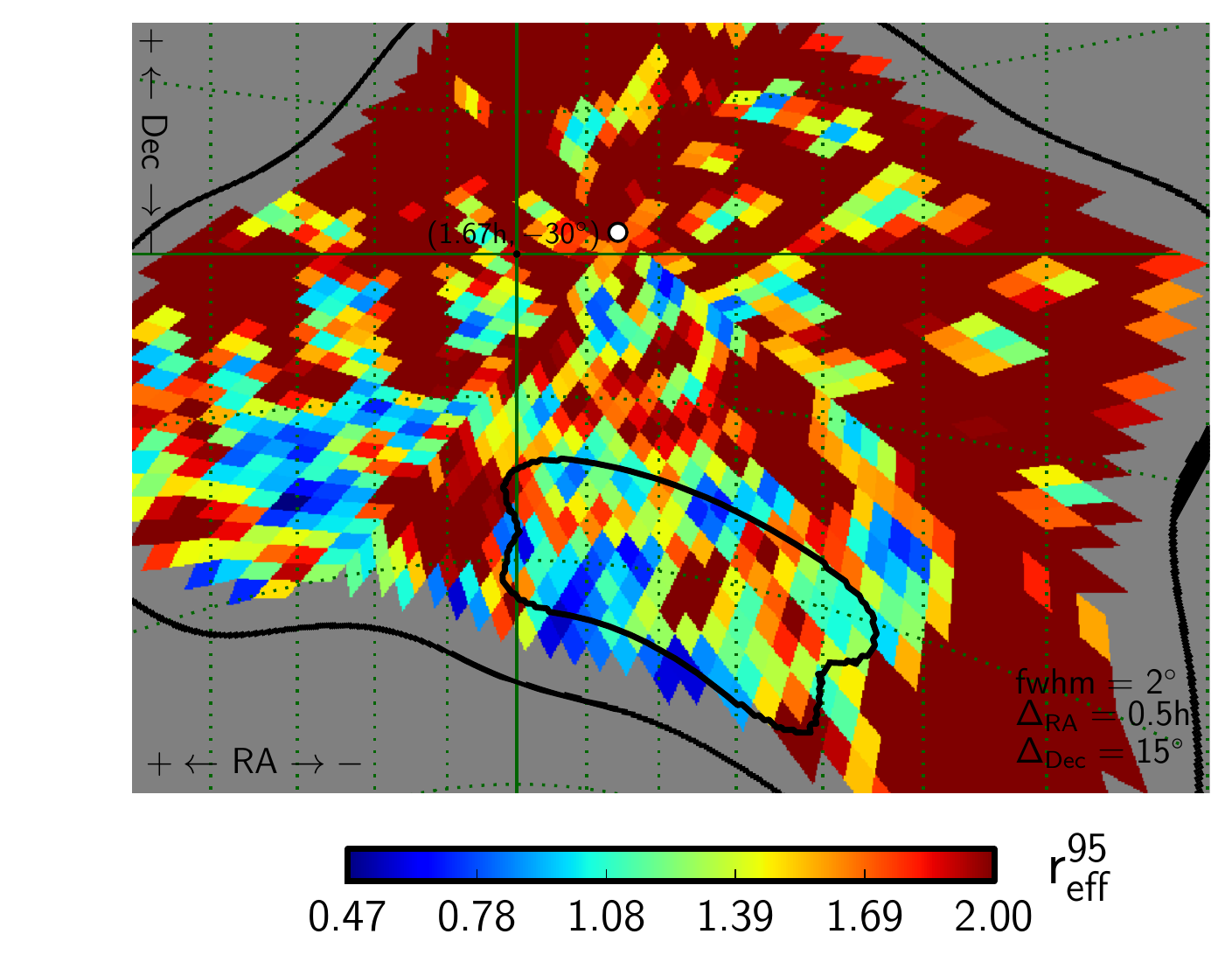}}
\subfigure{\label{fig:snr_south_smth2}\includegraphics[width=0.49\columnwidth]{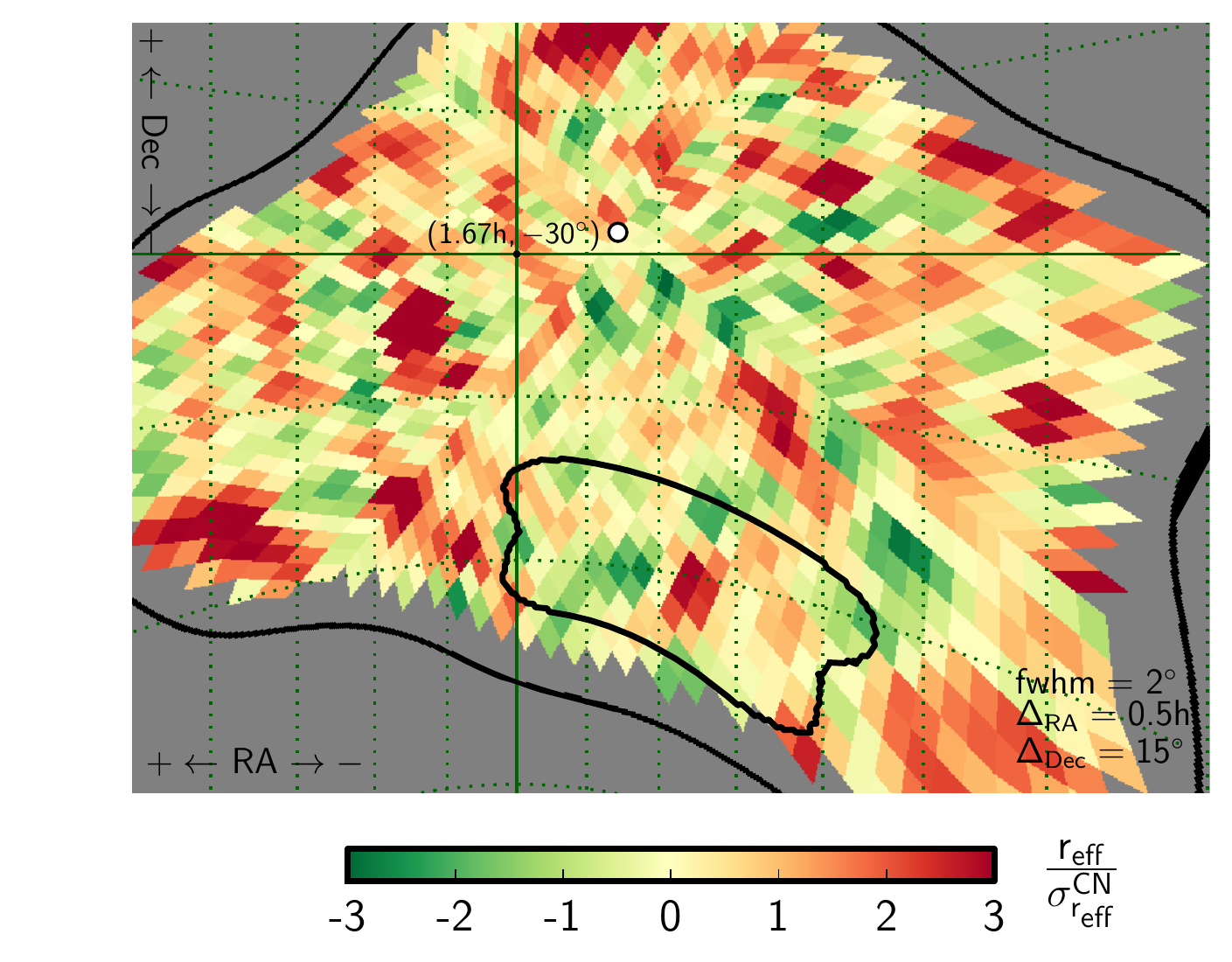}}
\subfigure{\label{fig:r95_south_smth4}\includegraphics[width=0.49\columnwidth]{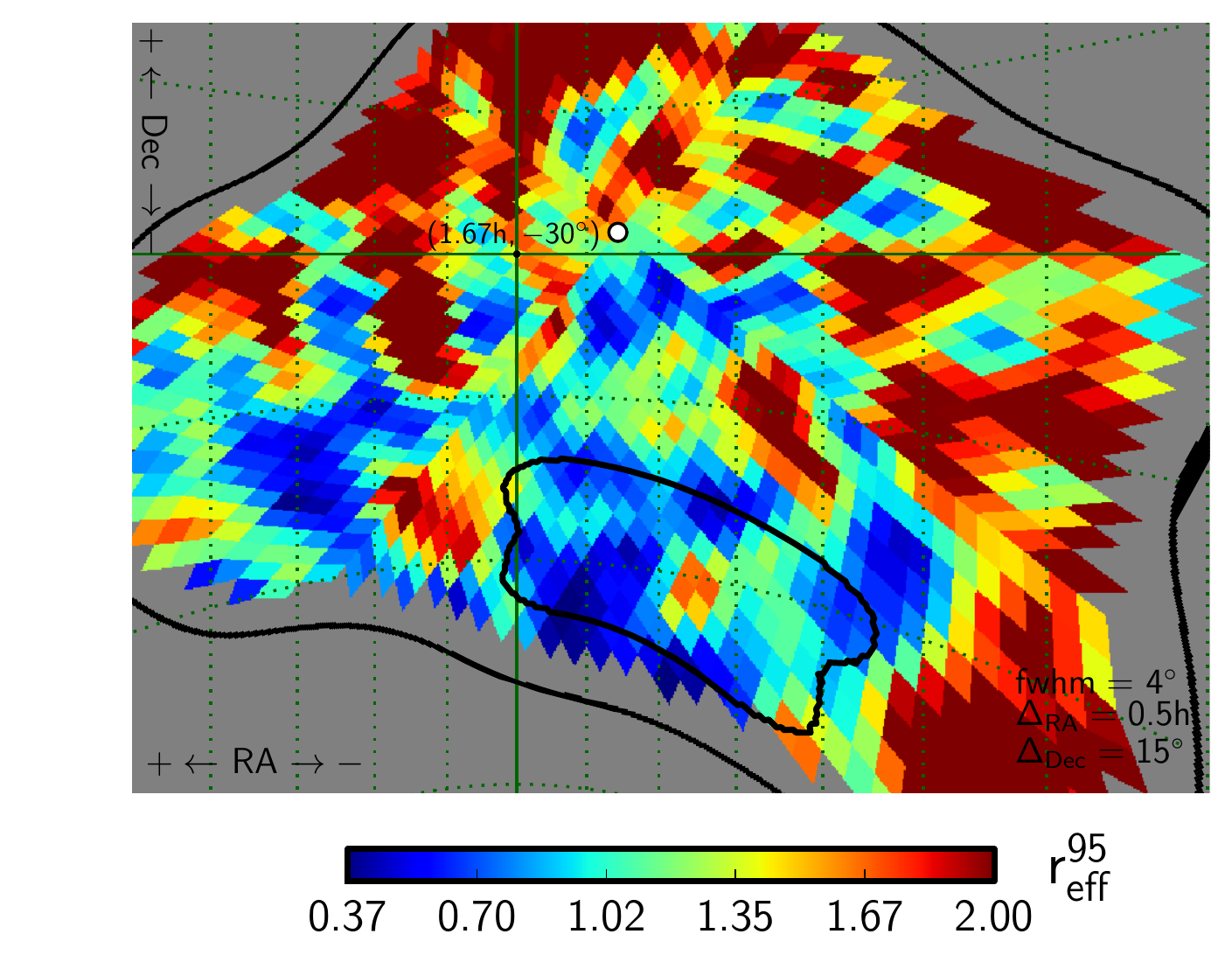}}
\subfigure{\label{fig:snr_south_smth4}\includegraphics[width=0.49\columnwidth]{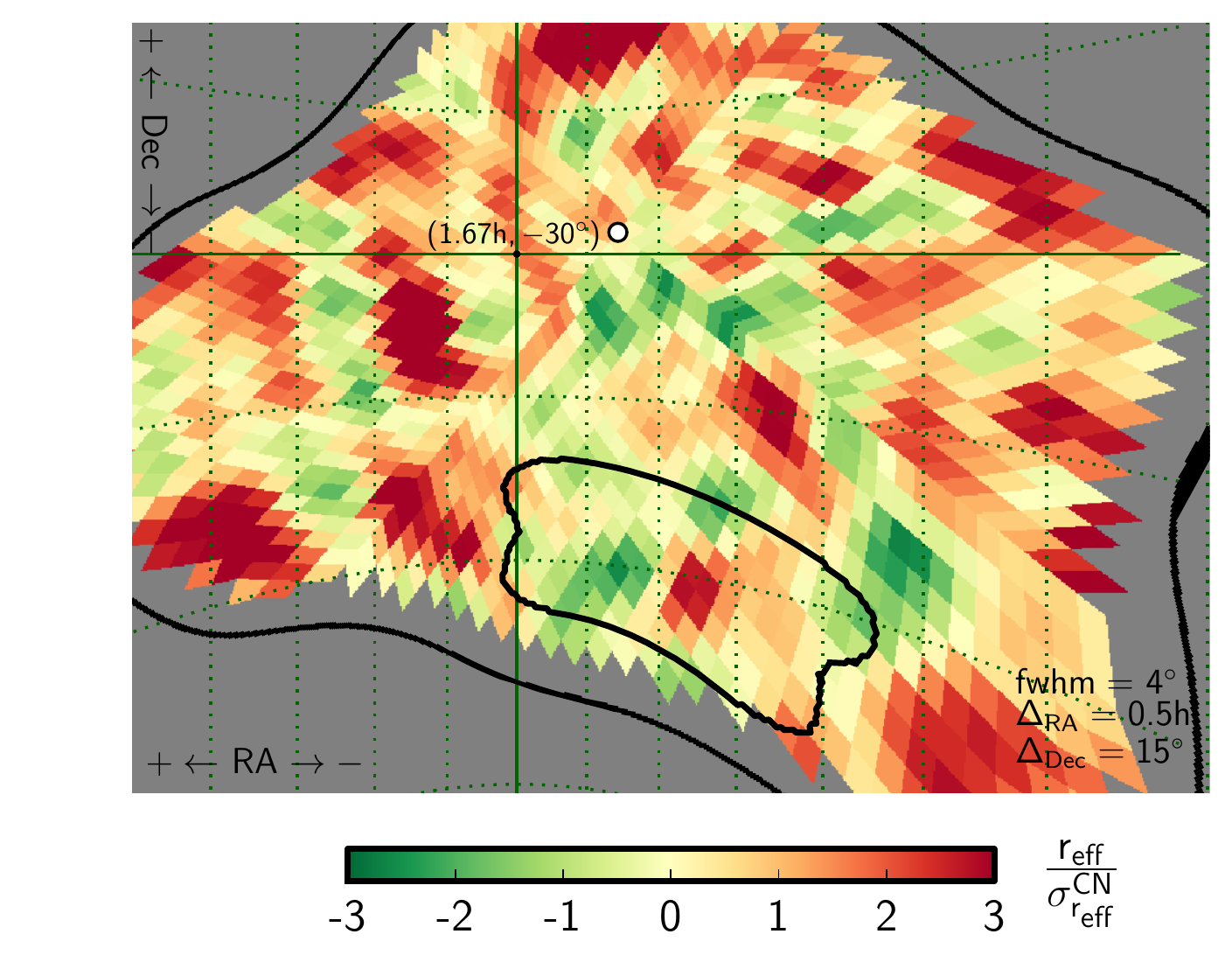}}
\subfigure{\label{fig:r95_south_smth6}\includegraphics[width=0.49\columnwidth]{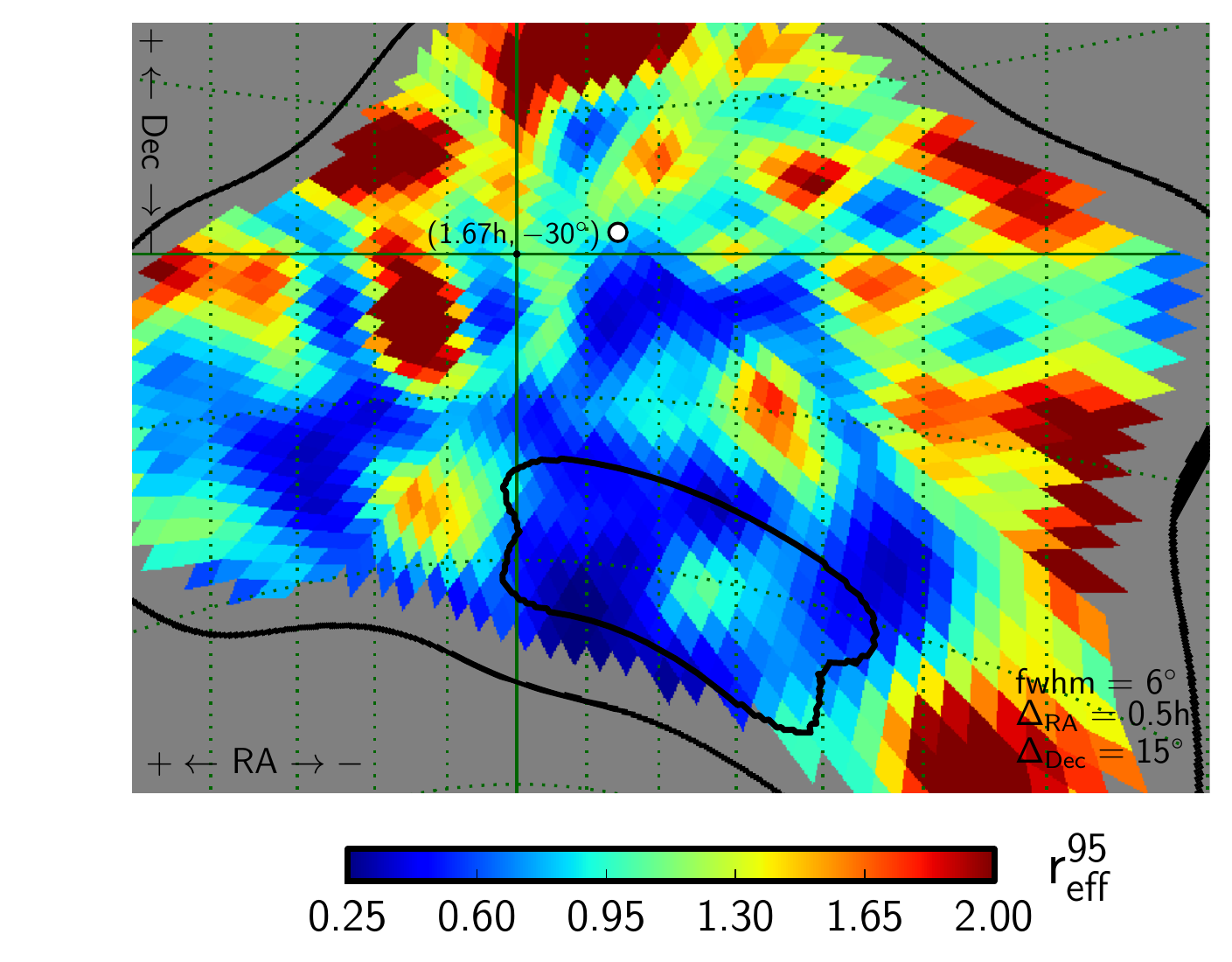}}
\subfigure{\label{fig:snr_south_smth6}\includegraphics[width=0.49\columnwidth]{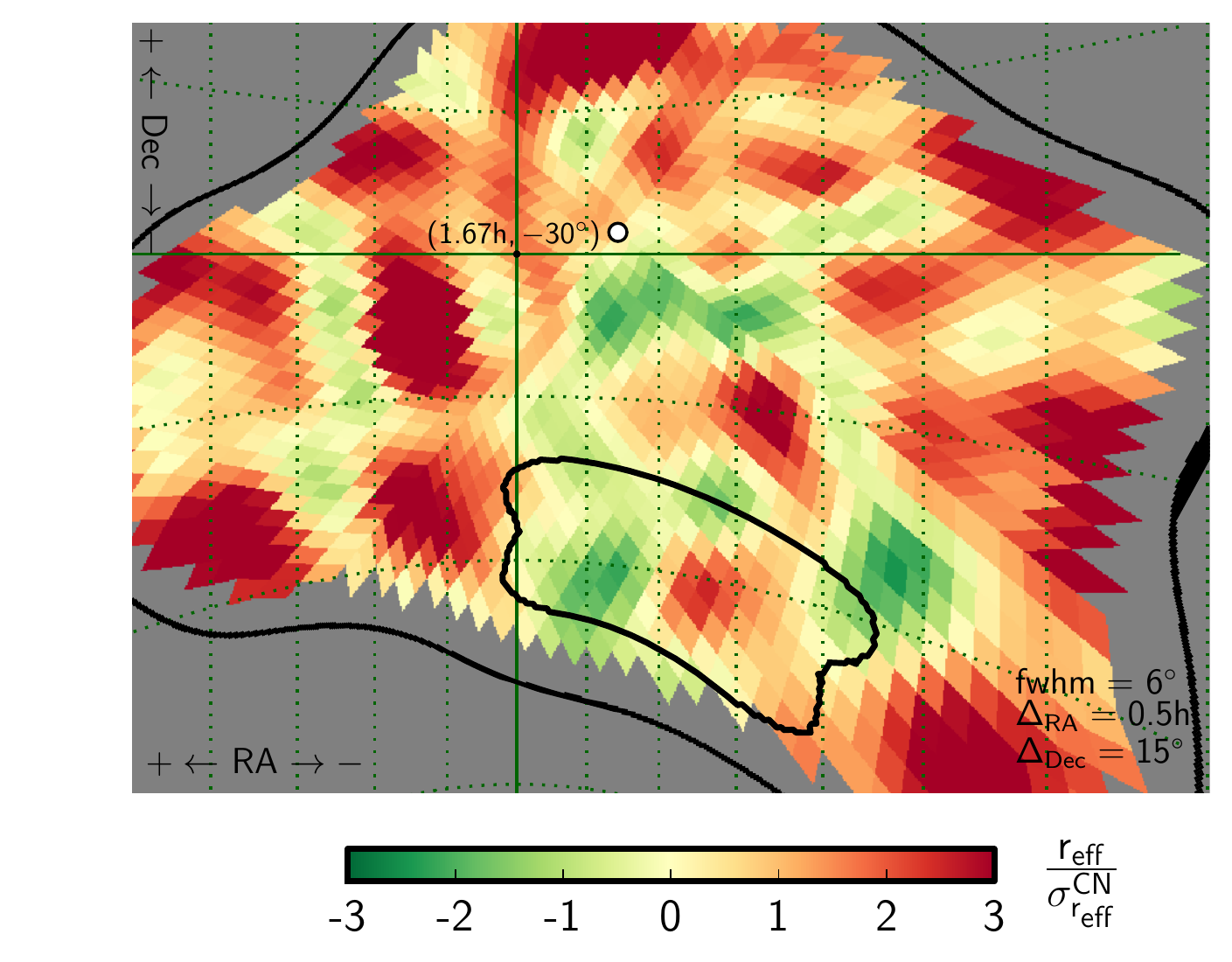}}
 \caption{Same as in Fig.~\ref{fig:fom_smth_north}, except that this figure zooms in on the potentially cleanest region near the Galactic south pole---marked by the white circle. Note that $r^{95}_{\rm eff}$ steadily decreases  on increasing the smoothing scale. Also note the significant detection in the BICEP region.} \label{fig:fom_smth_south}

 \end{figure}

Zooming in on these regions confirms this basic picture, as does smoothing the input $r_{\rm eff}$ maps.  We show the North in Fig.~\ref{fig:fom_smth_north} and the South in Fig.~\ref{fig:fom_smth_south}.  Since smoothing is an averaging operation, it tends to reduce noise in  noise-dominated regions and thus decrease the $r^{95}_{\rm eff}$ limit. On the other hand, smoothing does not much affect foreground-dominated regions, so long as the smoothing scale is smaller than scales over which the foreground changes.  How the statistics evolve with increasing smoothing scale thus provides insight. 
We use relative small Gaussian smoothing kernels with fwhm~$=2^\circ, 4^\circ,6^\circ$ (corresponding to $\sigma=0.85^\circ,1.7^\circ,2.55^\circ$), comparable to the pixel size, $\sim 1.8^\circ$ at $N_{\rm side}=32$. This choice of smoothing scales ensures that the statistical inferences made from the smoothed maps are still reliable. 

At the smallest smoothing scale (fwhm~=$2^\circ$), we find  the least contaminated region in the Northern Galactic Hemisphere has  $r^{95}_{\rm eff} = 0.28$ (see top left in Fig.~\ref{fig:fom_smth_north}).  This is smaller by almost a factor of 2 than the best region in the South, which has $r^{95}_{\rm eff}=0.47$ (see top left in Fig.~\ref{fig:fom_smth_south}, at the bottom edge of the BICEP patch).  On increasing the smoothing scale, the estimated $r^{95}_{\rm eff}$ limits in the North remains almost constant, though a steady increase is seen in the SNR, expected as a consequence of de-noising due to the smoothing operation.  This trend makes sense if $r^{95}_{\rm eff}$ is being dominantly determined by foregrounds in this region and the foreground do not vary much within the smoothing scale.  This bolsters our conclusion that here we are detecting a low level of foregrounds.

On the contrary, for the South, the $r^{95}_{\rm eff}$ limit drops  steadily as  the smoothing scale increases, by a factor of $\sim 2$ between fwhm~=$2^\circ$ and fwhm~=$6^\circ$. {The SNR does not rise steadily as in the North.} These trends makes sense if the lowest $r^{95}_{\rm eff}$ is being dominantly determined by the noise statistics, and hence steadily declines with smoothing due to de-noising.

Taken together, the implication is that higher sensitivity foreground ``quick-look'' measurements \cite{Kovetz2016} are unlikely to reduce the foreground upper limit in regions near the North ecliptic pole.
However, in and around the BICEP patch, the $r^{95}_{\rm eff}$ in the best regions is being determined by the noise fluctuations.  Hence it is very likely that the $r^{95}_{\rm eff}$ upper limit in these regions could be lowered further by making deeper measurements, as BICEP/Keck and other groups have continued to do.
%
%
These promising regions include the Eastern half (RA~$>0^\circ$) of the BICEP field, a region East of the BICEP field at 3 hr, a region at the northwest corner of the BICEP field, and a region southwest of South Galactic Pole.  Regions east  of the North Galactic Pole and west of the North Ecliptic Pole, at 14.5 hr and 11 hr, show similar denoising with smoothing.
The same metrics highlight nearly the same areas, with a coarser smoothing, when applied to the local power disc results.

We extend the smoothing kernels in steps of $2^\circ$ to find the scale beyond which $r^{95}_{\rm eff}$ in the southern sky stops declining. The South hits its minimum at fwhm~=$14^\circ (\sigma= 5.95^\circ)$, beyond which the minimum upper limit reaches $r^{95}_{\rm eff}=0.11$ in the BICEP patch.   This is an interesting level, close to the difference between the original detection at $r=0.2$ and the current, foreground-corrected upper limits on the tensor to scalar ratio $r \lesssim 0.09$ set by BICEP2/Keck \cite{bicepkeck2016}.  However, our limit must be treated cautiously because smoothing on these large scales makes the statistics more complicated: it generates additional correlations between pixels and so breaks our assumption of independent pixels.  Across all these smoothing scales the minimum $r^{95}_{\rm eff}$ upper limits for the northern sky hardly change.  

As noted, we find a foreground feature near the center of the BICEP patch, at RA~$\sim -5^\circ$, dec~$\sim -60^\circ$ (Fig.~\ref{fig:fom_smth_south}).  We see it at $3.16\sigma$ significance in the Planck 353 GHz data using our optimal estimator, although this does not take account of the likelihood of such a feature to be found anywhere in a patch the size of the BICEP field.  We also see  evidence for it in the non-optimal estimator and in an extremely simple pixel-based estimator (Sec.~\ref{sec:pixelspace}), at lower significance.  We see it in two multipole ranges, $l_1,l_2 \in [50,120]$ and $[120,370]$.  These give us confidence that this feature is present in the Planck data, {and is not simply a byproduct of our analysis method, even if the significance is ultimately found to have been boosted by a noise fluctuation.}
The $B$-mode map from the BICEP2/Keck team \cite{bicepkeck2015a} may show a hint of stronger $B$-modes at this same position.
This feature was not localized in the Planck team's disc analysis \cite{planck-intermediate-xxx} because of the coarseness of the disc size.  However, in \cite{planck-intermediate-xxx} they do note more power in the RA~$< 0^\circ$ portion of the BICEP field compared to the RA~$>0^\circ$ part, which our smoothing analysis shows is largely due to this feature.  In Fig.~\ref{fig:fom_smth_south} one can already notice the reduction in the SNR of this feature at fwhm=$6^\circ$, compared to the SNR seen at fwhm=$2^\circ$--$4^\circ$, as the power is being spread out by the wider kernel.
{Our assessment of its strength at 150 GHz also critically depends on the assumption that the spectral energy distribution for dust is common everywhere on the sky.}
We will continue to examine this feature in future work.

\section{Power spectrum of power anisotropy as a foreground diagnostic}
\label{sec:pam_power_spectrum}


Our estimators are constructed in harmonic space, so it is natural to study their power spectra---four-point statistics of the original map.  With them we can construct null tests for foreground contamination.  In a power spectrum the harmonic modes are averaged together, so this can be more sensitive than the map of the power anisotropy. 
The power spectrum also allows us to better assess the superiority of the optimal estimator $\hat P $ over the non-optimal estimator $\hat F^2$.

\begin{figure}[!t]
\centering
\includegraphics[width=0.8\columnwidth]{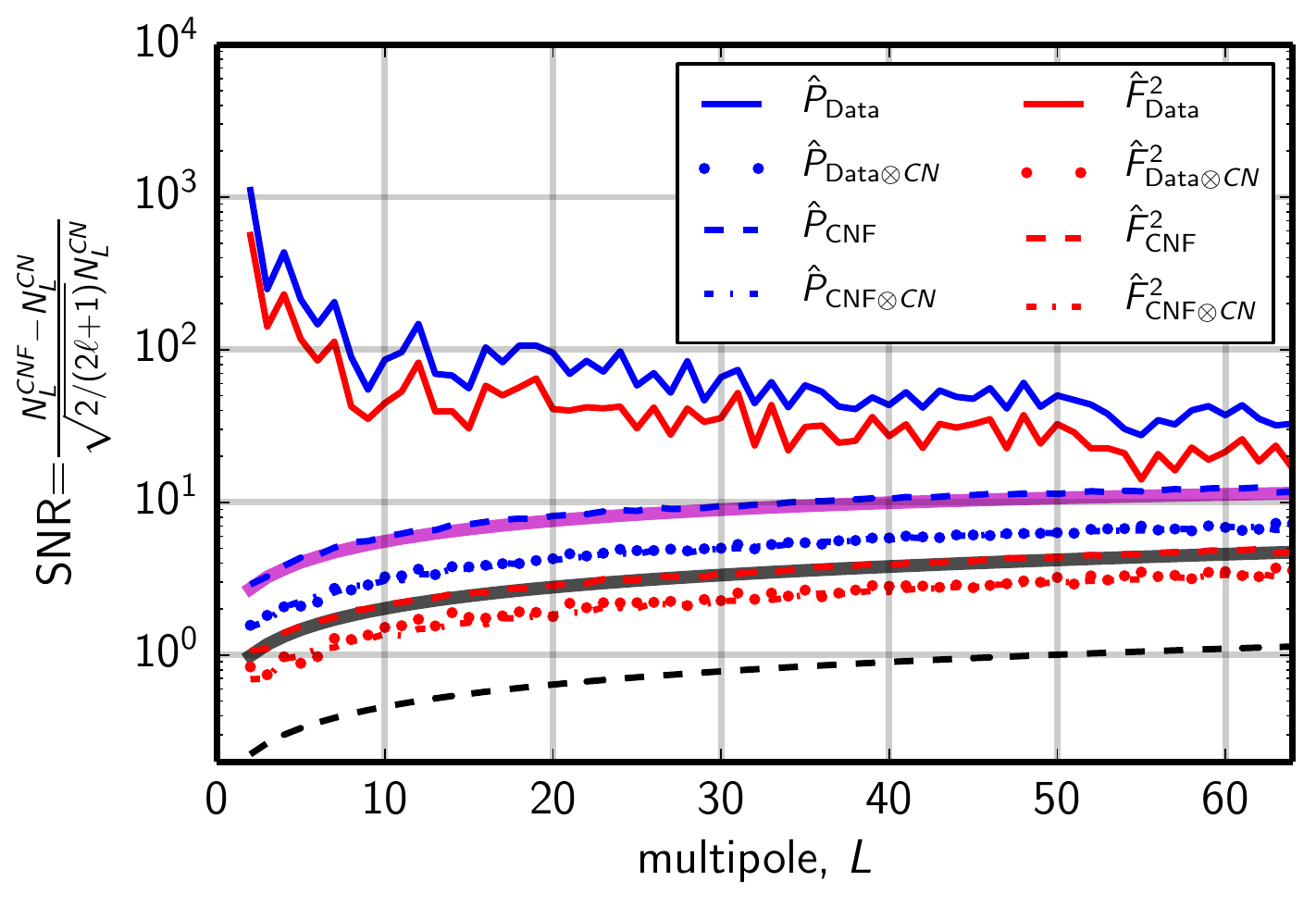}
\caption{Signal to noise ratio of the excess power in the power anisotropy due to the presence of foregrounds. Red curves depict the SNR per multipole for the power spectra derived from maps evaluated using the non-optimal estimator $\hat F^2$; blue curves depict the same for the optimal estimator $\hat P$. The solid lines show results derived from data, the dashed lines show results from ($CNF$) simulation, and the dotted and dashed-dotted lines show the contribution from cross correlation between noise and foregrounds in the respective maps. The magenta curve shows the analytic prediction of SNR per mode for the optimal estimator $\hat P$ while the black curve depicts the same for the non-optimal estimator $\hat F^2$; these agree well with the Gaussian and isotropic ($CNF$) simulations, but the non-isotropic and non-Gaussian data shows a significant excess. The dashed black line depicts the function $\sqrt{2L+1}$}
\label{fig:pam_spectra_snr}
\end{figure}
%

We ask the following question: \textit{Does any angular scale in the power anisotropy contain power in excess of expectations from CMB and noise alone?}
This is a null test for contamination.
In the case of Gaussian and statistically isotropic data, we can analytically express the power spectrum of the power anisotropy maps ($\hat F^2$ or $\hat P$).  It depends on the total power spectra of the components (CMB, noise, and foregrounds) and is given by,
\begin{equation}
N_L = \sum_{\ell_1 \ell_2} \left[\frac{w^L_{\ell_1 \ell_2}}{G^L_{\ell_1 \ell_2}} \right]^2 [C^{\rm noisy}_{\ell_1} C^{\rm noisy}_{\ell_2} + C^{\rm clean}_{\ell_1} C^{\rm clean}_{\ell_2}] \,,
\label{rec-noise}
\end{equation}
where $C_{\ell}^{\rm noisy}$  denote the auto correlation spectra for the noisy data maps ($CNF$ or $CN$) and $C_{\ell}^{\rm clean}$ denote the auto spectra of noiseless maps ($C$ or $CF$). {This expression assumes all components are Gaussian and isotropic, so does not accurately capture anisotropic noise in the data. These analytical estimates are only used to make comparative checks with simulations and hence this inadequacy is not a major concern.}

We can compare the analytic estimate to simulations (and later to data), expressing the result in terms of a power signal-to-noise, shown in Fig.~\ref{fig:pam_spectra_snr}.  We examine the average power spectrum of power anisotropy (of $\hat{F}^2$ and $\hat P$) from 1000 simulations.
The power in estimators based on CMB and noise alone is given by $N_L^{CN}= \langle N_{L}^{\hat E_{CN}- \langle \hat E_{CN} \rangle} \rangle$ (referred to as the reconstruction noise in the weak lensing context), while the larger power from including foregrounds is $N_{L}^{CNF}= \langle N_{L}^{\hat E_{CNF} - \langle \hat E_{CN} \rangle} \rangle$.  To quantify the excess from foregrounds,  we divide the difference of these, $(N_{L}^{CNF} - N_{L}^{CN})$, by the error per mode, $(2/(2L+1))^{1/2} N_{L}^{CN}$, yielding the power signal-to-noise ratio (SNR) per multipole. In the absence of foregrounds this measure should be consistent with zero.  When we consider realistic simulations and data, the reconstruction noise also carries information about the anisotropic noise, which constitutes a power bias. However, since both $N_L^{CNF}$ and $N_L^{CN}$ come from from maps with anisotropic noise, this bias cancels out.  This subtraction is thus important in the definition of this null diagnostic. }

The SNR curves shown in Fig.~\ref{fig:pam_spectra_snr} (dashed curves)  for power anisotropy estimators $\hat{F}^2$ and $\hat P$ are in excess of zero, and in the case of Gaussian foreground ($CNF$) simulations, match those predicted using Eq.~\ref{rec-noise}.  {Here we are simply detecting the power due to the addition of the foreground field.}  Because of the noise weighting, the optimal estimator ($\hat P$) has higher SNR than the non-optimal one ($\hat F^2$). 

When we also compare to the Planck 353 GHz $B$-mode data, we see larger SNR  (solid curves in  Fig.~\ref{fig:pam_spectra_snr}), and this is due to the isotropy-violating (and non-Gaussian) character of the foreground.  We see that the SNR is in excess of $10 \sigma$ at all multipoles, and again the optimal estimator outperforms the non-optimal one.  The optimal estimator is better by a factor of $\sim 2.7$ for the ($CNF$) simulations and by a factor of $\sim 1.8$ for 353 GHz data. 

{The SNR curves in Fig.~\ref{fig:pam_spectra_snr} for ($CNF$) simulations increases with $L$ for the isotropic Gaussian simulation, roughly as $(2L+1)^{1/2}$. This is because the variance of the field falls as the inverse of the number of modes accessible ($\propto (2L+1)^{-1}$) and because both $N_L^{CN}$ and $N_L^{CNF}$ from simulations are nearly constant functions in this case.  By contrast, for data the SNR is highest on the largest angular scales and drops as a function of $L$. Fig.~\ref{fig:op_data} shows that the southern sky has more foreground power than the North, and this is expected to show up as large scale power seen in  Fig.~\ref{fig:pam_spectra_snr}. On small angular scales, there are only a few obvious, discrete foreground sources seen in the map, indicating that large portions of the sky are noise dominated, and hence the SNR drops for higher $L$.}

We noted in the previous section that the western portion of the  BICEP field has more power than the East.  If this is true, by analogy this should show up as large scale power in the power spectrum of a power anisotropy map, if such a statistic were evaluated on the BICEP/Keck data.

The foregrounds contribute additional power primarily via auto-correlations, but also due to chance correlation with CMB and noise. We quantify this contribution by evaluating the SNR for the power spectrum of power anisotropy maps derived from cross-correlating either data or ($CNF$) simulations with random ($CN$) simulations.  The SNR contribution from chance correlations is shown in Fig.~\ref{fig:pam_spectra_snr},  represented by dotted lines for data and by dash-dotted lines for simulations. These curves are seen to lie of top of each other for the respective estimators, indicating that the excess power due to chance correlations is only determined by the power spectrum of foregrounds and is insensitive to their actual non-Gaussian \& non-isotropic nature.  The real foregrounds are concentrated into discrete features and hence one may expect a small contribution to the total power from chance correlations, unlike in the isotropic Gaussian foreground simulations, where the contribution of chance correlations to total power may be relatively higher. By studying the respective power spectra in Fig.~\ref{fig:pam_spectra_snr}, we note that for real data the chance correlations contribute approximately $10 \%$ of the total power, where as for isotropic foregrounds the chance correlations contribute up to $50-60 \%$ of the total power. 

Since the estimator maps themselves are derived from two-point correlators, these power spectra are measures of four-point functions, which could probe non-Gaussianity or the power associated with the isotropy violating component of the CMB map \cite{hajian2003}.  Since the actual foregrounds are strongly non-Gaussian and non-isotropic, we see what we expect: the power spectra of the foreground tracer maps derived from data are significantly greater from those derived from Gaussian simulations of data maps ($CNF$).




%

\section{Conclusions}\label{conclusions}


In the next decade, CMB experiments will search for primordial $B$-modes, but polarized emission from the Galaxy is perhaps strong enough everywhere to swamp the signal.  
Here we have explored isotropy-violation tests for foreground contamination.   The techniques are generic and can be applied to any CMB data set.  We focused on the $B$-mode maps derived from Planck 353 GHz data and used the bipolar spherical harmonic basis to construct estimators of isotropy violation. Specifically, we studied a non-optimal foreground-squared estimator ($\hat F^2$) and the related, noise-weighted optimal estimator ($\hat P$).  We can use such tests in concert with multi-frequency data.

Both these BipoSH-based estimators are closely related to local power spectrum analysis on discs \cite{planck-intermediate-xxx,Krachmalnicoff2015}, which we have repeated with minor but well-justified modifications.  {We broadly reproduced the features of those previous analyses, thus testing our pipeline and verifying those results.}  In the disc analysis, the resolution is limited by the disc size, but our new estimators can achieve higher resolution.  We cross-correlated two years of Planck data, assessing the statistical significance with an ensemble of simulations.  In our error estimates we accounted for the excess noise due to chance correlation of foregrounds with CMB and instrument noise.  

We set upper limits for foreground emission and examined signal-to-noise maps.  The lowest upper limits come from the North Ecliptic Pole where the Planck noise is low and from the Eastern portion of the BICEP field.  Other regions with potentially low foreground are near the South Galactic Pole, east and northwest of the BICEP field, {and a couple regions in the North}.  The declinations for these cleanest places have implications for siting telescopes at the South Pole, the Atacama in Chile, or Greenland \cite{Araujo2014}.

It is important to examine upper limits together with maps of the signal-to-noise.  Further measurements can improve the limits, but only in regions of low signal-to-noise.  We explored both as a function of smoothing scale, and we argued that more observations will help to characterize the foreground in the South and in some areas of the North, but at the North Ecliptic Pole the foregrounds (though low) are already being seen by Planck.

We found a potential foreground feature near the center of the BICEP field.  It needs further exploration of its significance.  Our assessment of its strength relies on the global extrapolation of dust from 353 GHz to 150 GHz; individual features may have a different spectrum.  An East--West power spectrum jackknife of the BICEP/Keck field may shed light on whether this feature is real and important.


{Beyond the blackbody spectrum and statistical isotropy, nearly-Gaussian fluctuations characterize the CMB; the power spectrum of our foreground tracer maps is sensitive to deviations. The excess power we see compared to  Gaussian, isotropic simulations may indicate non-Gaussianity of foregrounds present in the data.  However, we must be cautious since anisotropic but Gaussian foregrounds can also generate similar excess power; the power spectra of the foreground tracer maps by themselves cannot separately quantify these two characteristics, but can provide a null test since the CMB's genuine primordial $B$-mode should contain neither.
  
  This also suggests an exploration of estimators which measure the direction-dependent bispectrum and trispectrum. These will provide a direction-dependent measure of deviations from Gaussian statistics expected for CMB and noise alone. These are closely related to the estimators studied in this work \cite{Hu2001}. Detailed exploration of this idea is left for future work.}


Our methods are currently blind, and while blind searches for foregrounds are useful and conservative, they can be improved in two ways by incorporating more physics of the foregrounds.  First, with knowledge of the astrophysical processes underlying the emission, we can design weights that focus like a matched filter on the isotropy violations specific to the Milky Way's dust distribution and polarization.  The suggestion of  Kovetz and Kamionkowski \cite{kamionkowski_kovetz_2014}---look for coherent patches in the polarization direction---is one straightforward approach.
We may be guided by the structure in the interstellar medium, indicated by microwave data \cite{planckskymodel}, or other tracers like neutral hydrogen, where structures align with dust polarization \cite{Clark2015}.

Second, we can improve over blind searches by using models of the emission to establish a realistic prior for our assessment of the posterior distribution of foreground power.  We will need detailed, physical models of the interacting dust, gas, and magnetic fields to understand what is possible, and likely, for foregrounds.  This requires magnetohydrodynamics to probe all the physics, and a suite of galaxy-scale simulations to build statistical templates of the foreground.  This effort can also help us with the optimal weighting too, highlighting the isotropy-violating terms for dust emission, so that we can search for them explicitly. 

In this work we have focused on $B$-mode maps, but these estimators are easily generalized to cross-correlations  studies ($TB$, $EB$, etc.), which could be independent tracers of foregrounds.  We have also begun to cross-correlate maps from two detector frequency bands, taking advantage of the distinct frequency dependence of foregrounds and $B$-mode signals. {The estimators presented here only use the even parity BipoSH spectra.  The odd parity modes may contain orthogonal information.}

{ The isotropy-violation estimators presented here offer several advantages.  We can measure contributions to the isotropy violation from specific spectral modes, and target those where we expect primordial $B$-modes are largest.  We can control the spatial resolution of the reconstructed foreground tracer.  We can down-weight the contribution of noisy modes.  We can incorporate---in the future---additional physical knowledge about foregrounds to build matched filters and realistic priors for foreground isotropy violation. Finally, these estimators are closely connected to non-Gaussianity estimators, another characteristic distinguishing foregrounds from CMB.}

Primordial $B$-modes allow us to peer back to the first moments of the Universe. Thus they may hold keys to understanding the fundamental physical laws that govern inflation. This is why searching for $B$-modes is a core priority for CMB research.  Work on isotropy-violation now helps to identify the cleanest portions of the sky.  It will help to set diagnostic criteria to establish the robustness of future $B$-mode measurements.  Such foreground studies will be more and more important in the coming years as experimental designs approach the capability to reach $r = 10^{-3}$.


\section*{Acknowledgments}
We thank Ely Kovetz and Antonio Linero for useful discussions.
We acknowledge support from NASA and the US Planck project through Jet Propulsion Laboratory subcontract 1492773.  The Florida State University Office of Research supported this work through a Planning Grant from the Council on Research and Creativity. Some of the results in this paper have been derived using the HEALPix \cite{Healpix} package. This research used resources of the National Energy Research Scientific Computing Center, which is supported by the Office of Science of the U.S. Department of Energy.


\bibliographystyle{JHEP}
\bibliography{ref}

\appendix

\section{Mask dependence of estimates} \label{sec:mask_dependence}
\begin{figure}[!t]
\centering
\subfigure[GAL60 mask]{\label{fig:pam_gal60}\includegraphics[width=0.49\columnwidth]{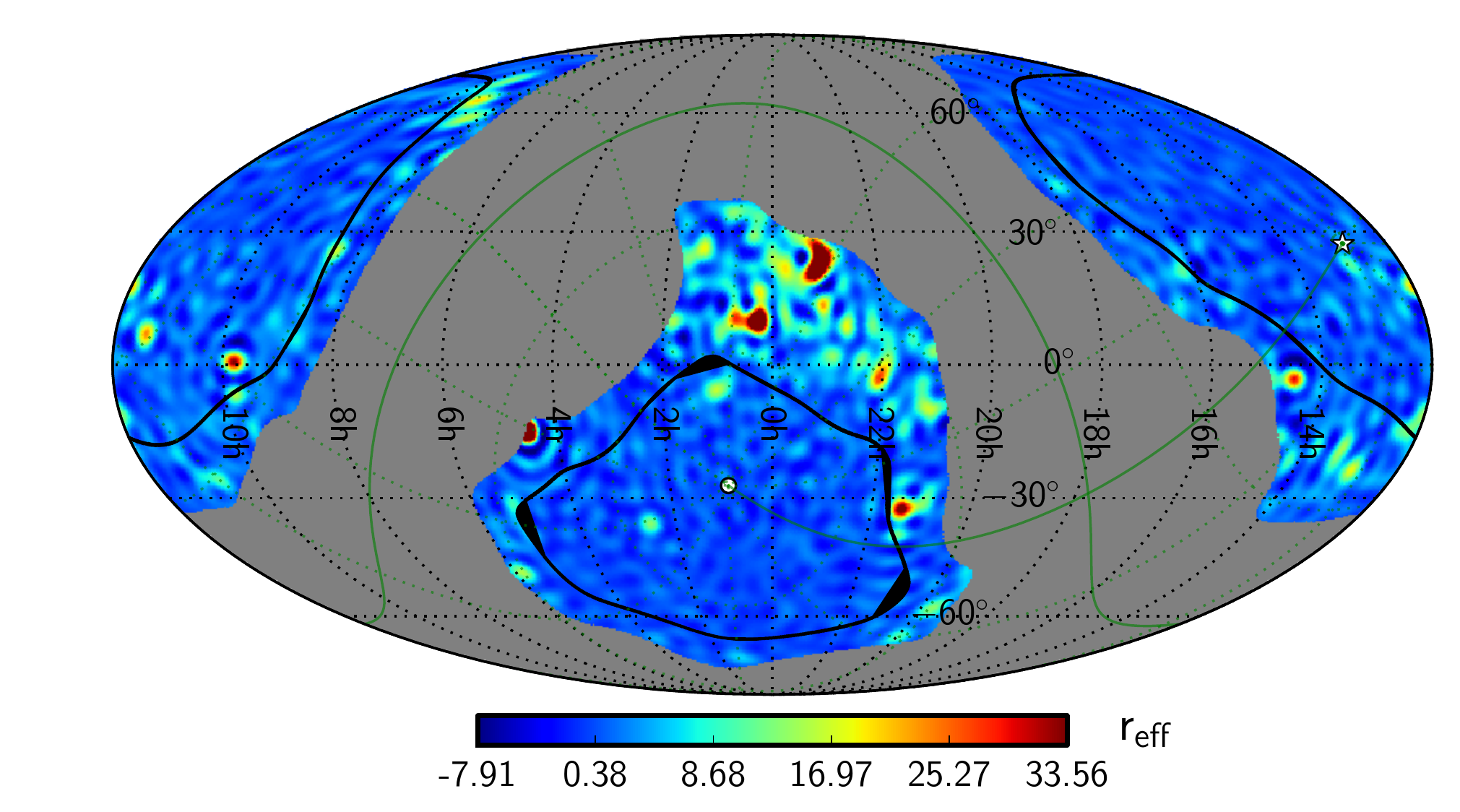}}
\subfigure[GAL40 mask]{\label{fig:pam_gal40}\includegraphics[width=0.49\columnwidth]{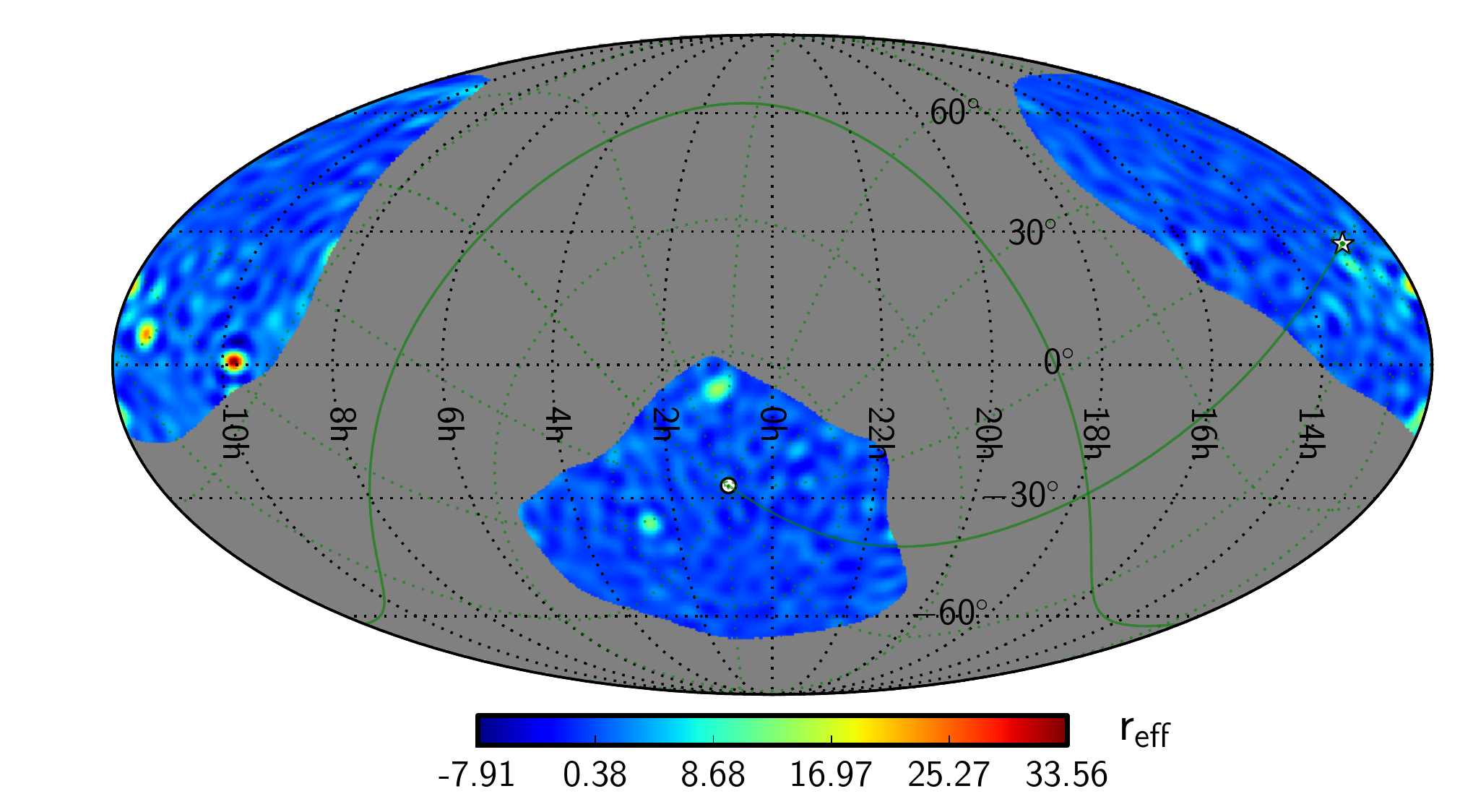}}
\subfigure[Difference]{\label{fig:pam_bias}\includegraphics[width=0.49\columnwidth]{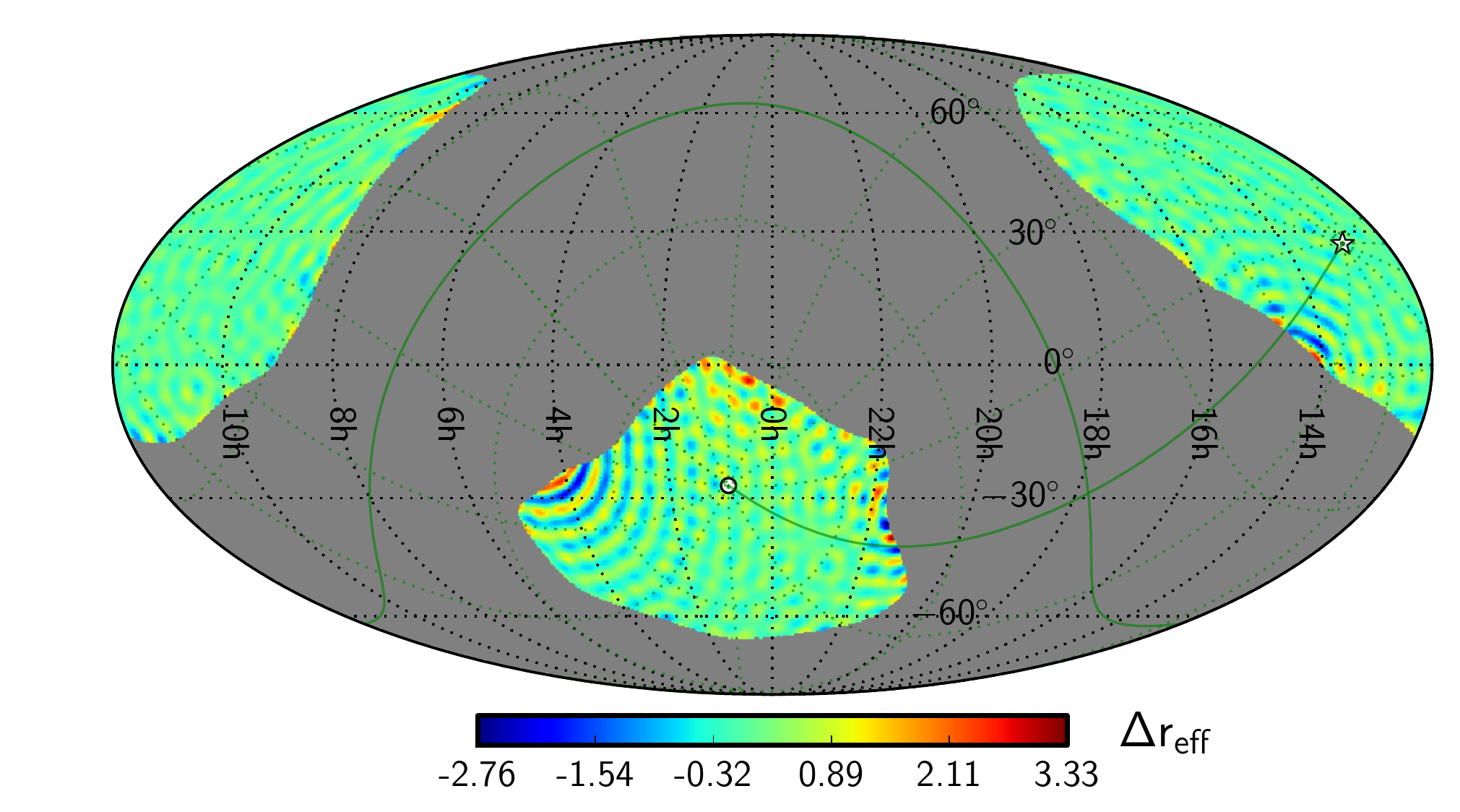}}
\subfigure[Difference normalized to error]{\label{fig:pam_bias_snr}\includegraphics[width=0.49\columnwidth]{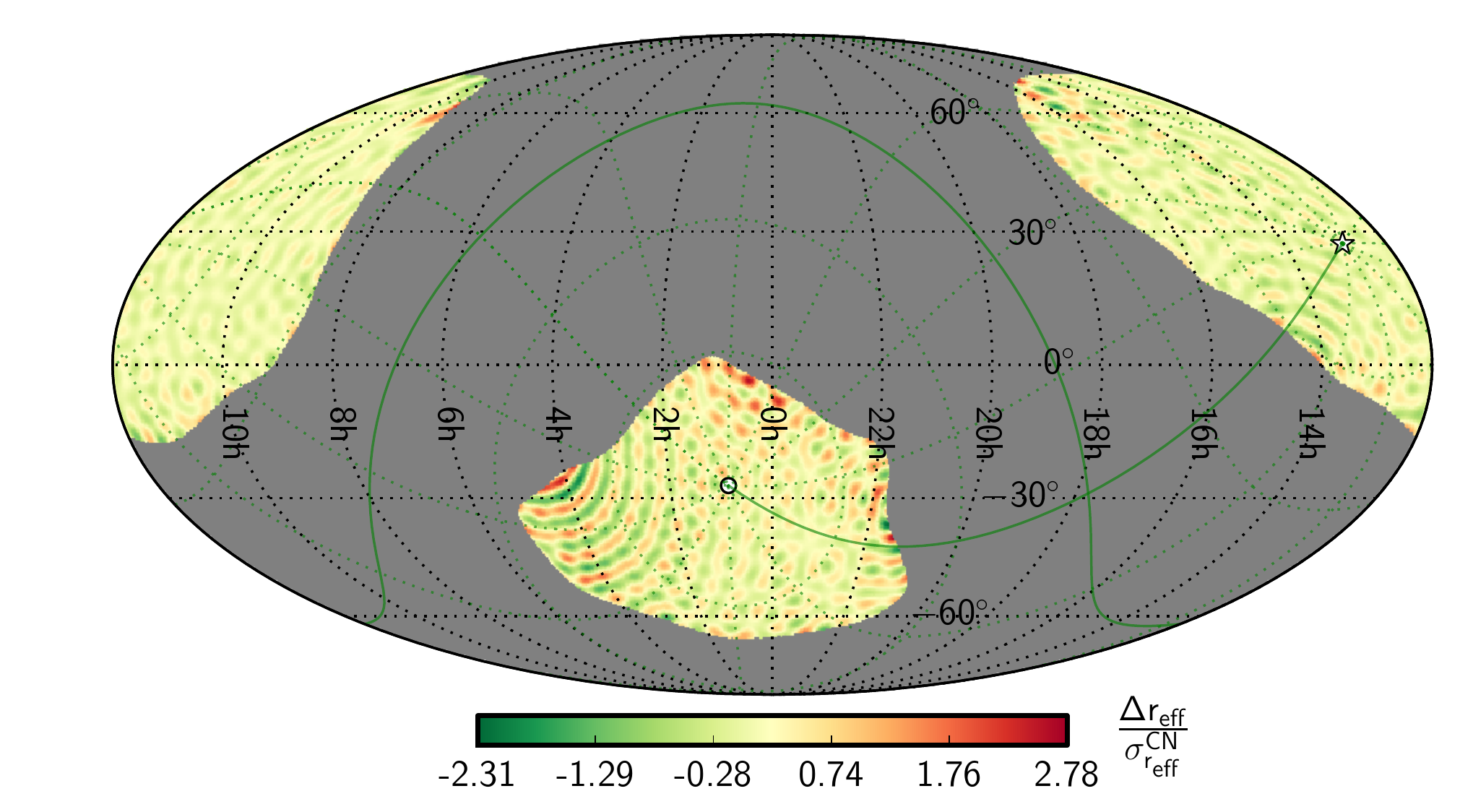}}
\caption{Dependence of optimal estimator results on mask, based on Planck $B$-mode maps at 353 GHz. \textit{Top left}: Using GAL60 mask for processing; we mark GAL40 for comparison. \textit{Top right}: Using GAL40 mask for both processing and display. The color scale is common between the top two plots. \textit{Bottom left}: Difference of the top two maps, showing that they contain the same information, except for ringing sourced by strong foregrounds close to the edges of the mask. \textit{Bottom right}:  Difference compared to the error per pixel (CMB plus noise simulations).  {Only foreground objects in the GAL60 mask, but which are excluded by the GAL40 mask, cause significant artifacts are due to ringing.  Objects within the GAL40 mask leave no significant residuals between the two masks, and hence we do not expect masking effects to alter our results.}}
\label{fig:pam-mask-dependence}
\end{figure}
The proposed harmonic space estimators operate on the full sky. However in practice they require a mask to exclude the most strongly foreground contaminated regions from the analysis. The measurement of the BipoSH spectra implicitly assumes the orthonormality of the spherical harmonic functions, a condition no longer satisfied on the masked sphere. Consequently we must check that the proposed estimators are reliable and not biased due to masking. 

We probe the consistency of harmonic space  foreground-squared estimators in the common area of two different masks.  Specifically we check if the reconstructed field in the GAL40 sub-region of the GAL60 mask is consistent with the maps estimated using the GAL60 mask. The results are summarized in Fig.~\ref{fig:pam-mask-dependence}. Many of the features that differ between the two maps are clearly remnant ringing artifacts from strongly peaked sources outside borders of the GAL40 mask.  These ringing artifacts are due to the band limit and not the masking.  Meanwhile, high-significance foreground features within the GAL40 mask appear the same in both the GAL40- and GAL60-derived maps, as compared to the errors in Fig.~\ref{fig:pam_bias_snr}, so we conclude mask-related biases are not a severe concern.


\section{Analogy with weak lensing reconstruction} \label{lensing_analogy}

The estimators here are closely related to weak lensing estimators \cite[e.g.][]{Hu2002a}. In the BipoSH basis, it can be shown that the BipoSH spectra for the weak lensed CMB sky is given by the following equation,
\begin{equation}
D^{LM}_{\ell_1 \ell_2} = X^{LM}_{\ell_1 \ell_2}  + \phi_{LM} G^{L}_{\ell_1 \ell_2} + \mathcal{O}(\phi^2) \,,
\end{equation}
where $X^{LM}_{\ell_1 \ell_2}$ are the coefficients of the unlensed CMB, $\phi_{LM}$ are the spherical harmonic coefficients of expansion for the projected lensing potential, and $G^{L}_{\ell_1 \ell_2}$ denotes some function  specific to lensing that is completely determined by the angular power spectrum of the lensed CMB sky. Keeping the lensing potential $\phi$ fixed, the lensed CMB sky is statistically anisotropic when averaged over CMB realizations, but the unlensed sky is isotropic: $\langle  X^{00}_{\ell_1 \ell_2}  \rangle= C_{\ell}, ~ \langle  X^{LM}_{\ell_1 \ell_2}  \rangle_{L \neq 0} = 0$. Hence the BipoSH spectra directly measure the cosmic lens: $\hat \phi_{LM} = D^{LM}_{\ell_1 \ell_2}/G^L_{\ell_1 \ell_2}$ \cite{aich_rotti2015}.

Analogously, in the case of the foreground contaminated CMB sky, the BipoSH spectra are given by the following expression,
\begin{equation}
D^{LM}_{\ell_1 \ell_2} = X^{LM}_{\ell_1 \ell_2}  + F^{LM}_{\ell_1 \ell_2} + [XF]^{LM}_{\ell_1 \ell_2} \,.
\end{equation}
In spirit similar to CMB weak lensing, keeping the foreground realization fixed and averaging over CMB realizations (so that $\langle  X^{00}_{\ell_1 \ell_2}  \rangle= C_{\ell}$, $\langle  X^{LM}_{\ell_1 \ell_2}  \rangle_{L \neq0}=0$, and $\langle  [XF]^{LM}_{\ell_1 \ell_2}  \rangle=0 $), the BipoSH spectra directly measure the foregrounds. The BipoSH spectra can be combined linearly to make an estimate of the foreground squared field as discussed in Section \ref{f2_estimator}. {From the measured foreground covariance  $F^{LM}_{\ell_1 \ell_2}$, we can also generate mock samples of realistic foreground skies using the technique of Cholesky decomposition.}

In the case of lensing reconstruction, the monopole is absent, since the gradient of a constant projected lensing potential vanishes and hence does not induce additional correlations in the CMB sky ($G^L_{\ell_1 \ell_2}$ vanishes for $L=0$.). In the case of foreground reconstruction however, there is important information in the monopole. In fact, this is the only mode ($L=0$) where statistically isotropic CMB and foregrounds are completely degenerate.  So working only with the angular power spectrum---the $L=0$ BipoSH spectra---one must resort to using multifrequency information or foreground models to subtract foregrounds from CMB \cite{Tegmark2003_ilc}.

\section{Optimal weights for estimator} \label{weights}
\begin{figure}[!t]
\begin{center}
\subfigure[]{\label{w_vary_ell}\includegraphics[width=0.5\columnwidth]{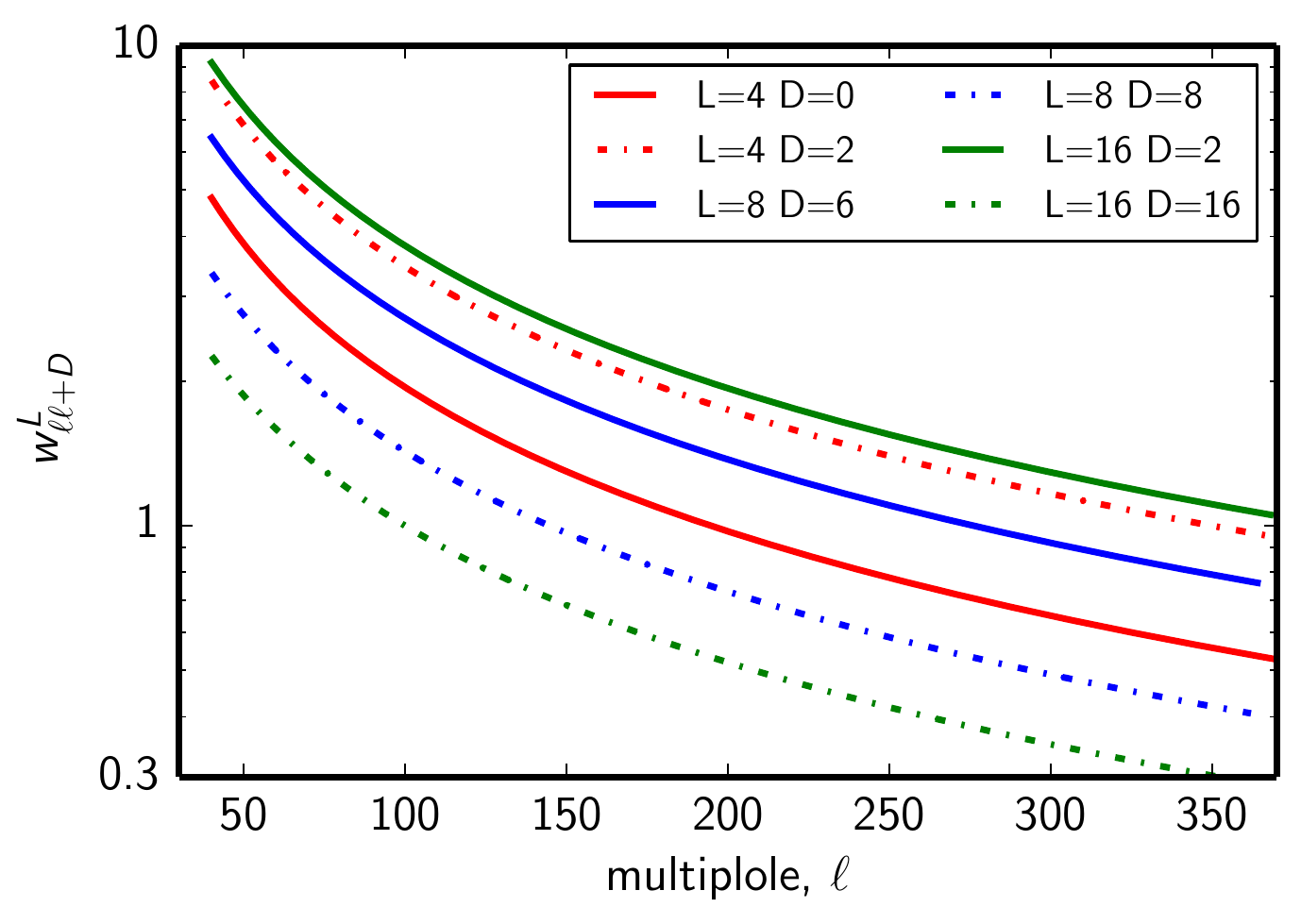}}
\end{center}
\subfigure[]{\label{w_vary_L}\includegraphics[width=0.5\columnwidth]{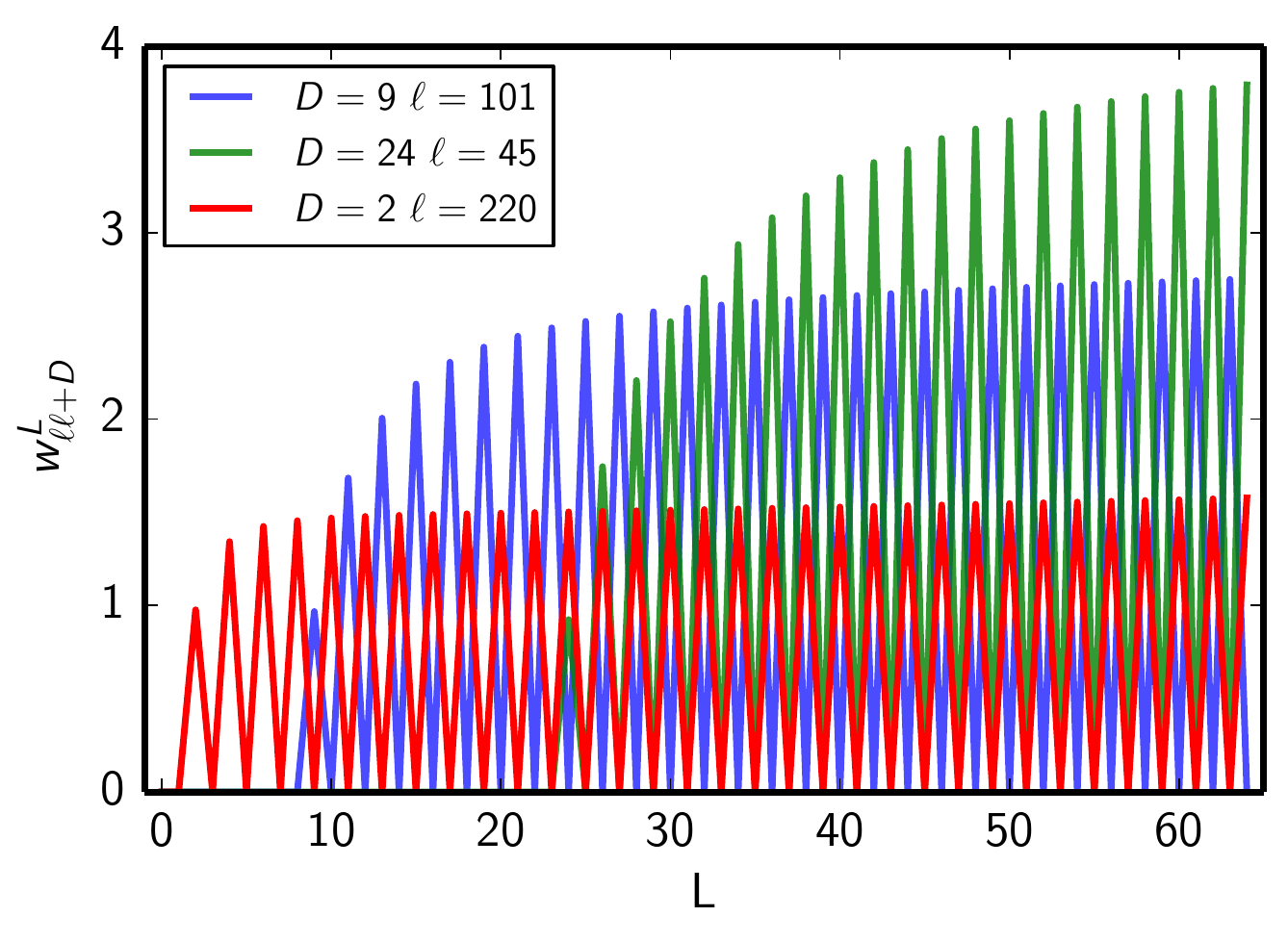}}
\subfigure[]{\label{w_vary_D}\includegraphics[width=0.5\columnwidth]{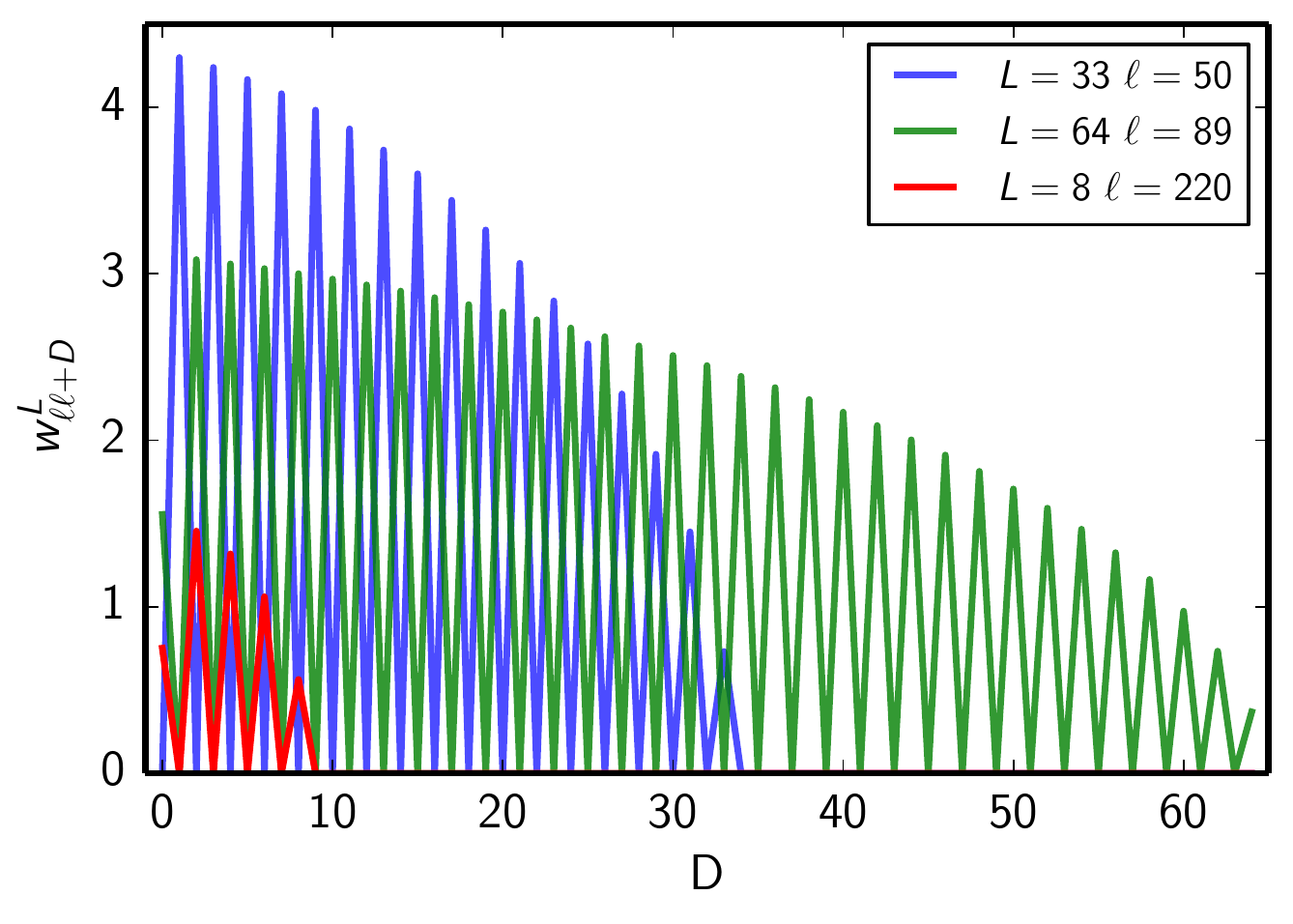}}
\caption{\textit{Top}: Weights as a function of $\ell$ while keeping $L$ and $D$ fixed. \textit{Bottom left}: Weights as a function of $L$ while keeping $\ell$ and $D$ fixed. \textit{Bottom right}: Weights as a function of $D$ while keeping $L$ and $\ell$ fixed.}
\label{op_pam_weights}
\end{figure}
In Section \ref{power_anisotropy_estimator}, we have already noted that the optimal power anisotropy estimator yields a biased estimate of the foreground-squared field, due to weights operating on the BipoSH spectra associated with foregrounds. Here we briefly describe the characteristics of the weights. The weights $w^L_{\ell_1 \ell_2}$ are equivalently denoted by $w^L_{\ell \ell+D}$, where $D$ is the difference between the two multipoles $\ell_1$ and $\ell_2$. In Fig.~\ref{op_pam_weights} we show a sampling of how different modes are weighted. 

The small angle modes on the CMB sky receive less weight (Fig~\ref{w_vary_ell}) since they are noise dominated. The oscillatory behavior seen in Fig~\ref{w_vary_L} and Fig.~\ref{w_vary_D} is because modes only contribute when $(L + D)$ is even.  When $(L+D)$ is odd, the weights vanish. In Fig~\ref{w_vary_L}, the weights are non-zero only when $D \le L$ which is representative of a part of the triangularity relation the Clebsch-Gordon coefficients need to satisfy.  In the specific analysis presented here, the noise power changes mildly across the multipole range.  Thus variations in the weights are dominantly determined by the geometric factor (compare Eq.~\ref{geom_factor}),
\begin{equation}\frac{\Pi_{\ell_1}\Pi_{\ell_2}}{\Pi_{L}} \mathcal{C}^{L0}_{\ell_1 0 \ell_2 0},\end{equation}
 since at each $L$ the CMB modes that contribute are fixed when we fix $\ell$ and $D$.  This condition on the noise may not hold in every situation.  For fixed $L$ and $\ell$ (Fig~\ref{w_vary_D}), when we increase $D$ we probe contributions from modes at smaller angular scales ($\ell + D$) in the CMB map.  These modes are noisier and so the weight declines with increasing $D$.

\section{Detecting B-mode foregrounds in a futuristic experiment} \label{forecast}
\begin{figure}[!t]
\centering
\includegraphics[width=\columnwidth]{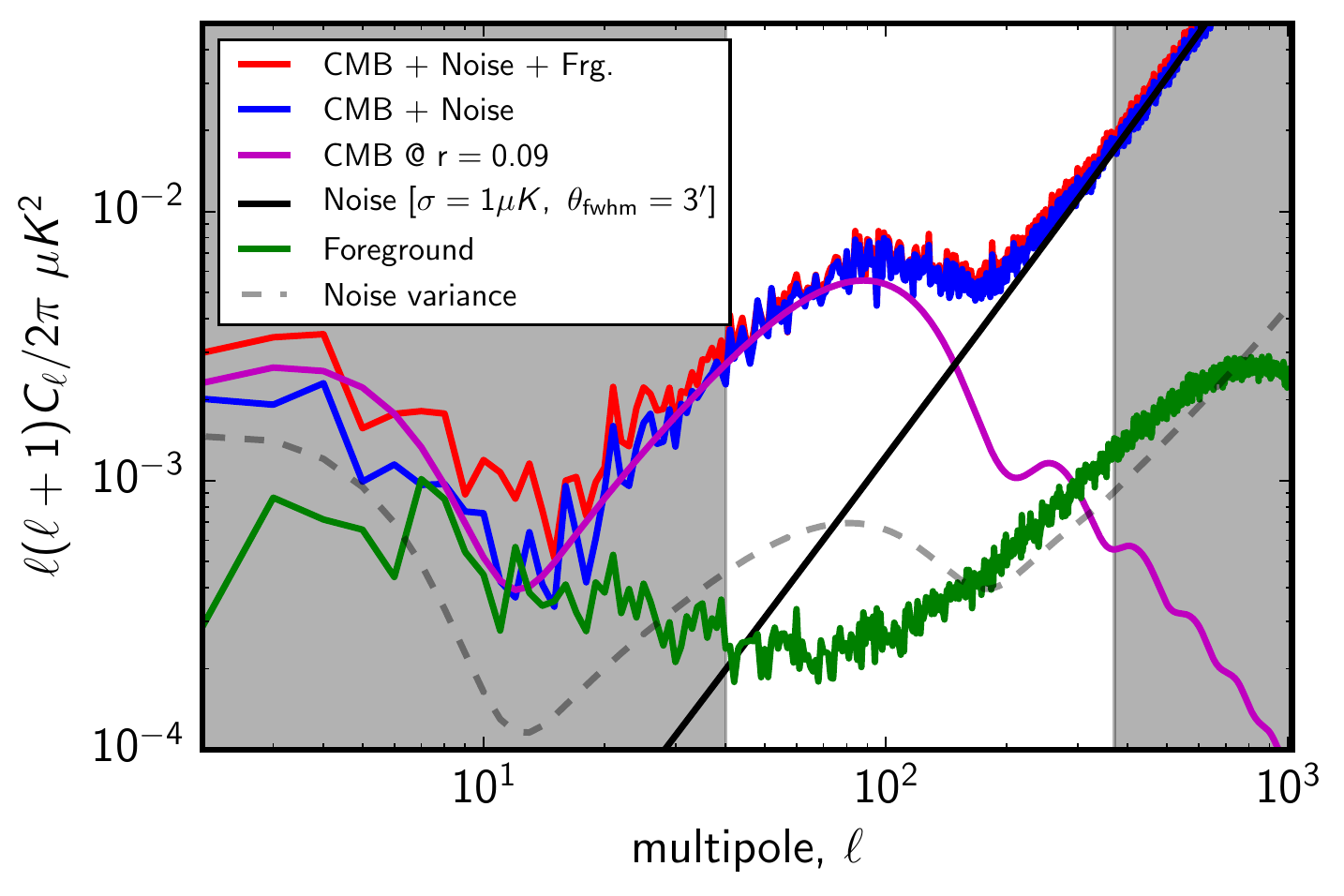}
\caption{This figure depicts the power spectrum of various components of the simulated maps. Only multipoles in the white band are used for the analysis.}
\label{fig:cmb_noise_frg_spectra}
\end{figure}
In the near future we will have significantly improved measurements of the B-mode CMB sky, made possible by improvements in detector sensitivity and frequency coverage, allowing for improved foreground cleaning.  In this section we demonstrate the usability of the proposed estimators in the case where the power spectrum of the foregrounds is below that of CMB and instrument noise. Here we intentionally consider the extreme case where the foregrounds have been suppressed to an extent where their power is comparable to the noise variance. 

\begin{figure}[!t]
\subfigure[CMB + Noise]{\label{fig:cn_map}\includegraphics[width=0.5\columnwidth]{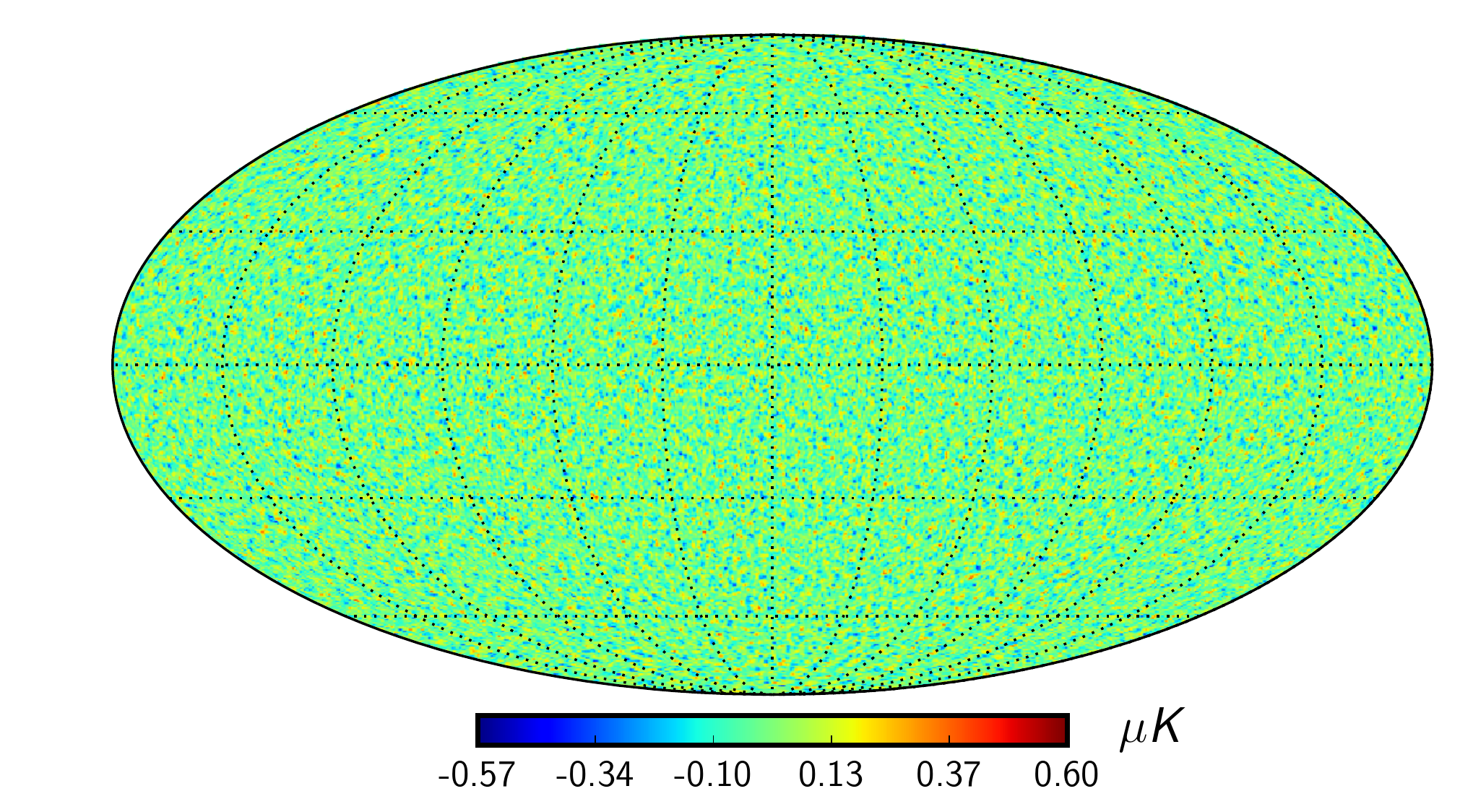}}
\subfigure[CMB + Noise + Foreground]{\label{fig:data_map}\includegraphics[width=0.5\columnwidth]{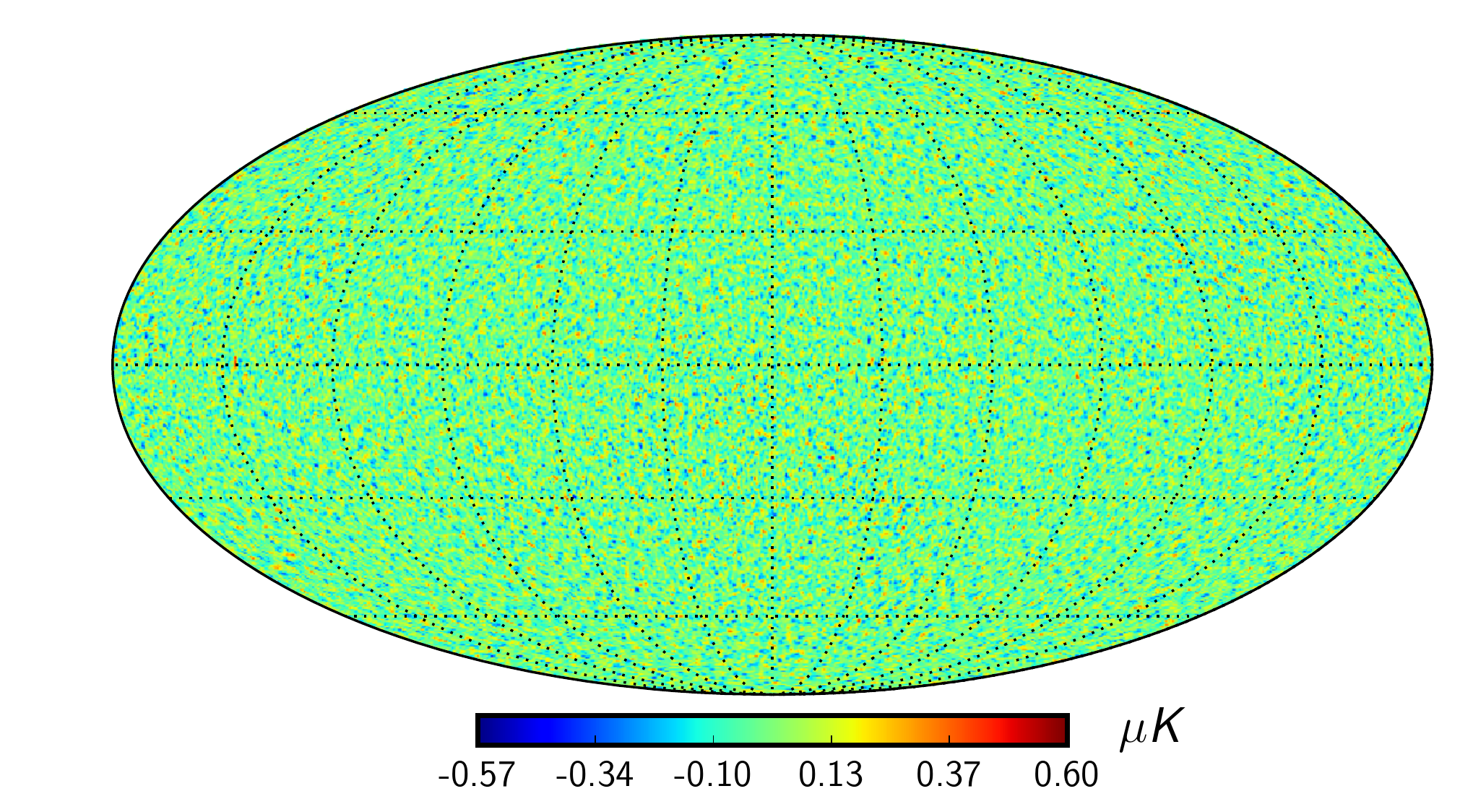}}
\caption{For a futuristic experiment, these figures depict the B-mode maps used for the analysis. These maps are filtered such that only multipoles in the band $\ell=[40,370]$ are retained, with apodization identical to that described in Section~\ref{power_anisotropy_estimator}. The map on the left contains only CMB and noise while the one on the right is foreground contaminated. The power spectra of the respective maps are depicted in Fig.~\ref{fig:cmb_noise_frg_spectra}}
\label{fig:simulated_data}
\end{figure}

For simulating the CMB sky we assume primordial B-mode corresponding to $\rm r =0.09$, but ignore the lensing B-modes. For noise we assume CMB-S4 \cite{cmbs4} characteristics, with detector noise of $\sigma_{\rm fwhm} = 1 \mu K$  and a  resolution of  $\theta_{\rm fwhm}= 3 {\rm ~arcmin}$. As polarized foreground template we use the maximum likelihood dust polarization maps derived from Commander \cite{planckdiffusefrgs2015a}. We mask the brightest portions of the foreground template using the GAL40 mask. Further,  we rescale the template foreground by a factor of  $1/\sqrt{300}$. This rescaling is chosen such that the foreground B-mode spectrum is comparable to the noise variance at all multipoles used for the analysis\footnote{Note that even if one masks the simulated CMB plus noise simulation using the GAL40 mask, the $\rm f_{\rm sky}$ suppressed power spectrum is still above that of foregrounds at all multipoles used in this analysis.}. The angular power spectra of various components of data, resulting from the choices described above, are shown in Fig.~\ref{fig:cmb_noise_frg_spectra}. The contaminated and uncontaminated B-mode maps used for the analysis are shown in Fig.~\ref{fig:simulated_data}. Note that, neither visual inspection of the maps, nor the power spectrum make the foregrounds obvious. 

\begin{figure}[!t]
\subfigure[CMB + Noise]{\label{fig:forecast_cn_pam}\includegraphics[width=0.5\columnwidth]{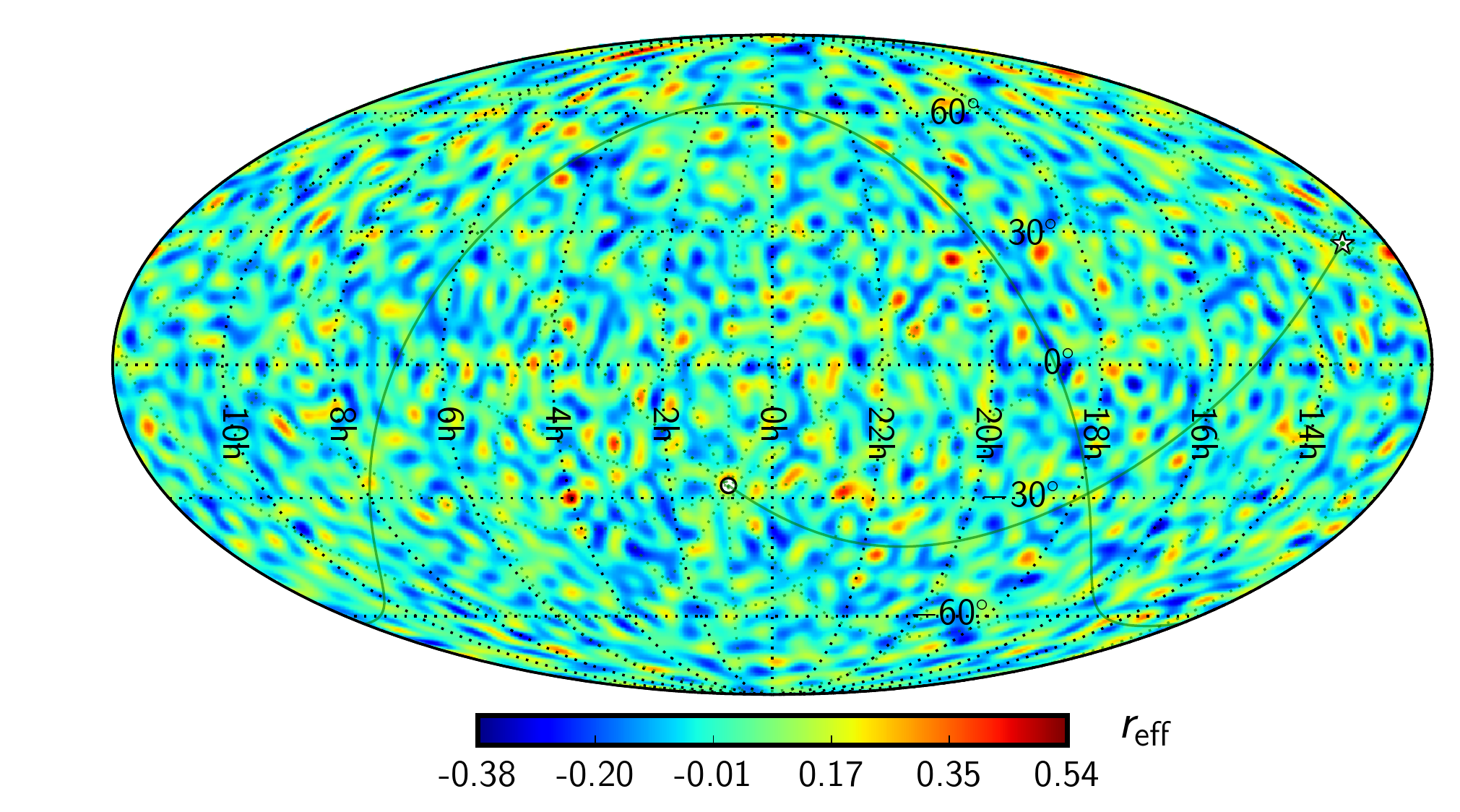}}
\subfigure[CMB + Noise  + Foreground]{\label{fig:forecast_data_pam}\includegraphics[width=0.5\columnwidth]{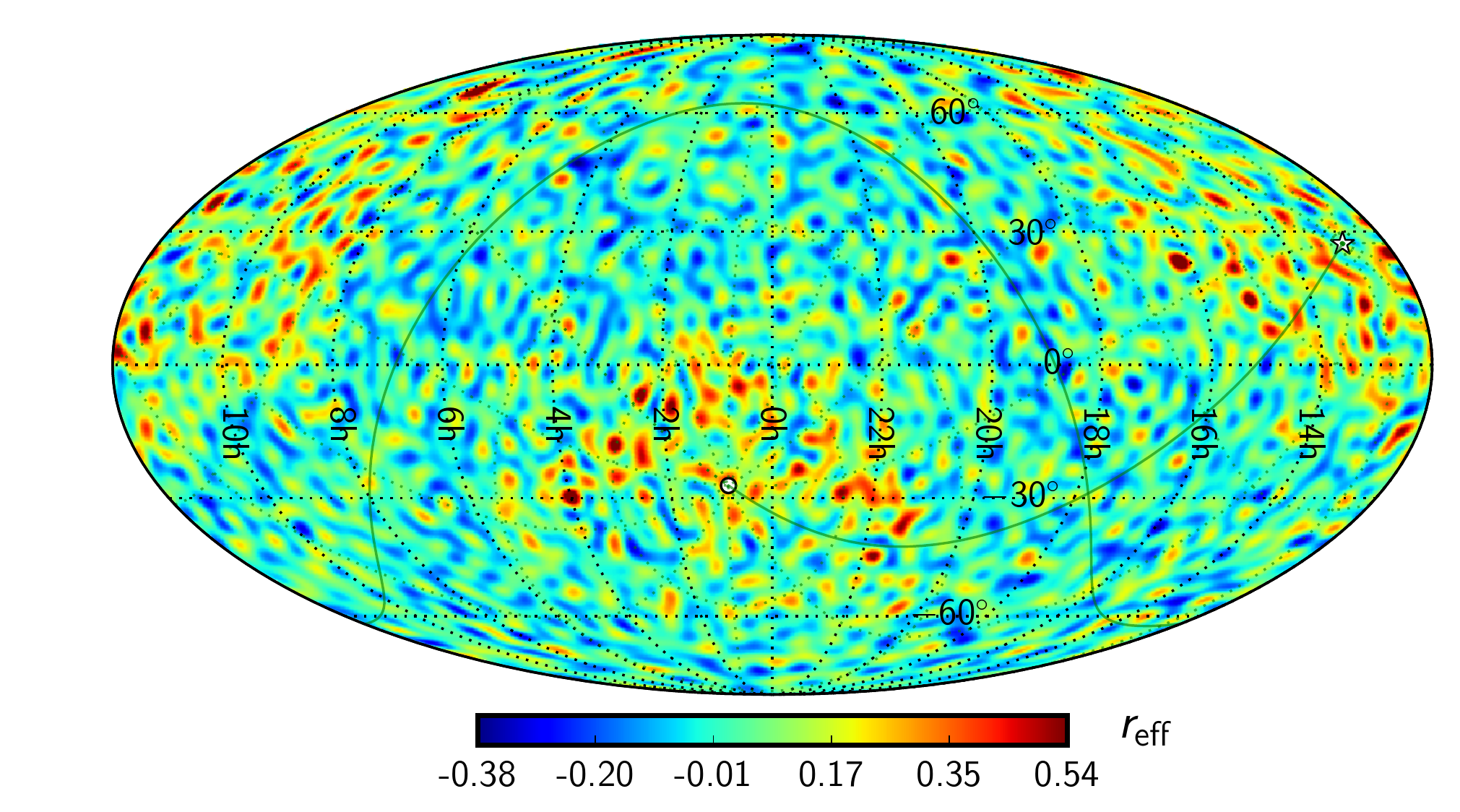}}
\subfigure[Foreground template]{\label{fig:forecast_frg_pam}\includegraphics[width=0.5\columnwidth]{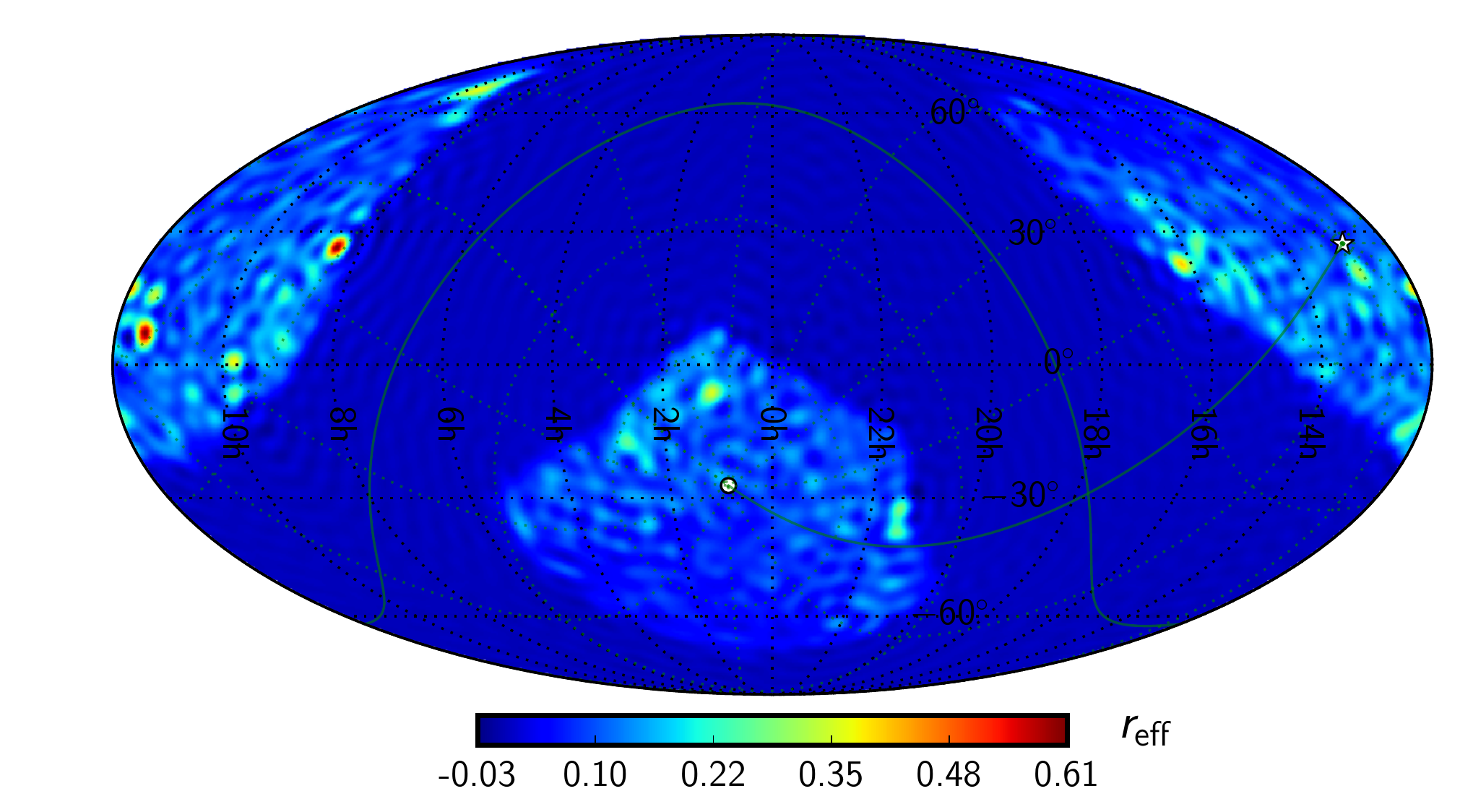}}
\subfigure[Spectra of reconstruction]{\label{fig:forecast_pam_spectra}\includegraphics[width=0.5\columnwidth]{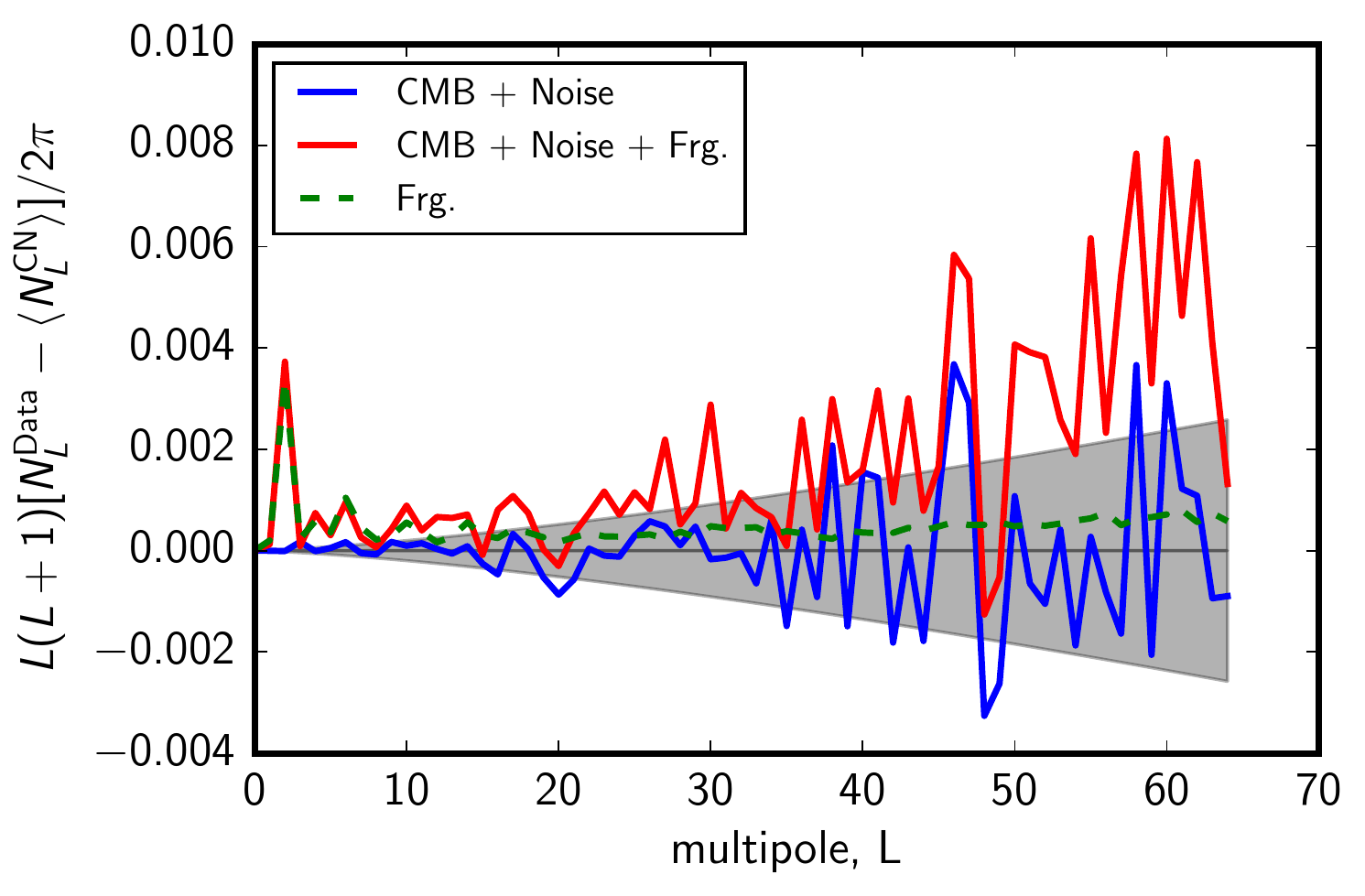}}
\caption{For a futuristic experiment, the power anisotropy maps and the corresponding spectra estimated from application of the optimal Bipolar Spherical Harmonic estimator $\hat P$ on simulated data. \textit{Top left:}  Estimated power anisotropy map from CMB $+$ noise map. \textit{Top right:} Estimated power anisotropy map from CMB $+$ noise $+$ foreground map. The color scale is set same as in the figure on the top left.  \textit{Bottom left:} Estimated power anisotropy map from the foreground template.   \textit{Bottom right:} Bias subtracted power spectra of the estimated maps. The grey band marks the $1\sigma$ Gaussian error bars given by $\sqrt{\frac{2}{2L+1}} \langle N_L^{CN} \rangle$. }
\label{fig:pam_forecast}
\end{figure}

{The foregrounds however are obvious when we evaluate the optimal $\hat P$ estimator on these maps, with details matching those discussed in Section \ref{power_anisotropy_estimator}. The resultant estimates of the foreground squared and the corresponding power spectra are presented in Fig.~\ref{fig:pam_forecast}. Note that the theoretically expected monopole is subtracted from the power anisotropy maps estimated from simulations. Specifically, Fig.~\ref{fig:forecast_frg_pam} depicts the power anisotropy map derived from the foreground template.  The inferred $\rm r _{\rm eff}$ is expectedly very noisy and consequently the reconstructed foreground squared map shown in Fig.~\ref{fig:forecast_data_pam} cannot uncover the exact features seen in Fig.~\ref{fig:forecast_frg_pam}. However note that the power anisotropy map is able to clearly discern the contaminated regions from the clean regions as apparent from comparing Fig.~\ref{fig:forecast_cn_pam} and Fig.~\ref{fig:forecast_data_pam}.}

We also compare the bias subtracted power spectra of these maps in Fig.~\ref{fig:forecast_pam_spectra}. Note that the blue spectra is consistent with zero within errors as expected while the red spectra is systematically shifted towards positive values owing to the excess power contributed by the presence of foregrounds. This excess power is particularly seen at high SNR at low multipoles, where it also matches the green spectra very well.

This exercise demonstrates that the proposed estimators can be effectively used even when the foreground power is below or comparable to CMB and noise.

\end{document}